\documentclass[draft]{agujournal2019}
\usepackage{url} 
\usepackage{lineno}
\usepackage[inline]{trackchanges} 
\usepackage{soul}
\usepackage{amsfonts,amsmath} 
\usepackage{multirow,hhline}
\usepackage{mathtools}
\usepackage{colortbl} 
\usepackage[table]{xcolor}     
\linenumbers
\usepackage[colorinlistoftodos]{todonotes}
\usepackage{subcaption}

\usepackage{graphicx}

\draftfalse

\journalname{Journal of Geophysical Research: Machine Learning and Computation}

\begin{document}
\nolinenumbers

\title{A Neural Operator Emulator for Coastal and Riverine Shallow Water Dynamics}

\authors{Peter Rivera-Casillas\affil{1,2}\thanks{peter.g.rivera-casillas@erdc.dren.mil}, Sourav Dutta\affil{3}, Shukai Cai\affil{3}, Mark Loveland\affil{1,3},
Kamaljyoti Nath\affil{4},
Khemraj Shukla\affil{4},
Corey Trahan\affil{1}, Jonghyun Lee\affil{2}, Matthew Farthing\affil{5}, Clint Dawson\affil{3}}

\affiliation{1}{Information Technology Laboratory, U.S. Army Engineer Research and Development Center, MS, USA}
\affiliation{2}{Civil, Environmental and Construction Engineering \& Water Resources Research Center, University of Hawai’i at Manoa, HI, USA}
\affiliation{3}{Oden Institute for Computational Engineering and Sciences, The University of Texas at Austin, TX, USA}
\affiliation{4}{Division of Applied Mathematics, Brown University, Providence, RI, USA}
\affiliation{5}{Coastal and Hydraulics Laboratory, U.S. Army Engineer Research and Development Center, MS, USA}

\correspondingauthor{Peter Rivera-Casillas}{peter.g.rivera-casillas@erdc.dren.mil}

\begin{keypoints}
\item We present MITONet, an autoregressive neural operator framework that emulates shallow-water dynamics in coastal and riverine regions. 
\item MITONet shows superior predictive skill and stability in comparison to other well-known neural operator models.
\item MITONet shows strong performance on two realistic shallow-water flow problems and has the potential to extend to other applications.
% \item The MITONet framework can be extended to other applications.
\end{keypoints}

\begin{abstract}
Coastal regions and river floodplains are particularly vulnerable to the impacts of extreme weather events. Accurate real-time forecasting of hydrodynamic processes in these areas is essential for infrastructure planning and climate adaptation. Yet high-fidelity numerical models are often too computationally expensive for real-time use, and lower-cost approaches, such as traditional model order reduction algorithms or conventional neural networks, typically struggle to generalize to out-of-distribution conditions. In this study, we present the Multiple-Input Temporal Operator Network (MITONet), a novel autoregressive neural emulator that employs latent-space operator learning to efficiently approximate high-dimensional numerical solvers for complex, nonlinear problems that are governed by time-dependent, parameterized partial differential equations. We showcase MITONet's predictive capabilities by forecasting regional tide-driven dynamics in the Shinnecock Inlet in New York and riverine flow in a section of the Red River in Louisiana, both described by the two-dimensional shallow-water equations (2D SWE), while incorporating initial conditions, time-varying boundary conditions, and domain parameters such as the bottom friction coefficient. Despite the distinct flow regimes, the complex geometries and meshes, and the wide range of bottom friction coefficients studied, MITONet displays consistently high predictive skill, with anomaly correlation coefficients above 0.9, a maximum normalized root mean square error of 0.011, and computational speedups between $100\text{x}-1,250\text{x}$, even for $175$ days of autoregressive rollout forecast from random initial conditions and with unseen parameter values.

\end{abstract}

\section*{Plain Language Summary}
Coastal and riverine hydrodynamics can be modeled accurately through numerical simulations. However, traditional numerical solvers are computationally intensive and require domain expertise. In this study, we introduce a machine learning framework, MITONet, which learns the governing physics from simulation data. Once trained, MITONet can quickly generate accurate predictions under similar or changing conditions, offering a faster, more accessible tool for planning and decision-making.

\section{Introduction}

Coastal regions are densely populated and host critical infrastructure vulnerable to the impacts of rising sea levels and increased frequency of extreme weather conditions \cite{oppenheimer2019sea}. River floodplains, which provide essential resources for society, are likewise susceptible to extreme weather, and human-induced disturbances \cite{arnell2016impacts, alifu2022enhancement}. High-fidelity numerical models of the two-dimensional (2D) depth-averaged shallow water equations (SWE) are typically employed to capture coastal circulation patterns and riverine hydrodynamics in order to aid in crucial decision-making and planning efforts \cite{pachev2023adcirce3sm,pevey2020adhcolumbia}. Despite recent technological and algorithmic advances, these methods can still be computationally expensive, time consuming, and technically challenging. Machine learning (ML) approaches have been successful in addressing some of these challenges due to their flexibility and computational efficiency \cite{abouhalima2024machine, giaremis2024storm, karim2023review,  pachev2023framework,song2023surrogate,liu2024bathymetry,dutta2021data}. 

Predicting behaviors of highly nonlinear systems remains a persistent challenge in computational modeling. Although the universal function approximation property of neural networks has allowed deep learning to make substantial progress in the development of surrogate models \cite{brunton_review_2019,karniadakis_review_2021,dutta_farthing_2021,dutta_rc_2021}, achieving consistent generalizability remains a challenge. To this end, operator approximation has recently emerged as an increasingly popular field of study in deep learning \cite{azizzadenesheli2024neural}. The earliest attempts included efforts like nonlinear integro-differential operator regression \cite{patel2018nonlinear}, where deep neural networks (DNNs) and Fourier pseudospectral methods were employed to recover complex operators from data, and physics-informed neural networks (PINNs), where DNNs were used to regress partial differential equation (PDE) models by exploiting physics knowledge \cite{raissi2019physics}. While useful in certain applications, these and other earlier approaches had limitations in scalability and generalizability across varying domains of physics, highlighting the challenges of learning high-dimensional systems and multiscale features.

The introduction of the Deep Operator Network (DeepONet) \cite{lu2021learning} marked a new era in operator approximation. DeepONet addressed many of the issues from earlier approaches by leveraging the universal approximation theorem for operators \cite{chen1995universal}, enabling the network to efficiently approximate maps between infinite-dimensional function spaces. Subsequent developments in neural operators include the Fourier neural operator (FNO) \cite{li2020fourier}, graph neural operators \cite{li2020neural}, transformer neural operators \cite{cao2021choose}, and the variationally mimetic operator network \cite{patel2024variationally}, among others, many of these with multiple extensions. 

Another critical challenge in data-driven modeling of nonlinear, real-world physical systems that has been an active area of research is the ability to forecast over long lead times \cite{10.1145/3533382, bodnar2024aurora, lam2022graphcast, lim2021time}. Recent advances, including the Diffusion-inspired Temporal Transformer Operator (DiTTO)\cite{ovadia2023ditto}, the Universal Physics Transformer \cite{alkin2024universal}, and OceanNet \cite{chattopadhyay2024oceannet}, demonstrate the effectiveness of neural operator-based autoregressive models in overcoming some of these challenges by addressing the spectral bias. DiTTO uses temporal bundling to jointly predict multiple future steps, which reduces recursive error accumulation and helps correct the small-scale spectral bias that typically appears in autoregressive rollouts. The Universal Physics Transformer conditions its prediction blocks through feature modulation, allowing the network to adapt across temporal and dynamical regimes and better represent multiscale behavior. OceanNet couples an FNO backbone with a PEC-style convergent integration scheme and an explicit spectral regularizer to mitigate autoregressive error growth and the small-scale spectral bias.

This research focuses on the DeepONet architecture and subsequent extensions \cite{jin2022mionet, kontolati2023learning, ovadia2023ditto, wang2022improved} that demonstrate improved performance. We specifically focus on the applications to modeling real-world coastal ocean circulation and river dynamics. An application to ocean dynamics was previously explored in \citeA{10337380}, where latent DeepONet and FNO are used for one-shot forecasting of ocean dynamics. Similarly, \citeA{choi2024applications} compared FNO variations and explored a recursive approach for one-shot forecasting of ocean dynamics given an initial condition (IC). Both efforts demonstrate the potential of neural operators to predict ocean dynamics; however, their usefulness is limited by their structure and data requirements, relying solely on initial ocean states without incorporating external forcings as boundary conditions (BC), which are crucial for real-world scenarios. Additionally, neither handles varying domain parameters, limiting their ability to generalize in dynamic environments. Another recent study developed an FNO-based digital twin model that autoregressively predicts the sea-surface height of the northwest Atlantic Ocean’s western boundary current \cite{chattopadhyay2024oceannet}, using IC and external BC forcings like wind stress and tides. The model performs comparably to traditional numerical models over long forecasting windows and offers a significant computational speed-up. However, its predictions are limited to sea-surface height and do not include currents, restricting its ability to fully address comprehensive hydrodynamic forecasting.

To our knowledge, prior applications of neural operators in modeling riverine hydraulics has largely been focused on predicting flood-inundation maps, water surface elevation (stage), and point/transect forecasts, rather than simultaneous space–time prediction of the shallow-water state variables. In \citeA{sun2023rapid}, a hybrid-FNO model that combines process-based modeling with data-driven learning predicts flood extent and inundation depth but not flow velocities. In \citeA{sun2024bridging}, DeepONet models are used for forward and inverse problems, predicting streamflow at specific gauge locations and its inverse mapping. In \citeA{pang2024efficient}, a FNO architecture combined with convolutional long short-term memory (ConvLSTM) networks yields a surrogate for the Hydrologic Engineering Center River Analysis System (HEC-RAS) model, that predicts time histories of water surface elevation and streamflow at specified transects. Similarly, \citeA{holmberg2025accelerating} proposes a hybrid autoregressive model combining Gated Recurrent Units (GRU) with a geometry-aware FNO as a fast surrogate for HEC-RAS, producing forecasts across cross-sections rather than full 2D fields. In another interesting recent work, \citeA{chen2025graph} uses a graph-enhanced neural operator to fill gaps in image-based river surface velocimetry. Overall, these approaches typically target specific quantities of interest (e.g., inundation extent, stage at gauges, or cross-section discharge) rather than learning the full space–time evolution of all shallow-water state variables. As a result, they function more as task-specific surrogates rather than as general hydrodynamic emulators, thereby limiting their ability to replace expensive, high-fidelity numerical solvers in engineering applications that require multiple full-field forecasts or the ability to generalize across changing forcings, boundary conditions, and parameter values. 

We introduce a new operator learning framework for space-time forecasting of shallow-water hydrodynamics in different flow regimes, denoted as the Multiple-Input Temporal Operator Network (MITONet), that is capable of parametric generalizability and long-horizon autoregressive forecasting. MITONet adopts several key architectural designs and implementation strategies to address some of the well-known limitations of neural network-based models, as described before, and ensure versatility and robustness across various applications. To improve computational efficiency during training and inference and improve convergence during training, MITONet leverages operator learning in a lower-dimensional latent space, defined by a pre-trained autoencoder network. To effectively learn the causal relationships underlying PDE solution operators that model the effect of multiple input functions (such as IC, BC and domain parameters) on the solution function, MITONet employs multiple branch networks to separately encode each relevant input function. To improve model expressivity and avoid the curse of vanishing gradients, MITONet uses additional so-called "encoder" networks that facilitate the propagation of input feature information between all the branch and trunk networks. Finally, to enable long-term autoregressive rollout prediction by mitigating accumulation of error, MITONet adopts temporal bundling during training and inference. The novelty of this work lies in the fusion of these different strategies to develop a structured neural operator framework that can be widely applicable in efficiently emulating the PDE solution operators of complex environmental flow problems.

The rest of the paper is organized as follows. In Section \ref{sec:methods}, we provide a brief review of neural operator methods, specifically the DeepONet model, and then introduce the proposed MITONet framework in detail. In Section \ref{sec:problem_desc}, we provide details about the two computational experiments used to evaluate the capabilities of the proposed framework. Section \ref{sec:results} features detailed discussions of the results of the computational experiments. Specifically, we outline the various error metrics used for quantitative assessments of model skill and compare the performance of the MITONet framework with other well-known DeepONet-based architectures for a tidal circulation problem in Section \ref{sec:shinn_model_comparison}. We also evaluate parametric extrapolation capabilities, long-term forecast skill, and the ability to forecast from unseen random as well as zero ICs in Sections \ref{sec:shinn_model_skill}, \ref{sec:shinn_long_term} and \ref{sec:shinn_random_init}. To demonstrate versatility of the framework, we repeat many of the above experiments with a second computational example involving riverine hydrodynamics in Section \ref{sec:redriver_results} and discuss the salient features and limitations of the proposed framework in Section \ref{sec:discussion}. Finally, in Section \ref{sec:conclusion}, the concluding remarks are presented and future work ideas are outlined.

\section{Methods}\label{sec:methods}
\subsection{Deep Operator Network}
Operator learning focuses on approximating maps between infinite-dimensional functions. In mathematical terms, given data pairs $(f,s)$, where $f \in \mathcal{U}\left(\mathcal{X};\mathbb{R}^{N_{d1}}\right)$ and $s \in \mathcal{V}\left(\mathcal{Y};\mathbb{R}^{N_{d2}}\right)$ are function spaces, and a potentially nonlinear operator $\cal G : \mathcal{U} \mapsto \mathcal{V}$ such that $\mathcal{G}(f) = s$, the objective is to find a neural network-based approximation $\mathcal{G}_{NN}$ such that for any new data $f^\prime \in \mathcal{U}$, we have $\mathcal{G}_{NN}( f^\prime; \boldsymbol{\theta}) \approx \mathcal{G}(f^\prime)$, where $\boldsymbol{\theta} \in \mathbb{R}^M$ are the parameters of the neural operator. DeepONet seeks to approximate this operator using an i.i.d collection of input-output function pairs $(f_{(i)}(\mathbf{x}), s_{(i)}(\mathbf{y}))_{i=1}^{N_{tr}}$, evaluated over discrete sets of spatial coordinates $\mathbf{x} \in \mathcal{X}$ and $\mathbf{y} \in \mathcal{Y}$, respectively. The standard (unstacked) DeepONet architecture \cite{lu2021learning} comprises two sub-networks denoted as the branch $\mathbf{B}$ and the trunk $\mathbf{T}$ that generate finite-dimensional encodings of the input function and the output coordinates, respectively, as $\mathbf{B}(f(\mathbf{x})) = [b_1, b_2, \dots, b_p]^T \in \mathbb{R}^p$ and $\mathbf{T}(\mathbf{y}) = [t_1, t_2, \dots, t_p]^T \in \mathbb{R}^p$. The output function is approximated as 

\begin{equation}
(\mathcal{G}_{NN}(f))(\mathbf{y}) \approx \mathbf{B}(f(\mathbf{x}); \theta_B) \odot \mathbf{T}(\mathbf{y}; \theta_T) + b_0,
\end{equation}
\noindent where $b_0$ is an optional bias term and $\odot$ is the Hadamard product (element-wise multiplication). Finally, the network parameters, $\theta_B$ and $\theta_T$ are optimized by minimizing a loss function $\mathcal{L}$ that measures the discrepancy between the data and the prediction in a desired norm,
\begin{equation}
\min_{\{\theta_B, \theta_T, b_0\}} \sum_{(f,s) \in \text{data}} \mathcal{L}\Big((\mathcal{G}_{NN}(f))(\mathbf{y}), \, s(\mathbf{y}) \Big).
\label{eq:operator_loss}
\end{equation}

In this work, we are interested in applying this framework to solve a parametric, initial boundary value problem i.e. modeling the time evolution of the solution operator of a PDE $(s(\mathbf{x},t) \mapsto s(\mathbf{x},t+\delta t)),\; \forall \mathbf{x}\in \Omega$, over a forecasting horizon, $t \in (0,T]$, given IC: $s(\mathbf{x},0)$, BC: $s(\mathbf{x},t)\vert_{\Gamma}$, and domain parameters, $\boldsymbol\mu$. Here $\Omega$ is a domain in  $\mathbb{R}^{N_{d2}}$ with boundary $\Gamma$.

Several variants of DeepONet have been proposed to improve the training robustness and enhance the prediction accuracy for unseen input functions. The Modified Deep Operator Network (M-DeepONet) \cite{wang2022improved} introduced two additional encoder networks to the architecture, one for the branch and the other for the trunk network. These encoders, uniquely interconnected, facilitated an information exchange between the two primary networks. This, coupled with slight adjustments to the forward pass, improved expressiveness and mitigated vanishing gradient problems , although at the cost of increased computational demands. The Latent Deep Operator Network (L-DeepONet) \cite{kontolati2023learning} was introduced to address the scaling challenges of high-dimensional data sets. In this method, autoencoders were employed to embed the data set in a lower-dimensional manifold, and the DeepONet was tasked with learning the functional relationships in this latent space. 

Another crucial limitation of using the standard DeepONets for approximating a PDE solution operator is its inability to handle multiple input functions, such as ICs and BCs. The multiple input operator network (MIONet) addressed this by combining multiple input branches into a single Operator Network \cite{jin2022mionet}. While MIONet provides a framework to handle different ICs, this is mostly useful in autoregressive predictions. Traditionally, DeepONet methods are designed to map an IC to all future times, and typically have limited skill over longer forecast periods. Techniques like temporal bundling can improve stability and enhance rollout time \cite{brandstetter2022message, ovadia2023ditto} by predicting multiple time steps synchronously during training and inference, which reduces solver calls and thus minimizes autoregressive error propagation. 

\subsection{Multiple Input Temporal Operator Network}

\begin{figure}[h]
    \centering
    \includegraphics[width=.9\textwidth]{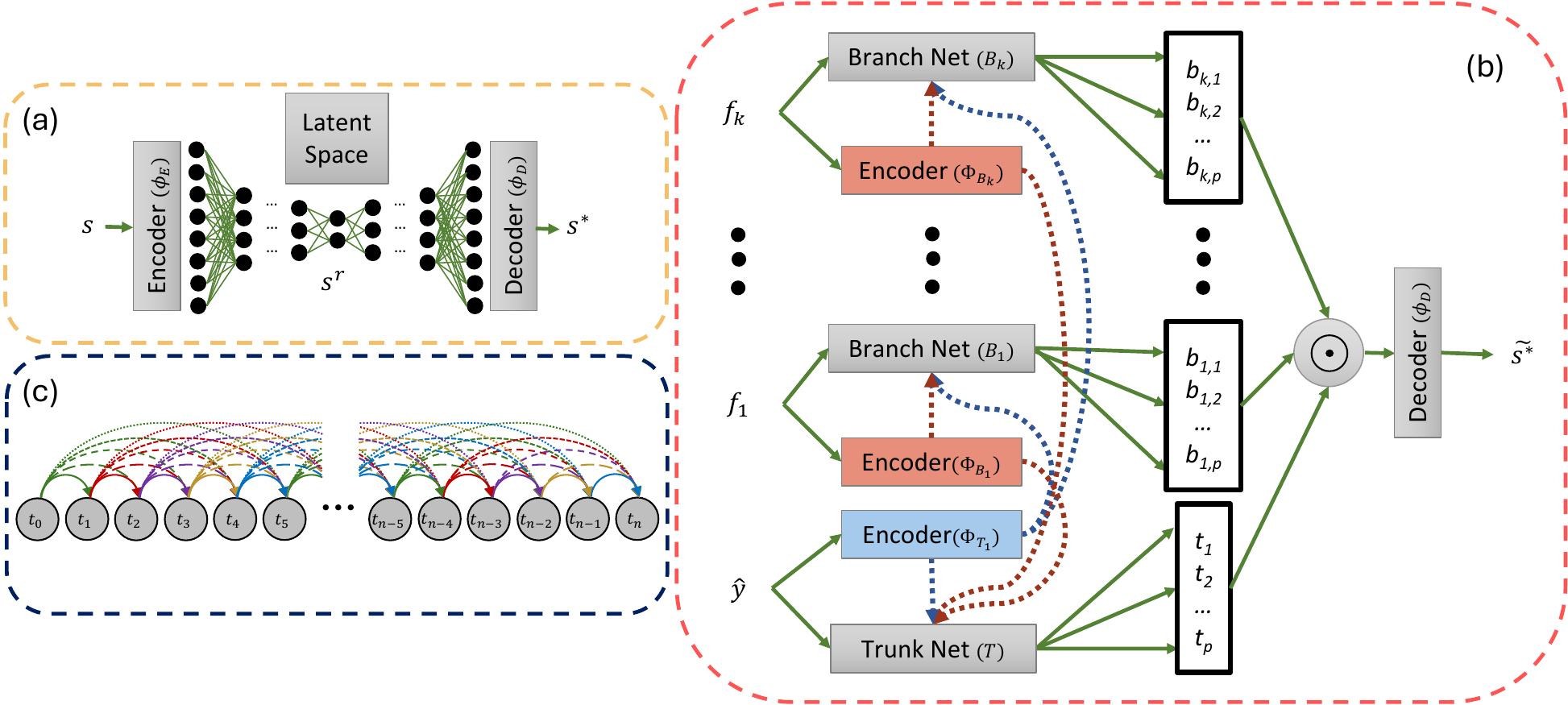}
    \caption{Schematic representation of the MITONet framework. First, (a) an autoencoder maps the high-dimensional solution snapshots to a low-dimensional latent space. Then, (b) the MITONet is provided with relevant input functions, such as domain parameters, initial conditions or their latent representation at a given time $t$, and boundary conditions for time $t+\beta \Delta t$ to predict the latent representation of the solution snapshot at time $t + \beta \Delta t$, where $\beta = 1, \ldots, \tau$ and $\tau$ is a chosen  window. This procedure is repeated autoregressively to generate a time series of snapshots in the latent space, which are then passed through the decoder network to recover the predictions in the high-dimensional space. To train the MITONet model, a temporal bundling approach is adopted to split the time series of training snapshots into sub-trajectories consisting of $\tau$ outputs for each input snapshot. For $\tau$=$5$, panel (c) illustrates the different sub-trajectories using different colors.}
    \label{fig:mitonet}
\end{figure}

The MITONet framework proposed here effectively integrates dimensionality reduction, multi-input handling, and temporal bundling, into a cohesive framework optimized for long-term hydrodynamic forecasting. Figure \ref{fig:mitonet} provides a schematic of the different components of the MITONet architecture. 
First, the discretized PDE solution, $\mathbf{s} \in \mathbb{R}^{N_s}$, defined on a high-dimensional computational mesh $\mathcal{M}_h$, is embedded into a lower-dimensional latent space using an autoencoder network, $\Phi_{AE}$ (see Figure \ref{fig:mitonet}(a)). Specifically, $\mathbf{s} \in \mathbb{R}^{N_s}$ is mapped to $\mathbf{s}^r \in \mathbb{R}^{N_r}$ such that $N_r \ll N_s$ using an encoder network, $\phi_{E}(\mathbf{s}; \theta_E): \mathbf{s} \rightarrow \mathbf{s}^r$, and a decoder network is used to reconstruct the high-dimensional solution,  
$\phi_{D}(\mathbf{s}^r;\theta_D): \mathbf{s}^r \rightarrow \mathbf{s}^*$ such that $\mathbf{s}^* \approx \mathbf{s}$. The autoencoder $\Phi_{AE}: \phi_{D} \circ \phi_{E}(\cdot)$ is trained by minimizing the mean square reconstruction error
\begin{equation}
\min_{\{\theta_E, \theta_D\}} \frac{1}{N_{tr}} \sum_{j=1}^{N_{tr}} \|\mathbf{s}_{(j)} - \mathbf{s}_{(j)}^*(\theta_E, \theta_D)\|_{L^2(\Omega)}^2.
\end{equation}

Following the L-DeepONet \cite{kontolati2023learning} philosophy, once the autoencoder is trained, a neural operator is designed to approximate the latent representation of the solution variable. Consequently, the choice of a dimension reduction approach, while not limited to autoencoders, should be made judiciously as the performance of MITONet will be limited by the latent representation error.

Second, multiple inputs are handled by introducing multiple independent branch networks, following the low-rank MIONet design \cite{jin2022mionet}. Each input function $f_k$ is encoded by a branch network as $\mathbf{B}_k(f_k(\mathbf{x}_{k}))=[b_{k,1}, b_{k,2}, \dots, b_{k,p}]^T \in \mathbb{R}^p$ where $\mathbf{x}_{k} \in \mathcal{X}_k$ are its discrete evaluation points (see Figure \ref{fig:mitonet}(c)). The trunk network is trained to encode only the temporal part of the output function domain, $\widehat{\mathbf{y}} = \mathbf{y}\vert_t$ as $\mathbf{T}(\widehat{\mathbf{y}}) = [t_1, \ldots, t_p]^T \in \mathbb{R}^p$ since the spatial encoding is performed via the pre-trained autoencoder. Combining the above, the latent solution vector is approximated as
\begin{equation}\label{eq:MITONet_ouput}
\mathbf{s}^r \approx (\mathcal{G}_{NN}(f_1, \ldots, f_k))(\widehat{\mathbf{y}}) = \mathbf{B}_1(f_1(\mathbf{x}_1)) \odot \ldots \odot \mathbf{B}_k(f_k(\mathbf{x}_k)) \odot \mathbf{T}(\widehat{\mathbf{y}}) + \mathbf{b}_0,
\end{equation}
where $\mathbf{b}_0 \in \mathbb{R}^p$ is an optional, trainable bias vector. In practice, $p$ is either chosen to be $dim(\mathbf{s}^r) = N_r$ or the right hand side of (\ref{eq:MITONet_ouput}) is multiplied by a trainable matrix, $\mathbf{P} \in \mathbb{R}^{N_r \times p}$ to achieve dimensional parity.

Third, the MITONet architecture is endowed with interconnected encoder networks following the M-DeepONet idea. However, unlike \citeA{wang2022improved}, to facilitate information exchange across multiple inputs, a branch encoder, $\Phi_{B_k}$ is introduced for each branch, $B_k$, and a trunk encoder $\Phi_{T}$ for the trunk network, $T$. These encoders are trained to generate embeddings for the corresponding branch and trunk input functions as given by, 
\begin{eqnarray}
\begin{split}
\mathbf{U}_k &\coloneq \Phi_{B_k}(f_k(\mathbf{x}_k)) = [u_{k,1}, u_{k,2}, \dots, u_{k,q}]^T \in \mathbb{R}^q, \\
\mathbf{W} &\coloneq \Phi_{T}(\mathbf{y}) = [v_1, v_2, \dots, v_q]^T \in \mathbb{R}^q.
\label{eq:MITONet_encoder}
\end{split}
\end{eqnarray}

The purpose of these encoder networks is to improve resilience against vanishing gradients during the training of MITONet by enhancing information propagation through the branch and trunk networks until the information fusion operation in (\ref{eq:MITONet_ouput}). This is enabled by modifying the forward pass of the MITONet as follows
\begin{eqnarray}\begin{split}
\mathbf{H}^{(l+1)}_{B_k} &= (\mathbf{1} - \boldsymbol{\Psi}^{(l+1)}_{B_k}(\mathbf{H}^{(l)}_{B_k})) \odot \mathbf{U_k} + \boldsymbol{\Psi}^{(l+1)}_{B_k}(\mathbf{H}^{(l)}_{B_k}) \odot \mathbf{W}, \\
\mathbf{H}^{(l+1)}_T &= (\mathbf{1} - \boldsymbol{\Psi}^{(l+1)}_{T}(\mathbf{H}^{(l)}_T)) \odot \mathbf{U}_1 \odot \mathbf{U}_2 \odot \dots \odot \mathbf{U}_k + \boldsymbol{\Psi}^{(l+1)}_{T}(\mathbf{H}^{(l)}_T) \odot \mathbf{W},
\label{eq:MITONet_hidden}
\end{split}\end{eqnarray}
where $\mathbf{H}^{(l)}_{B_k}, \mathbf{H}^{(l)}_{T}$ represent the output and $\boldsymbol{\Psi}^{(l)}_{B_k}, \boldsymbol{\Psi}^{(l)}_{T}$ denote the action of the $l^{th}$ hidden layer of the $k^{th}$ branch and the trunk networks, respectively. It is worth noting that (i) for a $L-$layer MITOnet, $\mathbf{H}^{(L)}_T, \mathbf{H}^{(L)}_{B_k} \in  \mathbb{R}^p$ using (\ref{eq:MITONet_ouput}), (ii) $\mathbf{H}^{(l)}_T, \mathbf{H}^{(l)}_{B_k} \in  \mathbb{R}^q \, \forall \, l=1,\ldots L-1$ using (\ref{eq:MITONet_encoder}) and (\ref{eq:MITONet_hidden}), and (iii) the forward pass in (\ref{eq:MITONet_hidden}) is designed to preserve the multi-input structure of the governing PDE, but can be easily modified based on any available information about the underlying input-output functional relationships. 

Finally, MITONet adopts a temporal bundling strategy \cite{brandstetter2022message} to enable long-term forecasting skill. As shown in Figure \ref{fig:mitonet}(b), the model is trained to learn the mapping $\mathbf{s}^r(\cdot, t) \rightarrow \mathbf{s}^r(\cdot, t+ \beta \Delta t) \, \forall \, \beta = \{1, \ldots, \tau\}$, where $\tau$ is a chosen  window such that $t + \tau \Delta t \leq T$. Hence, each training time series is split into $T/\Delta t - \tau + 1$ sub-trajectories of length $\tau$. During inference, MITONet generates $\tau$ predictions, $\hat{\mathbf{s}}^r(\cdot, t) \rightarrow \hat{\mathbf{s}}^r(\cdot, t+ \beta \Delta t) \, \forall \, \beta = \{1, \ldots, \tau\}$, and uses $\hat{\mathbf{s}}^r(\cdot, t+ \tau \Delta t)$ as IC to autoregressively predict the following $\tau$ time steps. 

\section{Problem Description}\label{sec:problem_desc}

The proposed MITONet framework is evaluated using high-fidelity numerical simulations of the 2D SWE in two distinct flow regimes: a coastal circulation problem driven by time-dependent tidal boundary conditions, and a riverine flow problem with time-dependent inflow and tailwater elevation boundary conditions, as described in the following section.

\subsection{Shinnecock Inlet}\label{sec:shinnecock_details}
\begin{figure}[h]
    \centering    \includegraphics[width=0.99\textwidth]{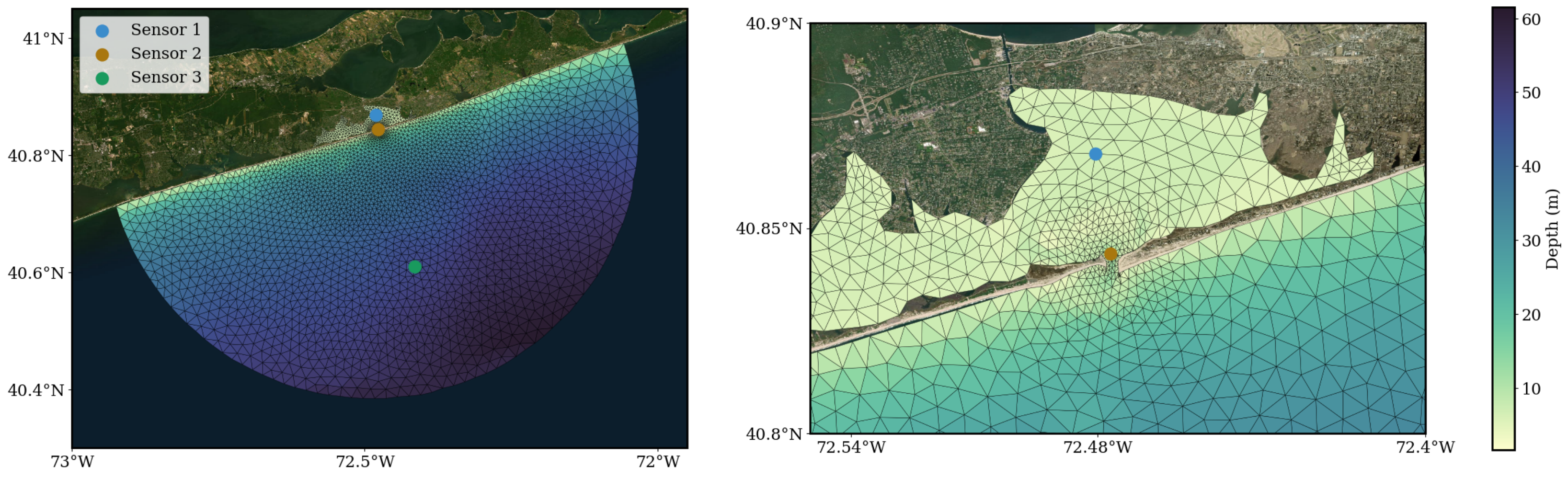}
    \caption{Full domain and zoomed-in view of the finite element mesh used by ADCIRC and bathymetry for the Shinnecock Inlet example. The model predictions are evaluated at three sensor locations that exhibit distinct flow dynamics. Sensor locations are marked with different colors: sensor 1 (blue) - inner bay, sensor 2 (orange) - inlet, and sensor 3 (green) - bay.}
    \label{fig:shinnecock_bathy}
\end{figure}

For the first computational example, the Advanced Circulation (ADCIRC) modeling suite (Text S1) \cite{luettich1992adcirc}, which solves the vertically-integrated Generalized Wave Continuity Equation (GWCE) for water surface elevation \cite{luettich2004formulation}, is used to simulate tidal hydrodynamics in the vicinity of Shinnecock Inlet, a geographical feature located along the outer shore of Long Island, New York. This example was derived from a study conducted by the US Army Corps of Engineers Coastal Hydraulics Laboratory \cite{morang1999shinnecock, militello2001shinnecock}, and is a commonly used test case for ADCIRC. The simulation utilizes a finite element grid containing $5780$ elements and $3070$ nodes, with the grid discretization varying from approximately $2$ km offshore to around $75$ m near the inlet and near-shore areas, which helps capture the challenging and complex circulation patterns near the inlet and the back bay. Figure \ref{fig:shinnecock_bathy} shows a visualization of the simulation grid overlaid on to the bathymetry field. 

The model uses an off-shore open boundary with prescribed elevations and a mainland boundary with zero normal flow. The model is forced using tidal elevations at the open boundary from $5$ tidal constituents (M2, S2, N2, O1, K1), ramped up over the first two days from a state of rest. 
Quadratic friction forcing is applied and wetting/drying is neglected. The simulation duration is $60$ days with a $6$-second time step, while water level and velocity fields are output every $30$ minutes, resulting in a total of $2,880$ time snapshots ($N_t$), henceforth denoted by $\{t_i\}_{i=1}^{N_t}$. A $60$-day simulation takes about $20$ minutes on two Frontera CPUs (Intel 8280, 2.7GHz) \cite{stanzione2020frontera}. 

The simulation data is parametrized by varying the scalar bottom friction coefficient ($r$), which enters the depth-averaged momentum equations via a quadratic bottom stress term (Text S1). Physically, $r$ reflects bed roughness (e.g., substrate, vegetation, or land cover) and acts as a momentum sink. Because bottom friction is a key control on momentum dissipation in shallow-water modeling, emulators that explicitly parameterize $r$ are especially useful for inverse problems (e.g., roughness calibration from observations) \cite{butler2015definition}. Its influence on the flow solution is evident in Figures S8-S10, which compare snapshots at the same timepoint for simulations rolled out from a common IC but with different values of $r$. To generate the full parametric dataset, $60$-day simulations are run for $14$ different values of $r$ ranging from $0.0025$ to $0.1$, such that the total wall-clock time to generate the entire dataset turns out to be about $14 \times 20\, \text{min} = 280\, \text{min} \approx 4.67\, \text{hrs}$. 

To enable model extrapolation across space and time, the data is carefully split into training, validation, and testing as described in Table \ref{tab:shinn_train_val_test}. The training set is used to learn the model parameters, the validation set is used for hyperparameter optimization and learning rate reduction, and the test set is reserved for the final evaluation. It is to be noted that the minimum and maximum parameter values in both the validation ($r = 0.00275, 0.075$) and the test set ($r=0.0025, 0.1$) are intentionally chosen to lie outside the range of parameter values in the training set ($0.003 \leq r \leq 0.05$), in order to minimize overfitting during training as well as allow for evaluation of the parametric extrapolation capabilities of the trained models.

\begin{table}[h]
\caption{Shinnecock Inlet Training, Validation, and Testing Data}
\centering
\begin{tabular}{l c c}
\hline
      & \textbf{Bottom Friction Coefficient,} $\boldsymbol{r}$                 & \textbf{Days (time steps)} \\
\hline
\textbf{Training}   & 0.003, 0.005, 0.01, 0.02, 0.03, 0.04, 0.05  & 15-30 (720) \\
\textbf{Validation} & 0.00275, 0.0075, 0.025, 0.075               & 5-15 (480) \\
\textbf{Testing}    & 0.0025, 0.015, 0.1                          & 5-60 (2,640) \\
\hline
\end{tabular}
\label{tab:shinn_train_val_test}
\end{table} %\todo{May be we can highlight 0.075 and 0.1 are extrapolation. I think $r$ is scalar.}

The architecture and training setup for MITONet proceeds as follows. First, three autoencoder networks are trained to encode $H$, $U$ and $V$ fields into a latent space. 

\begin{align}
\phi_E^S: S \rightarrow S^r; &\qquad \phi_D^S: S^r \rightarrow S^*, \quad S \in \{H,U,V \}
\label{eq:AE_setup}
\end{align}

Second, the MITONet for each variable is designed, as shown in eq. (\ref{eq:MITONet_setup}), with three branch networks: (a) IC in the latent space, (b) BC consisting of prescribed tidal elevations at the offshore boundary ($\Gamma$), and (c) scalar $r$. The trunk network, as described before, only encodes the temporal coordinates. This architecture enables MITONet to capture complex interactions between IC, BC, $r$, and the resulting hydrodynamic responses, similar to traditional numerical solvers but with enhanced flexibility and computational efficiency.
\begin{align}
\mathcal{G}_{NN}^S: S^r_t, H_{t+\beta}\vert_{\Gamma}, r,\beta \Delta t  \rightarrow S^r_{t+\beta}, \quad \forall \beta=\{1,\ldots,\tau\}, \; S \in \{H,U,V \}
\label{eq:MITONet_setup}
\end{align}

The autoencoder and MITONet models for each solution variable are constructed using multi-layer perceptrons (MLPs), with optimal configurations obtained from hyperparameter optimization studies (Text S2-S3). The autoencoders are trained to map the ADCIRC high-fidelity solution snapshots ($N_s = 3070$) to a latent space of dimension $N_r = 60$ for all hydrodynamic variables. The autoencoders contain between 9.5 and 12 million parameters, whereas the MITONet models contain between 1.3 and 3.5 million parameters. All models are trained for $20,000$ epochs using the Adam optimizer \cite{kingma2015adam} and a learning rate reduction algorithm (ReduceLROnPlateau), which decreases the learning rate by $10\%$ for every $100$ epochs without improvement. Furthermore, a  window of $\tau=5$ (Figure \ref{fig:mitonet}) is adopted for the temporal bundling technique after extensive trial-and-error experiments (Table \ref{tab:rmse_by_lfwindow}). For all variables, autoencoder training takes approximately 2 wall-clock hours on a single NVIDIA A40 GPU, whereas MITONet training takes approximately 7.5, 4.5, and 8 wall-clock hours for $H$, $U$, and $V$, respectively. However, MITONet generates predictions for a 60-day time period on a single NVIDIA A100 GPU, in $11$ seconds, thus achieving $109\text{x}$ speed-up over the physics-based, numerical solver, ADCIRC.

%on a standard laptop, equipped with a 2 GHz Quad-Core Intel Core i5 CPU, in $2$ to $4$ seconds depending on the model size, thus achieving a $300\text{x} - 600\text{x}$ speed-up over the physics-based, numerical solver, ADCIRC.

\subsection{Red River}\label{sec:red_river_details}
\begin{figure}[h]
  \centering\includegraphics[width=0.99\textwidth]{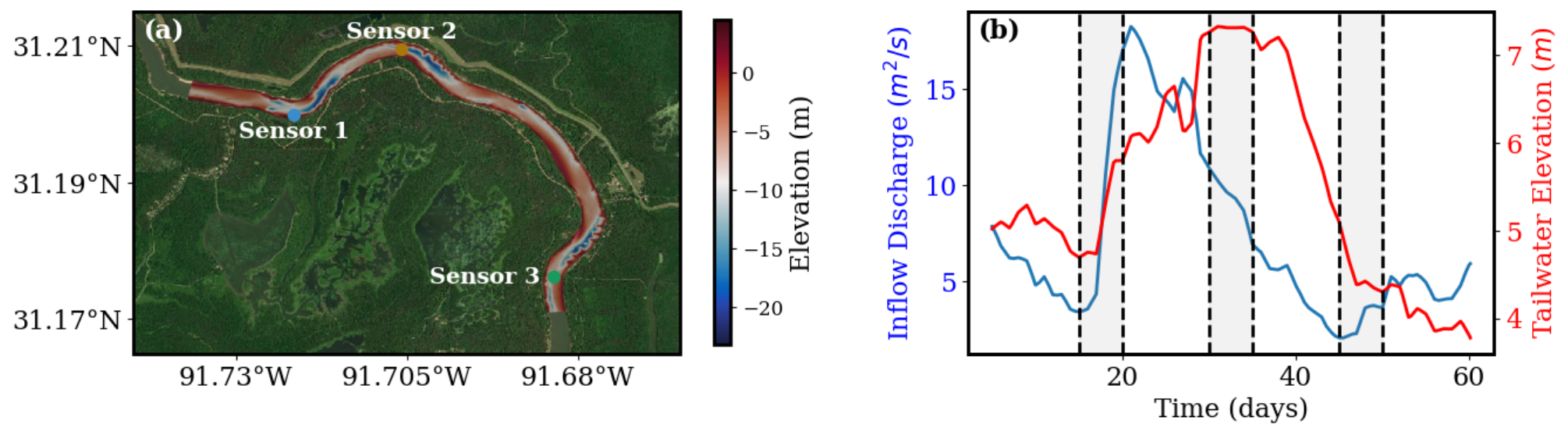}
  \caption{Domain and flow specifications for the Red River example. Panel (a) shows a representation of the bottom elevation of the flow domain obtained from in-situ surveys. The model predictions are evaluated at three sensor locations that exhibit distinct flow dynamics. The locations of the sensors are shown with different colors: sensor 1 (blue) - river bank (low flow), sensor 2 (orange) - channel center (high flow), and sensor 3 (green) - outflow. Panel (b) shows the time series of inflow discharge and tailwater elevation values used as upstream and downstream boundary conditions, respectively. The dashed vertical lines and shaded region represent the time windows used to create the training dataset.}
\label{fig:red_problem_details}
\end{figure}

The second computational example is an application of 2D SWE to simulate riverine hydrodynamics in a section of the Red River in Louisiana, USA (see Figure \ref{fig:red_problem_details}a). The 2D depth-averaged module of the Adaptive Hydraulics (AdH) finite element suite, which is a U.S. Army Corps of Engineers (USACE) high-fidelity, finite element resource for 2D and 3D dynamics \cite{trahan2018adh}, is used as the high-fidelity numerical solver. AdH is based on an implicit, stabilized finite element method, allows both spatial and temporal adaptivity, and supports a host of features such as baroclinic capabilities, surface wave and wind wave stress coupling, flow through hydraulic structures, and vessel flow interactions, that are crucial for most hydraulic and transport-engineering applications. More details about AdH's 2D SWE solver are provided in Text S1, and can also be found in \cite{martin2012adh,trahan2018adh}.

The AdH model for the Red River domain uses an unstructured triangular mesh consisting of $12291$ computational nodes and $23316$ elements, has a natural inflow velocity condition upstream, a tailwater elevation boundary downstream (as shown in Figure  \ref{fig:red_problem_details}b), and no flow boundary conditions along the river banks. The bathymetry data (as shown in Figure \ref{fig:red_problem_details}a) was obtained from USACE Hydrographic surveys while the inflow and tailwater boundary conditions were estimated based on observations from USGS gage 073556009 over a $60$-day period starting at $01/01/2017$:$0800$hrs. The bed topography measurements and water surface elevations are referenced with respect to the bathymetry data, following the NAVD88 convention. The simulation is run for a $60$ day period with a $5$-minute time step, while the state variables ($H, U, V$) are saved every $15$ minutes. After dropping the first 480 snapshots to neglect the model start-up effects, and skipping every $4$ snapshots, a total of $N_t = 1324$ high-fidelity snapshots are saved with an effective time step of $\Delta t = 1$ hour. A $60$-day simulation takes about $3.75$ hours using $12$ CPUs (AMD 7H12 Rome) on Narwhal, an HPE Cray EX system located at the U.S. Navy DoD Supercomputing Resource Center (DSRC).   

Similar to the tidal circulation example, the high-fidelity simulation is parametrized by varying the scalar bottom friction coefficient, $r$ between $0.02375$ and $0.02625$ to select $13$ values (Table \ref{tab:red_train_val_test}). A $60$-day AdH simulation is run for each parameter value, resulting in a total wall-clock time of $13 \times 3.75\, \text{hrs} = 48.75\, \text{hrs} \approx 2\, \text{days}$ for generating the entire parametric dataset. This is split into training, validation, and testing sets based on parameter values and subsampled time windows from the available simulation period, as described in Table \ref{tab:red_train_val_test}. Specifically, to add sufficient diversity in the training samples, the training set is constructed by choosing high-fidelity snapshots from three separate $5$-day time windows in the $60$-day period, namely days $15$-$20$, $30$-$35$, and $45$-$50$. As shown in Figure \ref{fig:red_problem_details}b, these windows represent distinct flow regimes. Based on our numerical experiments, adopting such a windowed training set yielded more expressive MITONet models than those obtained from using a contiguous $15$-day training window. %The training set is used to learn the model parameters, the validation set is used for hyperparameter optimization and learning rate reduction, and the test set is reserved for the final evaluation.

\begin{table}[h]
\caption{Red River Training, Validation, and Testing Data}
\centering
\begin{tabular}{l c c}
\hline
      & \textbf{Bottom Friction Coefficient,} $\boldsymbol{r}$                 & \textbf{Days (timesteps)} \\
\hline
\multirow{2}{*}{\textbf{Training}}   & 0.023875, 0.024375, 0.024875, 0.025125, & 15-20, 30-35, 45-50 (360) \\
                                     & 0.025625, 0.026125                      &  \\
\textbf{Validation}                  & 0.024125, 0.024625, 0.025375, 0.025875  & 20-30 (240) \\
\textbf{Testing}                     & 0.02375, 0.025, 0.02625                 & 5-60 (1320) \\
\hline
\end{tabular}
\label{tab:red_train_val_test}
\end{table}

The architecture and training for MITONet follows a similar setup as the one previously described for the Shinnecock case. First, three autoencoder networks are trained to encode $H$, $U$ and $V$ fields into a latent space, similar to eq. (\ref{eq:AE_setup}). Second, the MITONet for each variable, as shown in eq. (\ref{eq:MITONet_setup_red}), is designed with four branch networks: (a) IC in the latent space, (b) BC consisting of discharge ($Q$) at the upstream boundary, $\Gamma^{in}$, c) BC consisting of water level at the downstream boundary, $\Gamma^{out}$, and (d) scalar $r$. The trunk network, as described before, only encodes the temporal coordinates.
\begin{align}
\mathcal{G}_{NN}^S: S^r_t, Q_{t+\beta}\vert_{\Gamma^{in}}, H_{t+\beta}\vert_{\Gamma^{out}}, r,\beta \Delta t  \rightarrow S^r_{t+\beta}, \quad \forall \beta=\{1,\ldots,\tau\}, \; S \in \{H,U,V \}
\label{eq:MITONet_setup_red}
\end{align}

Similar to the Shinnecock Inlet example, the autoencoder and MITONet models for each solution variable are constructed using MLPs, with optimal configurations obtained from hyperparameter optimization studies (Table S5). The autoencoders are trained to map the high-dimensional snapshots for each variable ($N_s = 12291$) to a latent space of dimension $N_r = 96,64,96$ for $H, U, V$, respectively. The autoencoders all contain approximately 9.5 million parameters, whereas the MITONet models for $H, U, V$ contain 2.2, 0.6, and 2 million parameters respectively. All autoencoder models are trained for $20,000$ epochs using the AdamW optimizer \cite{loshchilov2019adamw} whereas the MITONet models are trained for $20,000$ epochs using the Adam optimizer \cite{kingma2015adam}. A learning rate reduction algorithm (ReduceLROnPlateau), which decreases the learning rate by $10\%$ for every $100$ epochs without improvement, is applied during the training of every model. Furthermore, a  window of $\tau=10$ is adopted for the temporal bundling technique after a series of trial-and-error experiments. For all variables, autoencoder training takes approximately $1.25$-$1.5$ wall-clock hours on a single NVIDIA A100 GPU, whereas MITONet training takes approximately $1.7$ wall-clock hours. During inference, generates predictions for a 60-day time period on a single NVIDIA A100 GPU, in $12$ seconds, thus achieving a $1,125\text{x}$ speed-up over the physics-based, numerical solver, AdH.

\section{Computational Experiments \& Discussion}\label{sec:results}
The effectiveness of the proposed MITONet framework is evaluated by considering two computational examples with varying degrees of complexity, as discussed in sections \ref{sec:shinnecock_details} and \ref{sec:red_river_details}. The first example exhibits quasi-periodic behavior, driven by tidal dynamics, and is used to demonstrate both the improved performance of MITONet over similar DeepONet-based architectures, as well as its robust long-term forecast and parametric extrapolation capabilities. The second example involves nonlinear, aperiodic flow features due to realistic hydrograph-induced riverine flow through an irregular channel geometry. Despite the challenging dynamical patterns, as evidenced by the extremely limited number of similar surrogate modeling efforts in the literature, the MITONet framework displays a high degree of expressibility and skill in capturing the multiscale, nonlinear features present in the high-fidelity numerical solution of the system. These results are presented in full detail in the remainder of the section. First, the various quantitative error metrics adopted to compare the performance of different models are discussed below. 
% \subsection{Error Metrics}\label{sec:error_metrics}

The spatial root mean square error ($RMSE$) is a widely used metric in computational science applications and is adopted here to evaluate MITONet's accuracy by measuring the square root of the average squared differences between the predicted solution, $\tilde{\mathbf{s}}$ and the true solution, $\mathbf{s}$. 
% This metric is commonly used in hydrodynamic applications as it provides a clear indication of the model's performance, with lower $RMSE$ values indicating higher accuracy and better alignment between the model's predictions and observed data. 
The $RMSE$ at any time instant $\{t_j\}_{j=1}^{N_t}$ is computed as,

\begin{equation}
RMSE_j = \sqrt{\frac{1}{N_s} \sum_{i=1}^{N_s} (s_{ij} - \tilde{s}_{ij})^2},
\end{equation}

\noindent where $N_s$ denotes spatial degrees of freedom, $N_t$ is the temporal degrees of freedom, $s_{ij} = \mathbf{s}(\mathbf{y}_{i}, t_{j})$ is the true solution at $(\mathbf{y}_{i}, t_j)$, where  $i = \{1, \ldots, N_s\}$, $j = \{1, \ldots, N_t\}$, and $\tilde{s}_{ij} = \tilde{\mathbf{s}}(\mathbf{y}_{i}, t_{j})$ is the predicted solution at $(\mathbf{y}_{i}, t_j)$. In the following discussion, the temporal and spatial means of any variable or metric are computed as $\overline{e_j} = \frac{1}{N_t}\sum_{j=1}^{N_t} e_{ij}$ and $\overline{e_j}= \frac{1}{N_s} \sum_{i=1}^{N_s} e_{ij}$, respectively, where $e$ is any given spatio-temporally varying quantity or metric.

The normalized root mean square error ($NRMSE$), which normalizes the $RMSE$ by the range of the true solution, is computed to provide a relative measure of error that accounts for variations in the true data. The $NRMSE$ at $\{t_j\}_{j=1}^{N_t}$ is given by,

\begin{equation}
NRMSE_j = \frac{RMSE_j}{s_{j}^{max} - s_{j}^{min}},
\end{equation}

\noindent where $s_{j}^{max} = \max\limits_{1 \leq i \leq N_s} s_{ij}$ and $s_{j}^{min} = \min\limits_{1 \leq i \leq N_s} s_{ij}$ are the maximum and the minimum value of the true solution $\mathbf{s}$ at time $t_j$, respectively. 

% To quantify pattern error independent of a mean offset, an unbiased (centered) RMSE is also computed by removing the spatial mean at each time step as given by,

% \begin{equation}
% shaded_j \;=\; \sqrt{\frac{1}{N_s}\sum_{i=1}^{N_s}
% \Big[\big(s_{ij}-\overline{s_j}\big) - \big(\tilde{s}_{ij}-\overline{\tilde{s}_j}\big)\Big]^2},
% \end{equation}

% \noindent where $\overline{s_j}$ and $\overline{\tilde{s}_j}$ are the temporal means of the true solution and the predicted solution, respectively.

% The bias at time $t_j$ is computed by

% \begin{equation}
% Bias_j \;=\; \overline{\tilde{s}_j} - \overline{s_j}.
% \end{equation}

% The temporal means of the above metrics are reported as

% \begin{align}
% \overline{RMSE} \;=\; \frac{1}{N_t}\sum_{j=1}^{N_t} RMSE_j,  \quad & \quad
% \overline{NRMSE} \;=\; \frac{1}{N_t}\sum_{j=1}^{N_t} NRMSE_j, \\
% \overline{uRMSE} \;=\; \frac{1}{N_t}\sum_{j=1}^{N_t} uRMSE_j, \quad & \quad
% \overline{Bias} \;=\; \frac{1}{N_t}\sum_{j=1}^{N_t} Bias_j.
% \end{align}

A temporal mean absolute error (MAE) is defined to quantify the average error over time at each spatial node, and is computed as
\begin{equation}
MAE_i \;=\; \frac{1}{N_t}\sum_{j=1}^{N_t}\big|\,s_{ij}-\tilde{s}_{ij}\,\big|.
\end{equation}

Finally, we compute the Anomaly Correlation Coefficient ($ACC$) for every bottom friction coefficient, $r$. The $ACC$ is used to evaluate the accuracy of the prediction model by measuring the correlation between the predicted solution and the true solution, after removing their respective spatial means. It is widely applied in hydrodynamic studies as it quantifies the alignment of spatial patterns, with values closer to 1 indicating better performance.

\begin{equation}
ACC = \frac{1}{N_t}\left\{ \sum_{j=1}^{N_t}\left[\frac{\sum_{i=1}^{N_s} (\tilde{s}_{ij} - \overline{\tilde{s_j}})(s_{ij} - \overline{s_j})}{\sqrt{\sum_{i=1}^{N_s} (\tilde{s}_{ij} - \overline{\tilde{s_j}})^2 \sum_{i=1}^{N_s} (s_{ij} - \overline{s_j})^2}} \right] \right\},
\end{equation}

\noindent where $\overline{s_j}= \frac{1}{N_s} \sum_{i=1}^{N_s} s_{ij}$ is the spatial mean of the true solution, and 
$\overline{\tilde{s_j}} = \frac{1}{N_s} \sum_{i=1}^{N_s} \tilde{s}_{ij}$ is the spatial mean of the predicted solution.

\subsection{Shinnecock Inlet}\label{sec:shinnecock_results}

For the first computational example involving tide-driven coastal circulation in the Shinnecock Inlet Bay, NY, we demonstrate MITONet's accuracy both by computing various summary statistics as well as by analyzing predictions at three specific sensor locations (Figure  \ref{fig:shinnecock_bathy}). These locations were strategically selected to evaluate MITONet's capability to emulate varying dynamical features that occur in different regions of the domain: the bay, the inlet, and the back bay, which are typically modeled by ADCIRC using numerical grids of variable resolutions. Unless otherwise specified, all MITONet predictions are computed in a roll-out manner starting from an unseen IC at day 5.

\subsubsection{Cross-Model Comparison and Look-Forward Sensitivity}\label{sec:shinn_model_comparison}

In this section, various benchmarking results are presented comparing MITONet with well-known operator-learning architectures from the DeepONet family, including vanilla DeepONet (DON) \cite{lu2021learning}, Modified DeepONet (M-DON) \cite{wang2022improved}, MIONet \cite{jin2022mionet}, and Latent DeepONet (L-DON) \cite{kontolati2023learning}. To conduct a fair comparison, all the models considered here are fine-tuned using Optuna in the same manner as MITONet (Text S2), with the resulting optimal configurations detailed in Table S4. Additonally, all the models are set up to generate $\tau=5$ future state predictions from a given IC, where $\tau$ is also the look-forward window size used by the corresponding MITONet model. To this end, for DON, M-DON, and L-DON, the branch inputs (IC, BC at $\tau$ future time steps, and $r$) are concatenated into a single input vector. MIONet, on the other hand, uses multiple branches like MITONet, with the key difference being that MIONet does not encode the IC into a latent space. Additionally, for DON, M-DON, and MIONet, the trunk network incorporates both space and time coordinates, whereas L-DON maps only the temporal coordinates like MITONet.

Table \ref{tab:rmse_by_model_and_windows} shows the parametric mean of the $\overline{RMSE}$ of each model for every state variable, computed over three different prediction window sizes: days $5$-$15$, $5$-$25$ and $5$-$45$. MITONet obtains the lowest errors in long-horizon autoregressive rollouts across all the different prediction windows and parameter settings, whereas none of the other models uniformly outperforms any of the others across the entire time-parameter range. Note that in some cases the mean $\overline{RMSE}$ for the $10$-day prediction window exceeds that of the $20$- and $40$-day windows. This likely reflects the fact that the latter two prediction windows have, at least, a partial overlap with the training window consisting of days $15-30$, which potentially leads to a reduction in the prediction errors. In addition to the superior prediction accuracy across different time windows and parameter variations, the MITONet framework offers the flexibility of generating predictions independent of the size of the look-forward window used during training. In other words, even if a MITONet model is trained with a look-forward window size, $\tau$ of say, $\tau=5$, it is capable, by design, of using the state snapshot at a random time instant $t_j$ as IC and the relevant BCs at any future time instant $t^F = t_j + k\Delta t$ for $k = 1, \ldots, 5, \ldots $ to predict the future state at $t^F$. In contrast, all of the other models considered here are limited by the input structure adopted during training, that is, if the BCs corresponding to $5$ future time steps were used as input features for training, the model is only capable of generating predictions for exactly those $5$ future time steps during inference. While this may not be a significant advantage in every application, for certain use cases such as generating predictions from an unknown or zero IC, the ability to evolve the solution in time, in an autoregressive fashion, using single time step predictions, allows the MITONet model to ramp-up faster, as will be shown in Section \ref{sec:shinn_random_init}.

% \definecolor{LtGrn}{rgb}{0.58, 0.9, 0.58}
\definecolor{celadon}{rgb}{0.67, 0.88, 0.69}
\begin{table}[h]
\centering
\caption{Performance comparison of different architectures based on the mean $\overline{RMSE}$ over all bottom friction coefficient values for each variable and over $10$-, $20$-, and $40$-day prediction windows using the Shinnecock Inlet example. Lowest $\overline{RMSE}$ for each window is highlighted.}
\label{tab:rmse_by_model_and_windows}
\resizebox{0.9\textwidth}{!}{
\begin{tabular}{l c c c c c c}
\hline
\textbf{Variable} & \textbf{Days} & \textbf{DON} & \textbf{M-DON} & \textbf{L-DON} & \textbf{MIONet} & \textbf{MITONet} \\
\hline
\multirow{3}{*}{$\boldsymbol{H}$}
 & \textbf{5--15} & 4.19 & 5.62 & 4.04 & 3.71 & \cellcolor{celadon}\textbf{0.39} \\
 & \textbf{5--25} & 4.22 & 5.39 & 3.99 & 3.68 & \cellcolor{celadon}\textbf{0.28} \\
 & \textbf{5--45} & 4.35 & 5.10 & 4.08 & 3.68 & \cellcolor{celadon}\textbf{0.28} \\ \hline
\multirow{3}{*}{$\boldsymbol{U}$}
 & \textbf{5--15} & 3.31 & 2.70 & 2.62 & 2.35 & \cellcolor{celadon}\textbf{1.27} \\
 & \textbf{5--25} & 3.66 & 2.86 & 2.31 & 1.64 & \cellcolor{celadon}\textbf{0.96} \\
 & \textbf{5--45} & 3.57 & 3.12 & 2.32 & 2.06 & \cellcolor{celadon}\textbf{0.95} \\ \hline
\multirow{3}{*}{$\boldsymbol{V}$}
 & \textbf{5--15} & 2.02 & 2.66 & 1.17 & 2.05 & \cellcolor{celadon}\textbf{1.10} \\
 & \textbf{5--25} & 2.23 & 2.90 & 1.09 & 2.19 & \cellcolor{celadon}\textbf{0.90} \\
 & \textbf{5--45} & 2.28 & 2.96 & 1.34 & 2.23 & \cellcolor{celadon}\textbf{0.83} \\
\hline
\end{tabular}
}
\end{table}

For a detailed qualitative analysis, Figures S1-S4 present additional results. These include side-by-side snapshots of the full solution fields, zoomed-in comparisons of key regions, and maps showing the spatial distribution of the Mean Absolute Error (MAE) over time.

Table~\ref{tab:rmse_by_lfwindow} presents a study of the effect of different look-forward window sizes, $\tau$, used for temporal bundling on MITONet performance using the parametric mean $\overline{RMSE}$ across all bottom friction coefficients for a $55$-day prediction of each variable. MITONet models with identical architectures are trained with $\tau= 5, 10, 15, 20$. Results for models with $\tau = 5–15$ are mixed and variable-dependent, whereas $\tau=20$ models consistently performed worst (Table~\ref{tab:rmse_by_lfwindow}). Given the comparable accuracy for models with  $\tau = 5, 10,$ and $15$, and the additional computational cost of using larger look-forward windows during training, a look-forward window size of 5 ($\tau=5$) is adopted for all the models used in the first computational example and discussed in the remainder of this section.

\begin{table}[h]
\centering
\caption{Comparison of the parametric mean $\overline{RMSE}$ of autoregressive MITONet predictions between days $5$-$60$ using models trained with different look-forward window sizes ($\tau$) for the Shinnecock Inlet example.}
\label{tab:rmse_by_lfwindow}
\resizebox{\textwidth /2}{!}{
\begin{tabular}{c c c c c}
\hline
\textbf{Variable} & \multicolumn{4}{c}{\textbf{Look-forward Window} ($\boldsymbol{\tau}$)} \\
 & \textbf{5} & \textbf{10} & \textbf{15} & \textbf{20} \\
\hline
$\boldsymbol{H}$ & 0.25 & 0.25 & 0.39 & 0.55 \\
$\boldsymbol{U}$ & 0.90 & 0.88 & 1.19 & 0.97 \\
$\boldsymbol{V}$ & 0.79 & 1.01 & 0.77 & 0.93 \\
\hline
\end{tabular}
}
\end{table}

\subsubsection{Model Skill Assessment}\label{sec:shinn_model_skill}

\begin{figure}[h]
    \centering
    \includegraphics[width=.9\textwidth]{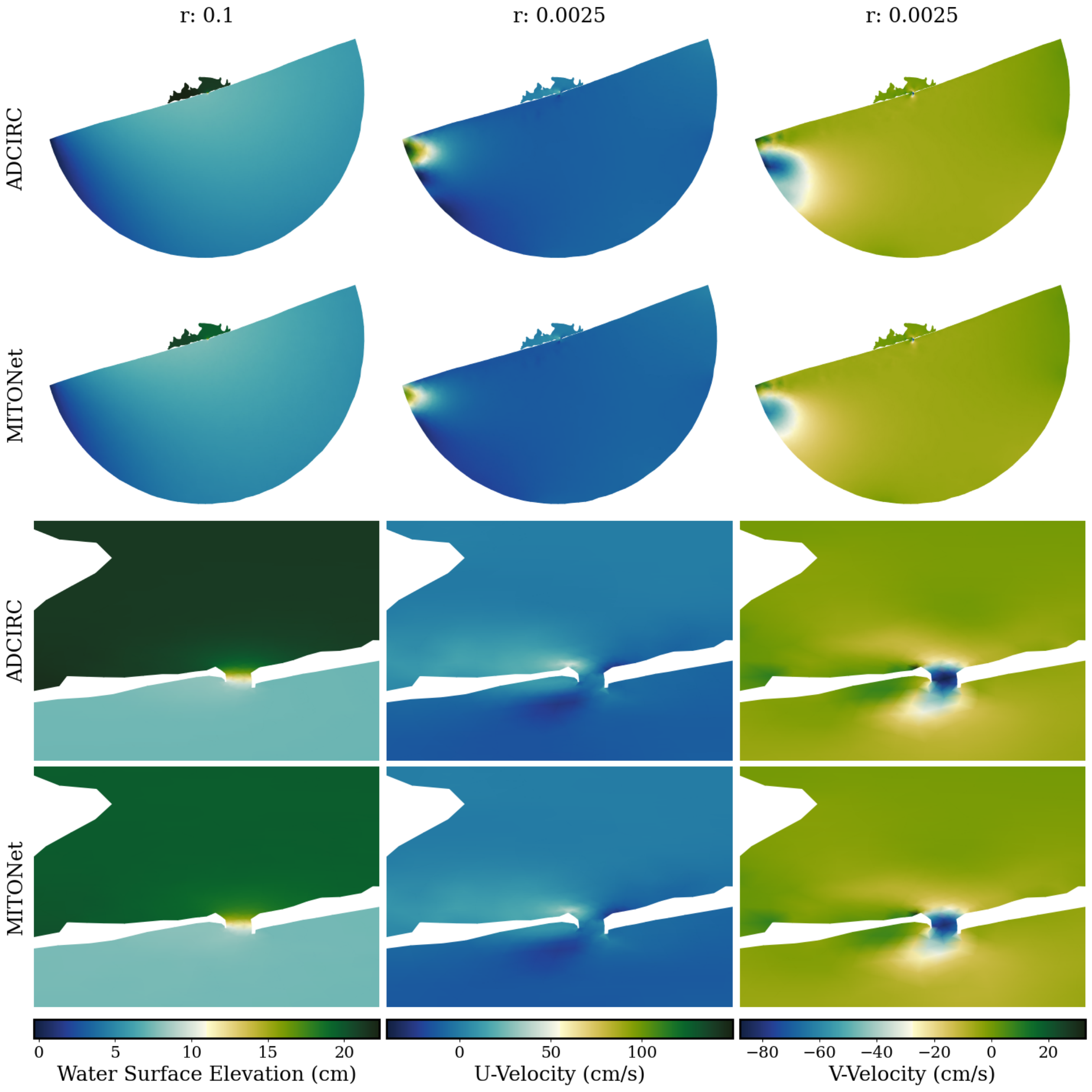}
    \caption{Snapshots of the ADCIRC solutions (Rows 1 and 3) and MITONet predictions (Rows 2 and 4) for the the most challenging test parameter values of the Shinnecock Inlet example: $r=0.1$ for $H$ (column 1), and $r=0.0025$ for $U$ and $V$ (columns 2 and 3) on day 60. Both full domain (Rows 1 and 2) and zoomed-in views of the inlet (Rows 3 and 4) are shown to highlight important features.}
    \label{fig:snaps_mixed}
\end{figure}

Figure \ref{fig:snaps_mixed} shows snapshots of MITONet and ADCIRC predictions at day $60$. Higher $r$ values i.e., higher frictional resistance introduces a lag in the arrival time of tidal waves from the off-shore boundary to the inner bay which results in more complicated dynamics for water surface elevation and thus, more challenging dynamics for $H$ in these cases. In contrast, for $U$ and $V$ fields, a lower $r$ value results in a more advection-dominated flow regime with complex, nonlinear, localized flow features especially in the inlet region and higher average magnitude of the solution fields (see colorbars in Figures S8-S10), thus leading to more challenging dynamics. Therefore, we evaluate the model predictions using $r=0.1$ snapshots for $H$ (column 1) and $r=0.0025$ for $U$ and $V$ (columns 2 and 3, respectively). These are the most challenging cases for each variable and are also extrapolatory cases, not included in the training set.  We show the full domain, and a zoomed-in view of the inlet to highlight important features. 

Overall, we see great agreement between MITONet and ADCIRC at all scales, demonstrating that the model can emulate different types of flow regimes and accurately learn both large-scale behavior as well as fine-scale features. The differences between MITONet and ADCIRC occur primarily near the corners of the domain and are noticeable, especially for $U$ and $V$ (see Figure \ref{fig:snaps_mixed}), near the north-west corner of the domain where the off-shore open boundary meets the zero normal flow mainland boundary. These are highly nonlinear, and most importantly, non-physical artifacts, often observed in numerical simulations using ADCIRC and other solvers, that primarily arise from numerical and mesh instabilities induced by the advective transport terms, boundary treatments, and grid specifications \cite{RTI2015_adcirc}. Efforts to address these numerical instabilities with techniques such as the modification of the mesh near the boundary, smoothing bathymetry, turning off boundary forcing of several smaller tidal harmonics, ignoring advection terms and/or its derivatives, and many more usually only help in delaying the formation of the instability. Hence, if possible, the simulation domain is designed in a way such that these effects do not influence the solution near the region of interest (such as the inlet area of the Shinnecock Bay). Therefore our emphasis when analyzing the performance of the MITONet models is the area in and around the inlet. For the interested readers, snapshots for all variables and test cases are presented in Figures S8-S10.

\begin{figure}[h]
    \centering
    \includegraphics[width=.9\textwidth]{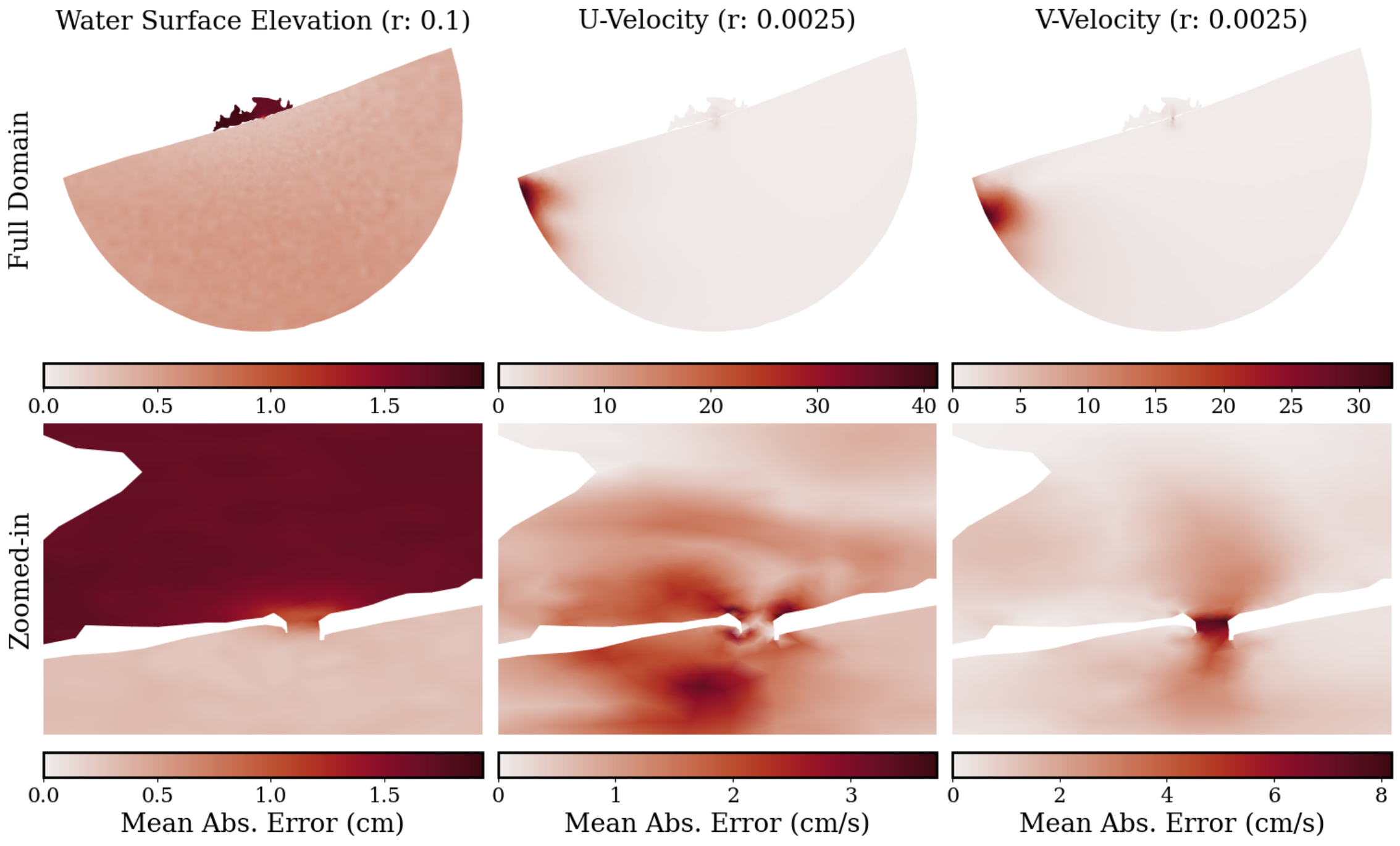}
    \caption{Spatial distribution of the temporal MAE of MITONet predictions for the Shinnecock Inlet example using $r=0.1$ for $H$ (column 1), and $r=0.0025$ for $U$ and $V$ (columns 2 and 3) on day $60$. Full domain (Row 1) and zoomed-in views of the inlet (Row 3) are shown to highlight important features.}
    \label{fig:mae}
\end{figure}

The spatial distribution of errors for $r=0.1$ in $H$ and for $r=0.0025$ in $U$ and $V$ is shown in Figure \ref{fig:mae}, which plots the $MAE$ over time at each spatial point. For $H$, the largest errors occur in the inner bay, the shallowest part of the domain, and the region most affected by the previously noted lag in water elevation. For $U$ and $V$, the largest errors are concentrated near the north-west domain corner, which is consistent with spurious dynamics from numerical instabilities in this region. For $V$, the errors are also relatively higher in the inlet channel, where the velocity magnitudes are generally higher than the rest of the domain, as can also be seen in the bottom two rows (right column) of Figure  \ref{fig:snaps_mixed}.

\begin{figure}[h]
    \centering
    \includegraphics[width=.9\textwidth]{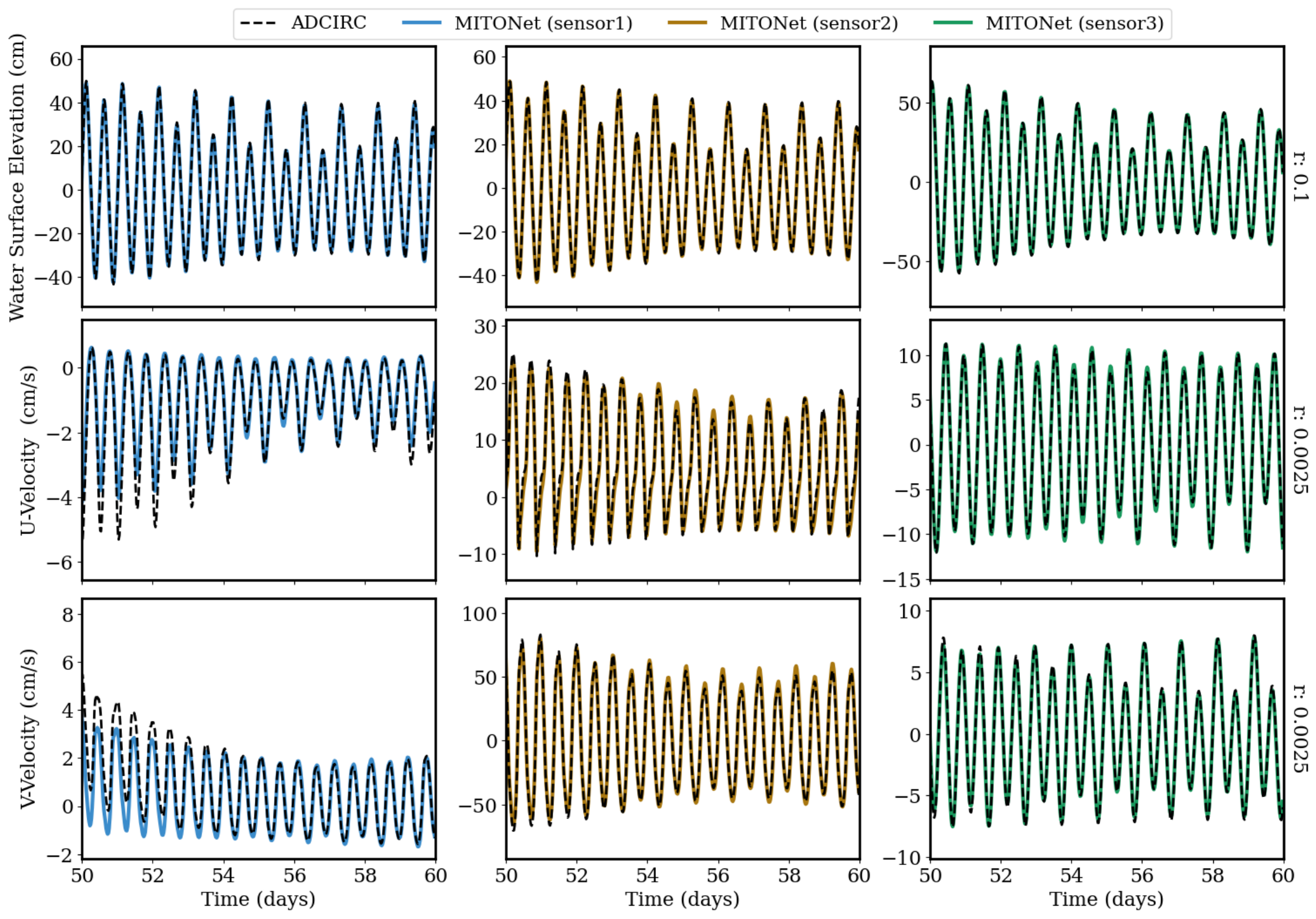}
    \caption{Comparison of the ADCIRC solution (black dashed lines) with the MITONet predictions (blue, yellow, and green lines) for the Shinnecock Inlet example at three different sensor locations (see Figure\ref{fig:shinnecock_bathy}) between days $50$ to $60$ using the most challenging test parameter values: $r=0.1$ for $H$ (Row 1), and $r=0.0025$ for $U$ (Row 2) and $V$ (Row 3).}
    \label{fig:signals_mixed}
\end{figure}

Time series comparison of MITONet and ADCIRC predictions for each variable at the sensor locations is shown in Figure \ref{fig:signals_mixed}. We show the cases with the highest $RMSE$ values i.e., $r=0.1$ for $H$ and $r=0.0025$ for $U$ and $V$, to demonstrate the model's capability to extrapolate in both parametric space and time. It is worth noting that the MITONet predictions are generated autoregressively, starting from an unseen IC at day 5, whereas Figure \ref{fig:signals_mixed} shows the comparison between day 50 to day 60 to highlight MITONet's performance after 45 days (2,160 time steps) of autoregressive rollout. This demonstrates that even when extrapolating in parameter space, MITONet's long-term temporal forecast is stable and in excellent agreement with the numerical solutions for $H$, $U$ and $V$ at the sensor locations. Full comparisons for all variables and test cases are shown in Figures S5-S7.

\begin{figure}[h]
    \centering
    \includegraphics[width=.9\textwidth]{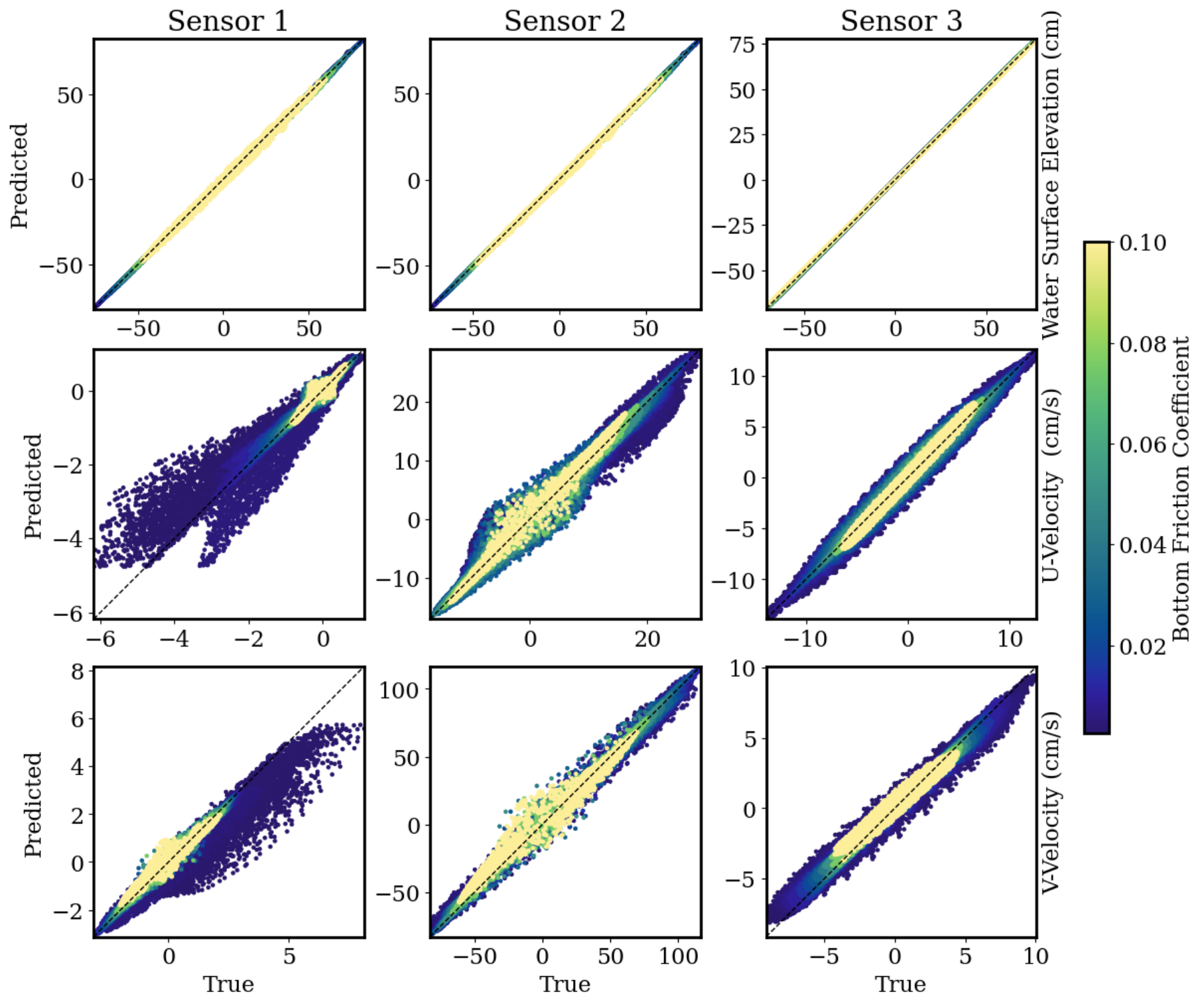}
    \caption{Parity plots for the Shinnecock Inlet example comparing MITONet predictions with ADCIRC solutions of $H, U,$ and $V$ at three sensor locations, over days $5–60$ and for all values of the bottom friction coefficient, $r$.}
    \label{fig:parity}
\end{figure}

To assess MITONet’s performance at the sensor locations over the entire simulation period, Figure~\ref{fig:parity} shows parity plots (predicted vs. ADCIRC snapshots) for each variable across the entire time series. For $H$, agreement with ADCIRC is excellent at all sensors and across all $r$ values. For $U$ and $V$, agreement remains strong at sensors 2 and 3, whereas for the smallest bottom friction coefficient ($r=0.0025$), noticeable discrepancies appear at sensor 1, which is located in the inner bay, the shallowest region of the domain. Although these deviations look pronounced, it is worth noting that the range in signal magnitudes at sensor 1 is significantly smaller than the ranges at sensors 2 and 3. For sensor 1 and the $r=0.0025$ case, the scatter points for $U$ fluctuate around $1:1$ line, indicating alternating under and over prediction with little systematic offset. In contrast, the scatter points for $V$ at sensor 1 and for $r=0.0025$ predominantly lie below the $1:1$ line, indicating that for $r=0.0025$ MITONet tends to under predict $V$ in the inner bay region.

\begin{figure}[h]
    \centering
    \includegraphics[width=.9\textwidth]{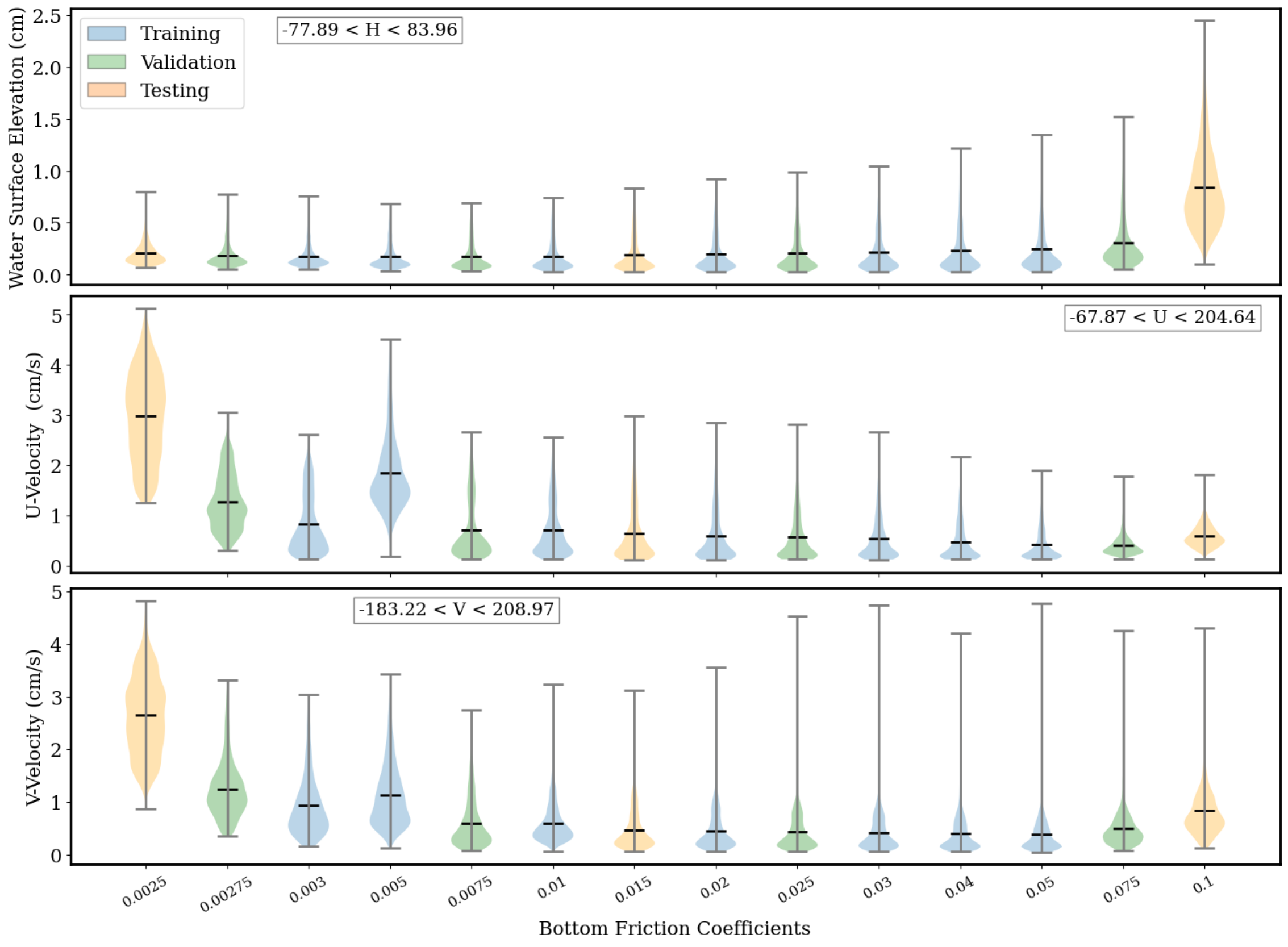}
    \caption{Violin plot of the $RMSE$ for MITONet predictions of $H$ (Row 1), $U$ (Row 2), and $V$ (Row 3) for the Shinnecock Inlet example, between days $5-60$ and for all bottom friction coefficient ($r$) values. Each subplot includes a textbox indicating the range of the corresponding ground truth solution over the spatio-temporal domain. Colors denote training (blue), validation (green), and testing (orange) values of $r$.}
    \label{fig:violin}
\end{figure}

Figure \ref{fig:violin} summarizes the results for all parameter values, with a violin plot of the $RMSE$ between the MITONet prediction and the ADCIRC solution. The $RMSE$ values for $H$ are the lowest. This is expected because it is likely easier to predict the water surface elevation field given a tidal elevation BC, than it is to predict currents from the same BC. As previously discussed, higher $r$ values represent the most challenging cases for $H$ and thus, higher $RMSE$ values are observed as $r$ increases. In fact, the median $RMSE$ for $H$ when $r=0.1$ is more than double that of the other cases. Nonetheless, even in this case, the maximum $RMSE$ for $H$ remains below $2.5$~cm, which is relatively small compared to previous modeling efforts in this region \cite{militello2001shinnecock, lin2022numerical}. For $U$ and $V$ fields, the lower $r$ values represent more challenging dynamics, and thus the $RMSE$ increases as $r$ decreases, with the highest $RMSE$  corresponding to $r=0.0025$. On the other hand, higher $r$ results in lower mean $RMSE$ but increases the number of outliers, with some outliers reaching $5$~cm/s. Once again, these errors remain relatively small compared to previous modeling efforts in this region \cite{militello2001shinnecock}.

The increase in $RMSE$ values for the challenging extreme values of $r$ can be primarily attributed to increase in nonlinear and localized flow features. This is further supported by the violin plots in Figure S11 which are obtained by training MITONet models for each variable using all values of $r$ and computing prediction $RMSE$ for each value of $r$. The trend of almost monotonic increase in $RMSE$ with an increase in $r$ for the $H$ model and a decrease in $r$ for the $U$ and $V$ models indicate that these errors are not due to parametric extrapolation, but due to the change in flow physics as discussed above.   

\begin{figure}[h]
    \centering
    \includegraphics[width=.9\textwidth]{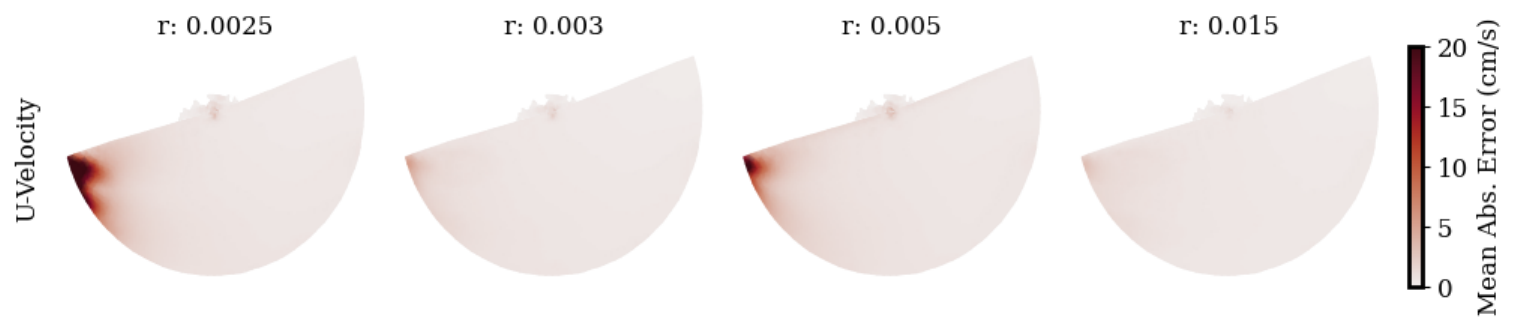}
    \caption{Spatial distribution of temporal MAE of the MITONet predictions for $U$ using the Shinnecock Inlet example. Two test values of bottom friction coefficient, $r = 0.0025$ and $0.015$, and two training values, $r=0.003$ and $0.005$ are shown.}
    \label{fig:corner}
\end{figure}

Furthermore, for $r=0.005$ the $RMSE$ is higher than for neighboring cases, even though $r=0.005$ is included in the training set. The spatial distribution of errors in terms of $MAE$ (Figure~\ref{fig:corner}, top row) shows that this increase is concentrated in the upper-left corner of the domain. As discussed above, these larger errors arise from numerical instabilities in the ADCIRC solution that can induce spurious local dynamics. For the various $r$ values, errors remain localized to the corner region. For $r=0.005$ the corner artifact is sufficiently pronounced to elevate the domain-mean $RMSE$, whereas for other $r$ values it primarily appears as outliers with limited impact on the mean.

Finally, the $\overline{RMSE}$, $\overline{NRMSE}$, and $ACC$ for a 55-day rollout beginning from an unseen IC at day 5 are reported in Table \ref{tab:transposed_metrics_simplified}. As expected, the test cases (highlighted) exhibit the lowest $ACC$ values and the highest $\overline{RMSE}$ values, especially in the extrapolation cases. Nonetheless, accuracy remains high, with the minimum $ACC$ equal to 0.91. Because $\overline{RMSE}$ is not scale-invariant, it can be misleading, particularly for $U$ and $V$, where the spatial mean varies substantially across bottom friction coefficients. In contrast, $\overline{NRMSE}$, which normalizes by the range of the true field at each time, shows consistently strong performance across all cases. As $r$ increases, $\overline{NRMSE}$ for $H$ increases, whereas for $U$ and $V$ it decreases with increasing $r$, consistent with the earlier discussion of the most challenging regimes for each variable. Finally, $ACC$, which removes the spatial mean, further confirms strong agreement between predictions and reference solutions.

% \definecolor{gray50}{gray}{0.9}
\definecolor{gray50}{gray}{0.8}
\definecolor{gray60}{gray}{0.9}
\renewcommand{\arraystretch}{1.25}
\setlength{\tabcolsep}{8pt}
\begin{table}[h]
\caption{$ACC$, $\overline{RMSE}$ (m/s), and $\overline{NRMSE}$ of MITONet predictions for $H$, $U$, and $V$ with a $55$-day rollout starting at day $5$, across all bottom friction coefficient ($r$) values of the Shinnecock Inlet example. Test and validation scenarios are shown in darker and lighter shades of gray, respectively.}
\label{tab:transposed_metrics_simplified}
\centering
\resizebox{\textwidth}{!}{
\begin{tabular}{c c c c c c c c c c}
\hline
& \multicolumn{3}{c}{$\boldsymbol{H}$} & \multicolumn{3}{c}{$\boldsymbol{U}$} & \multicolumn{3}{c}{$\boldsymbol{V}$} \\ \hline
{$\boldsymbol{r}$} & {$\boldsymbol{ACC}$} & $\boldsymbol{\overline{RMSE}}$ & $\boldsymbol{\overline{NRMSE}}$
                   & {$\boldsymbol{ACC}$} & $\boldsymbol{\overline{RMSE}}$ & $\boldsymbol{\overline{NRMSE}}$
                   & {$\boldsymbol{ACC}$} & $\boldsymbol{\overline{RMSE}}$ & $\boldsymbol{\overline{NRMSE}}$ \\ \hline
\rowcolor{gray50}\textbf{0.0025}  & 0.998 & 0.211 & 0.00131 & 0.916 & 2.98 & 0.0109 & 0.945 & 2.65 & 0.00676 \\ \hline
\rowcolor{gray60}\textbf{0.00275} & 0.999 & 0.184 & 0.00114 & 0.984 & 1.28 & 0.00519 & 0.989 & 1.25 & 0.00321 \\ \hline
\textbf{0.003}   & 0.999 & 0.179 & 0.00111 & 0.992 & 0.830 & 0.00359 & 0.994 & 0.934 & 0.00242 \\ \hline
\textbf{0.005}   & 0.999 & 0.173 & 0.00107 & 0.962 & 1.85 & 0.0110 & 0.992 & 1.13 & 0.00302 \\ \hline
\rowcolor{gray60}\textbf{0.0075}  & 0.999 & 0.173 & 0.00107 & 0.992 & 0.718 & 0.00512 & 0.996 & 0.587 & 0.00161 \\ \hline
\textbf{0.01}    & 0.999 & 0.179 & 0.00111 & 0.992 & 0.706 & 0.00520 & 0.994 & 0.599 & 0.00168 \\ \hline
\rowcolor{gray50}\textbf{0.015}   & 0.999 & 0.192 & 0.00119 & 0.990 & 0.650 & 0.00508 & 0.995 & 0.467 & 0.00137 \\ \hline
\textbf{0.02}    & 0.999 & 0.203 & 0.00127 & 0.989 & 0.599 & 0.00493 & 0.993 & 0.448 & 0.00138 \\ \hline
\rowcolor{gray60}\textbf{0.025}   & 0.999 & 0.212 & 0.00133 & 0.989 & 0.571 & 0.00493 & 0.992 & 0.428 & 0.00138 \\ \hline
\textbf{0.03}    & 0.999 & 0.220 & 0.00138 & 0.989 & 0.548 & 0.00495 & 0.991 & 0.422 & 0.00141 \\ \hline
\textbf{0.04}    & 0.999 & 0.233 & 0.00146 & 0.991 & 0.478 & 0.00466 & 0.991 & 0.395 & 0.00142 \\ \hline
\textbf{0.05}    & 0.999 & 0.246 & 0.00154 & 0.993 & 0.420 & 0.00437 & 0.990 & 0.382 & 0.00147 \\ \hline
\rowcolor{gray60}\textbf{0.075}   & 0.998 & 0.309 & 0.00194 & 0.994 & 0.413 & 0.00469 & 0.980 & 0.504 & 0.00222 \\ \hline
\rowcolor{gray50}\textbf{0.1}     & 0.998 & 0.839 & 0.00526 & 0.991 & 0.590 & 0.00724 & 0.955 & 0.832 & 0.00410 \\ \hline
\end{tabular}
}
\end{table}

\subsubsection{Long-Term Forecasting}\label{sec:shinn_long_term}

\begin{figure}[h]
    \centering
    \includegraphics[width=.9\textwidth]{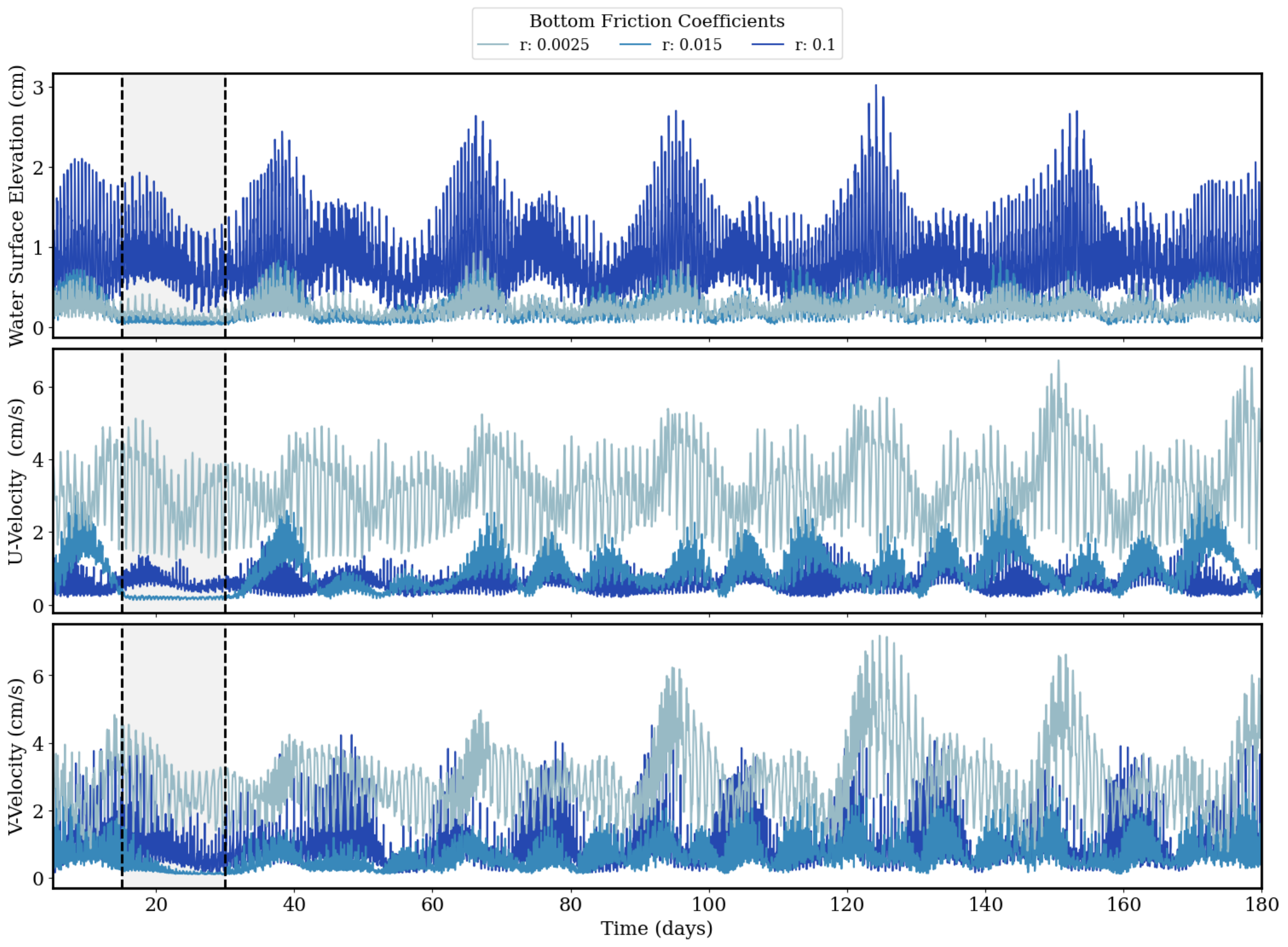}
    \caption{Time series of the $RMSE$ of MITONet predictions for $H$ (Row 1), $U$ (Row 2), and $V$ (Row 3), between days $5$ and $180$ and for all test $r$ values in the Shinnecock Inlet example. The shaded region between days $15$ and $30$ represents the time series window used in the training data.}
    \label{fig:rmse}
\end{figure}

The $RMSE$ time series for all test cases, as shown in Figure \ref{fig:rmse}, support the previous observations while further demonstrating MITONet's stability during a 175 day (8,400 time steps) rollout prediction starting from an unseen IC. Notably, MITONet maintains the level of prediction accuracy throughout the entire prediction horizon, showing minimal signs of deterioration in accuracy due to error accumulation, thus providing further evidence of its ability to perform long-term autoregressive forecasting. In fact, the $\overline{RMSE}$ for a $175$-day rollout prediction using the most challenging test parameters for $H$, $U$, and $V$ increases by only 3.27\%, 5.78\%, and 8.83\%, respectively from the corresponding values for a $55$-day rollout prediction.

\subsubsection{Random Initialization}\label{sec:shinn_random_init}

\begin{figure}[h]
    \centering
    \includegraphics[width=.9\textwidth]{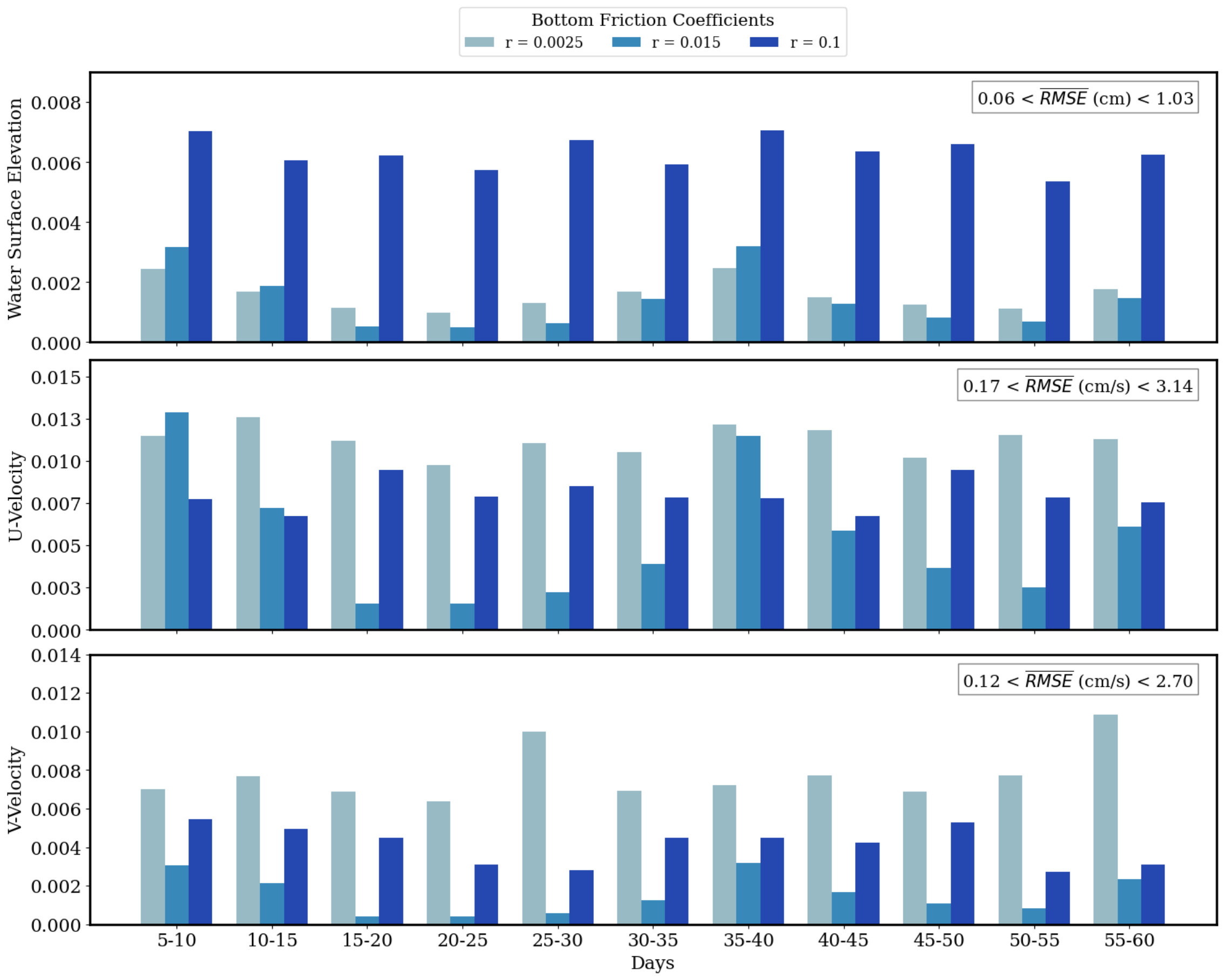}
    \caption{Bar plots of $\overline{NRMSE}$ for MITONet predictions of $H, U,$ and $V$ using three test values of $r$ for the Shinnecock Inlet example. The x-axis represents different 5-day predictions, each initialized with a unique IC. Maximum and minimum values for the $NRMSE$ calculations are computed individually for each $r$ value and $5$-day segment. The range of $\overline{RMSE}$ values for each variable are displayed in separate text boxes in each subplot.}
    \label{fig:random_hot}
\end{figure}

Aside from its robust long-term forecasting capabilities, MITONet can seamlessly handle unseen ICs and provide accurate results. To assess MITONet’s ability to generalize across initial conditions, we generate 5-day predictions from unique ICs. The results of this analysis are presented in Figure \ref{fig:random_hot}, where the computed $\overline{NRMSE}$ values are shown for all $r$ test cases. For most $r$ values, the $\overline{NRMSE}$ is lowest for test cases initialized within the time range of training days 15–30. This likely occurs because BCs within this time period are identical to the training data and hence, the solution variables might exhibit similar features to those encountered during training, even though the $r$ values simulated here were not included in the training set. Although noticeable variations exist in the $\overline{NRMSE}$ across different segments and $r$ values, their small magnitudes suggest that MITONet maintains stable performance across varying ICs. Furthermore, the range of the $\overline{RMSE}$ is included in each plot to provide a sense of their magnitudes for each hydrodynamic variable, which are relatively close to those observed in Table \ref{tab:transposed_metrics_simplified}. This analysis highlights MITONet’s ability to handle different IC and produce consistent results, commonly referred to as a ``hotstart'' in numerical simulations. 
% as confirmed by the relatively uniform $\overline{NRMSE}$ values observed in Figure \ref{fig:random_hot}. 

\begin{figure}[h]
    \centering
    \includegraphics[width=.9\textwidth]{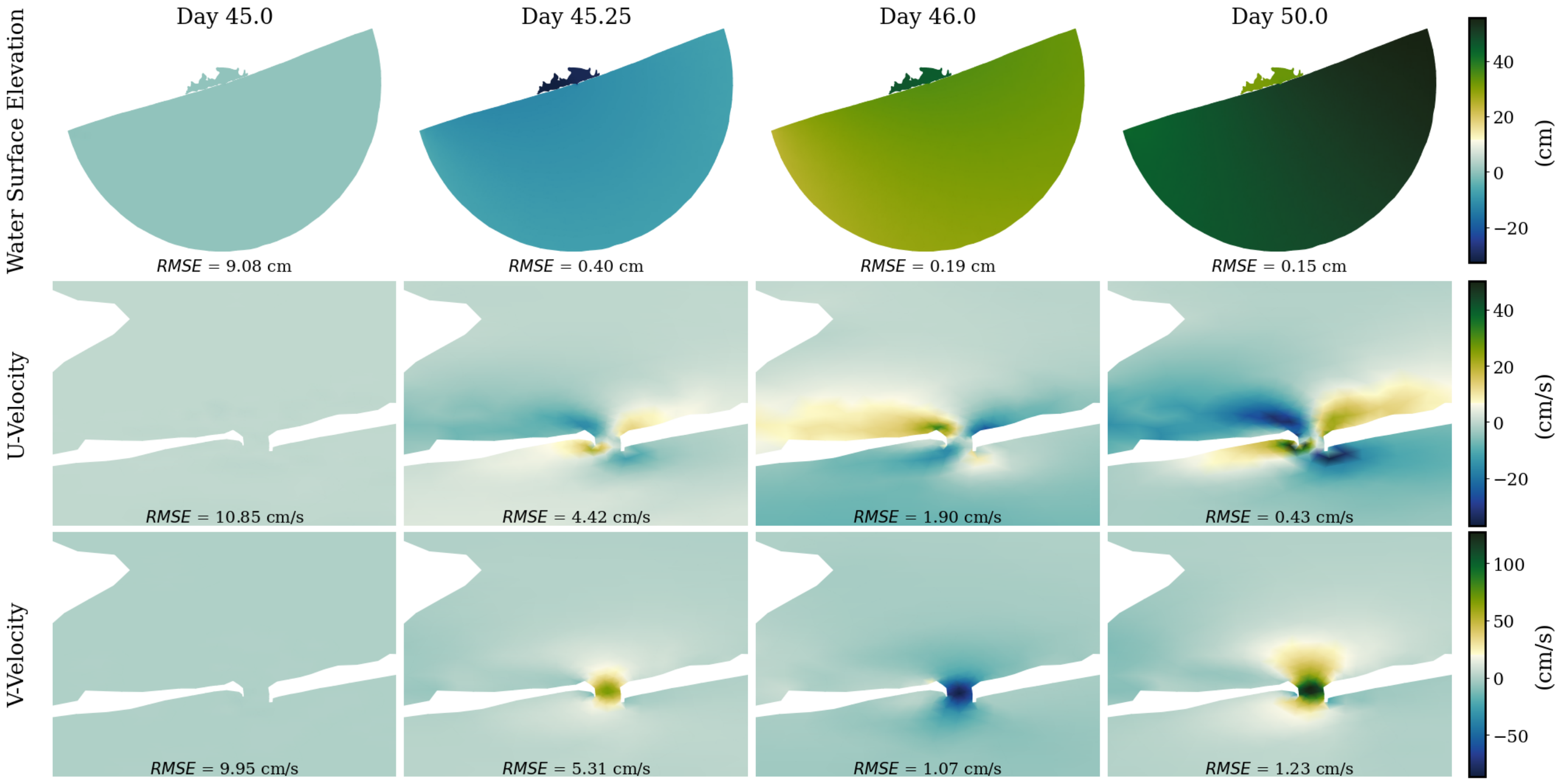}
    \caption{Snapshots of MITONet predictions of $H$ (row 1, full domain), $U$ and $V$ (rows 2 and 3, zoomed in at the inlet) for the Shinnecock Inlet example, with zero as IC, $r=0.015$, and using a $1$-step autoregressive rollout prediction.}
    \label{fig:var_rest}
\end{figure}

Another important advantage of MITONet is the ability to make predictions starting from rest (zero IC). Figure \ref{fig:var_rest} shows the MITONet predictions for $H$, $U$, and $V$ at different time points over a period of $15$ days, starting from zero IC at day $45$ and using $\tau=1$. The $RMSE$ values for each snapshot, as shown in Figure \ref{fig:var_rest}, demonstrate that, similar to any traditional numerical model, the errors diminish progressively over a ramp-up period (roughly 2 days), after which the model is able to accurately emulate all relevant features of the high-fidelity solution (Figure \ref{fig:var_rest}). Besides demonstrating MITONet’s stability and flexibility, these findings strongly indicate that MITONet is learning the governing physics of the system and functioning as an efficient neural emulator.

\subsection{Red River}\label{sec:redriver_results}
To demonstrate the ability of the proposed MITONet framework to emulate PDE solution operators for different types of problems, flow conditions and geometries, an example of realistic riverine dynamics controlled by variable inflow and tailwater elevation conditions in a section of the Red River in Louisiana is chosen as the second computational experiment. As stated, the flow primarily depends on two non-periodic, time-varying boundary conditions imposed at the upstream and downstream boundaries of the study domain (see Figure \ref{fig:red_problem_details}b), thus making the Red River example inherently more complex than the tide-driven dynamics of the Shinnecock Inlet example. As a result, a non-trivial reduction in the model skill in comparison to the previous example can be expected. Nevertheless, MITONet produces stable and accurate results in emulating space–time river dynamics, a task not previously demonstrated in the neural-operator literature, as noted in the Introduction.

As in the Shinnecock Inlet example, MITONet’s accuracy is evaluated using a combination of various error metrics and summary statistics as well as by analyzing predictions at three sensor locations (Figure \ref{fig:red_problem_details}a). These sites were chosen to probe distinct flow regimes and dynamical features. Unless otherwise specified, all MITONet predictions are rolled out from an unseen initial condition at day~$5$.

\begin{figure}[h]
    \centering
    \includegraphics[width=.9\textwidth]{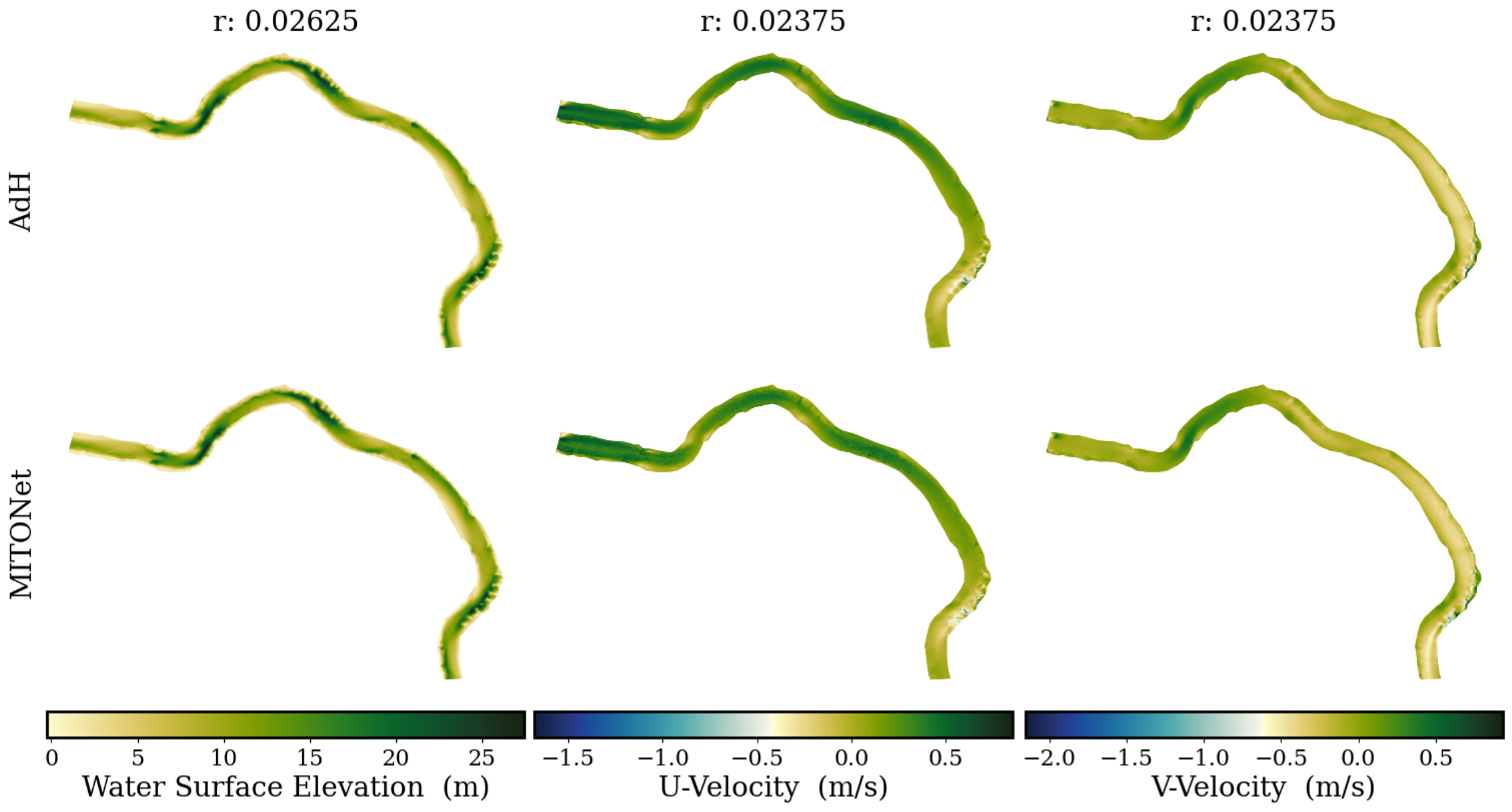}
    \caption{Snapshots of the AdH solution (top row) and MITONet predictions (bottom row) for the Red River example on day 60, using  test values of the bottom friction coefficient: $r=0.02625$ for $H$ (column 1), and $r=0.02375$ for $U$ and $V$ (columns 2 and 3). }
    \label{fig:red_snapshots}
\end{figure}

Figure \ref{fig:red_snapshots} shows snapshots at day~60 following 55 days of autoregressive prediction starting from day 5. For $H$ we show the case $r=0.02625$; for $U$ and $V$ we show $r=0.02375$. In the Shinnecock Inlet example, an exaggerated range of $r$ values is used to introduce significant variability in the parametric solution space, in order to evaluate MITONet's skill in extrapolatory prediction with significant out-of-distribution shifts in the data. In contrast to that, a more natural and practical range of $r$ values have been adopted in this example that represents the expected flow conditions in the chosen section of the Red River. This implies that changes in the value of $r$ do not have extremely pronounced effects in the flow characteristics, unlike the Shinnecock example, and hence there is very minimal deterioration in model skill with change in parameter values. However, keeping in mind the physical behaviors observed in the system variable in the previous example, the highest and lowest values of test bottom friction coefficient (i.e., $r=0.02625$ and $r=0.02375$) are chosen to visualize the model predictions for $H$ and $U,V$, respectively, in all the subsequent domain plots. Overall, the predictions agree well with the AdH solution for all variables, indicating that MITONet can produce stable and accurate predictions of river dynamics over long prediction horizons, even for unseen test parameters.

\begin{figure}[h]
    \centering
    \includegraphics[width=.9\textwidth]{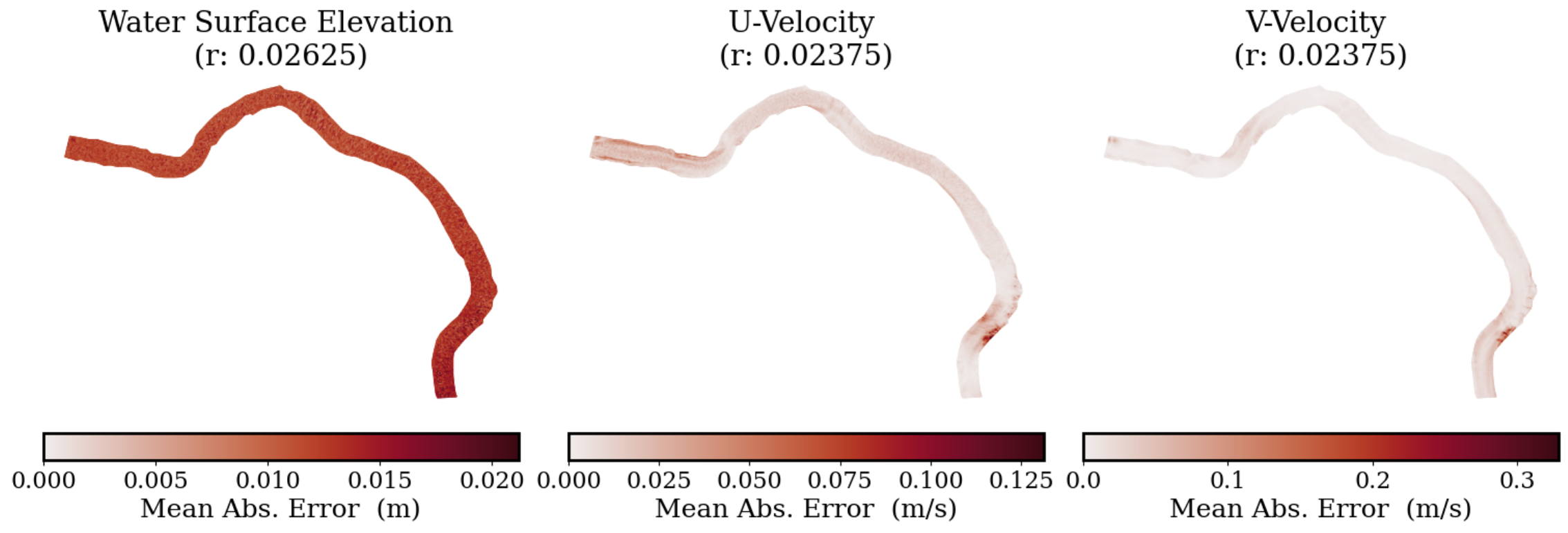}
    \caption{Spatial distribution of the temporal MAE of MITONet predictions for the Red River example, using $r=0.02625$ for $H$ (column 1), and $r=0.02375$ for $U$ and $V$ (columns 2 and 3). }
    \label{fig:red_mae}
\end{figure}

The spatial distribution of errors is summarized in Figure~\ref{fig:red_mae}, which plots the $MAE$ over time at each spatial point for $r=0.02625$ in $H$ and $r=0.02375$ in $U$ and $V$. For $H$, errors are relatively uniform across the domain. For $U$ and $V$, the largest errors occur in shallow bends, particularly toward the tail of the river, where velocities and shear intensify. Overall magnitudes remain within an acceptable range for riverine applications.

\begin{figure}[h]
    \centering
    \includegraphics[width=.9\textwidth]{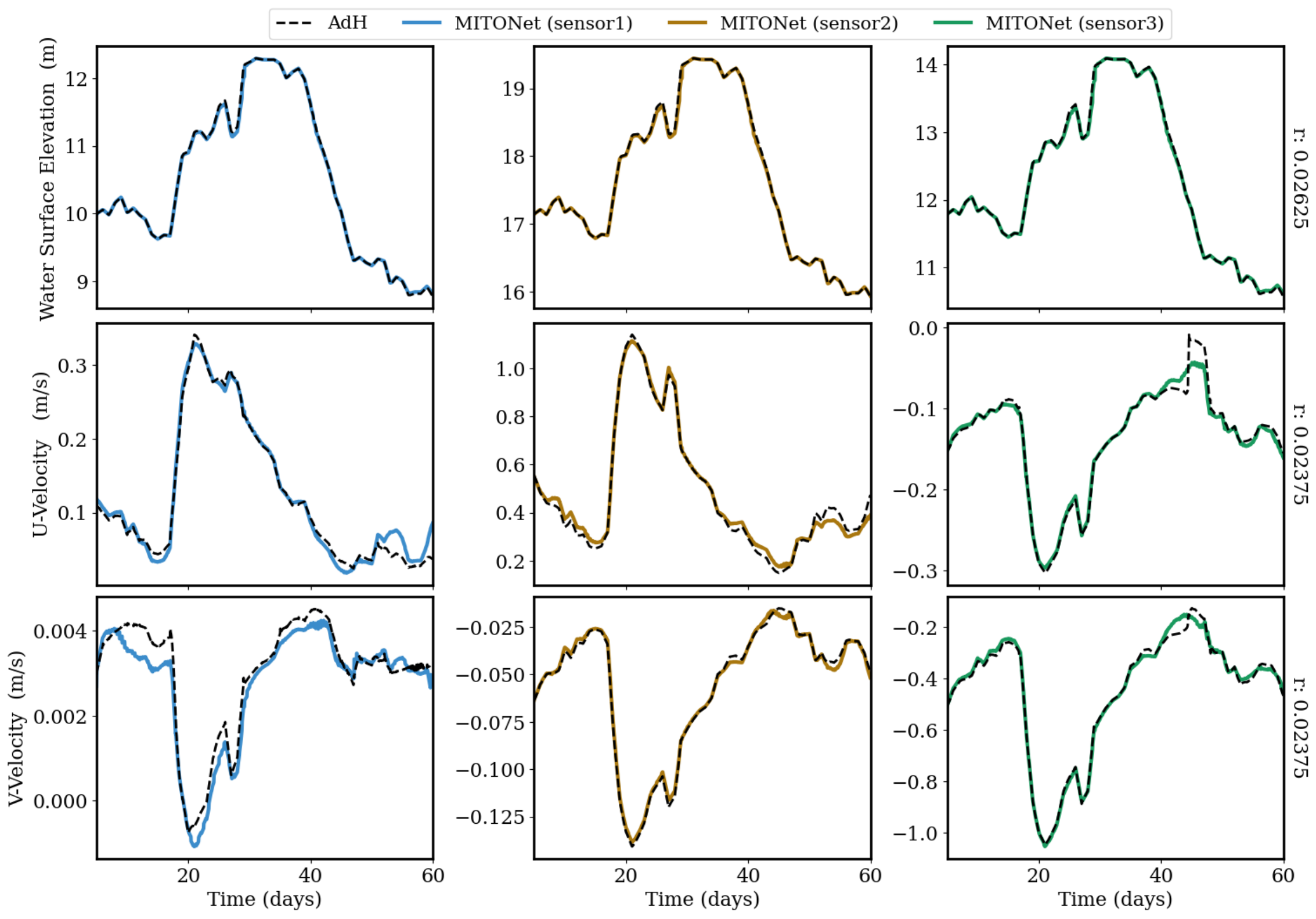}
    \caption{Comparison of the AdH solution (black dashed lines) with the MITONet predictions (blue, yellow, and green lines) for the Red River example at three different sensor locations (see Figure \ref{fig:red_problem_details}a) using test values, $r=0.02625$ for $H$ (Row 1), and $r=0.02375$ for $U$ (Row 2) and $V$ (Row 3).}
    \label{fig:red_mito_sensors}
\end{figure}

Comparisons of AdH and MITONet predictions from day $5$ to day $60$ for all three variables at the three sensors are shown in Figure \ref{fig:red_mito_sensors}, using $r=0.02625$ for $H$ and $r=0.02375$ for $U$ and $V$. The MITONet predictions are rolled out autoregressively from an unseen initial condition at day 5. For $H$, MITONet agrees closely with AdH across all sensor locations. For $U$ and $V$, discrepancies appear at sites with shallower depths, but the model captures the dominant, non-periodic trends at all sensors.

\begin{figure}[h]
    \centering
    \includegraphics[width=.9\textwidth]{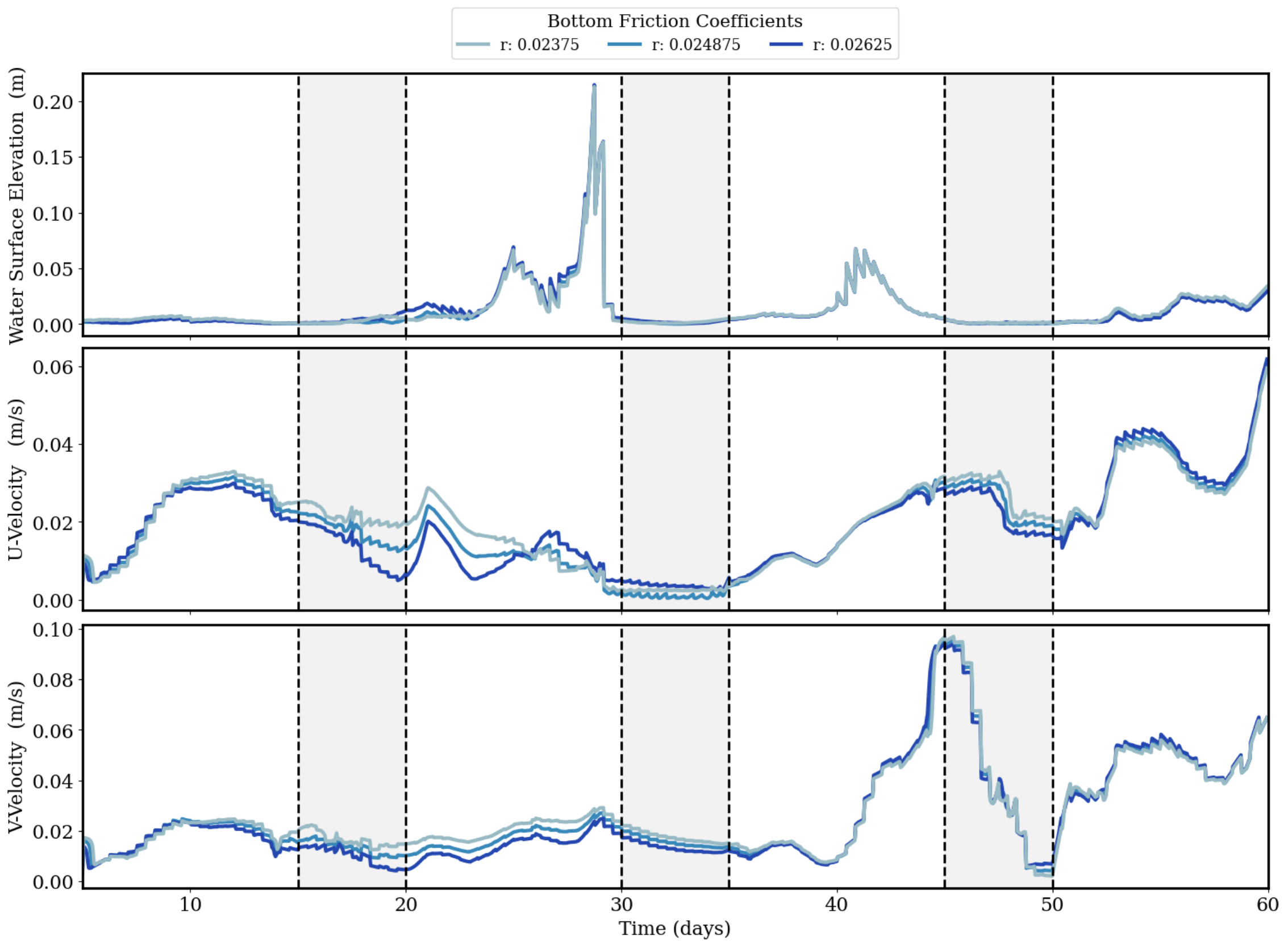}
    \caption{Time series of the $RMSE$ of MITONet predictions for $H$ (Row 1), $U$ (Row 2), and $V$ (Row 3), between days $5$ and $60$ and for all test $r$ values in the Red River example. The shaded regions between days $15$-$20$, $30$-$35$, and $45$-$50$ represent the segments of the time series used in the training data.}
    \label{fig:red_mito_rmse}
\end{figure}

Figure \ref{fig:red_mito_rmse} presents the $RMSE$ time series for all test $r$ values where shaded bands indicate the training windows. Performance is similar across different test $r$ values for $H$, $U$, and $V$, though $H$ exhibits larger errors outside the training window. For $U$ and $V$, peaks in $RMSE$ align with boundary-condition transients—periods of changing inflow discharge and tailwater elevation (see Figure \ref{fig:red_problem_details}b). 

The $\overline{RMSE}$, $\overline{NRMSE}$, and $ACC$ for a $55$-day rollout from an unseen IC (day 5) are summarized in Table~\ref{tab:transposed_metrics_simplified_rr}. As expected for the limited range in $r$, the metrics vary only slightly across rows. Water-surface elevation ($H$) achieves the lowest errors and near-perfect correlations, while velocities ($U$, $V$) show comparatively higher, but still relatively small errors with consistently high $ACC$. The $\overline{NRMSE}$ is $\mathcal{O}(10^{-4})$ for $H$ and $\mathcal{O}(10^{-3})$ for $U$ and $V$, with only minor shifts across $r$, indicating strong performance relative to the dynamic range of the true fields. The highlighted test settings ($r=0.02375$, $0.025$, $0.02625$) follow the same pattern, reinforcing that MITONet maintains strong skill under small parametric shifts.

\renewcommand{\arraystretch}{1.25}
\setlength{\tabcolsep}{8pt}
\begin{table}[h]
\caption{$ACC$, $\overline{RMSE}$ (m/s), and $\overline{NRMSE}$ of MITONet predictions for $H$, $U$, and $V$ with a $55$-day rollout starting at day $5$, across all bottom friction coefficient ($r$) values of the Red River example. Test and validation scenarios are shown in darker and lighter shades of gray, respectively.}
\label{tab:transposed_metrics_simplified_rr}
\centering
\resizebox{\textwidth}{!}{
\begin{tabular}{c c c c c c c c c c}
\hline
& \multicolumn{3}{c}{$\boldsymbol{H}$} & \multicolumn{3}{c}{$\boldsymbol{U}$} & \multicolumn{3}{c}{$\boldsymbol{V}$} \\
\hline
{$\boldsymbol{r}$} & {$\boldsymbol{ACC}$} & $\boldsymbol{\overline{RMSE}}$ & $\boldsymbol{\overline{NRMSE}}$
                   & {$\boldsymbol{ACC}$} & $\boldsymbol{\overline{RMSE}}$ & $\boldsymbol{\overline{NRMSE}}$
                   & {$\boldsymbol{ACC}$} & $\boldsymbol{\overline{RMSE}}$ & $\boldsymbol{\overline{NRMSE}}$ \\
\hline
\rowcolor{gray50}\textbf{0.02375}  & 1.000 & 0.0130 & 0.000417 & 0.985 & 0.0203 & 0.00558 & 0.968 & 0.0274 & 0.00859 \\ \hline
\textbf{0.023875} & 1.000 & 0.0130 & 0.000415 & 0.985 & 0.0201 & 0.00554 & 0.968 & 0.0272 & 0.00854 \\ \hline
\rowcolor{gray60}\textbf{0.024125} & 1.000 & 0.0128 & 0.000411 & 0.985 & 0.0198 & 0.00547 & 0.968 & 0.0270 & 0.00847 \\ \hline
\textbf{0.024375} & 1.000 & 0.0127 & 0.000407 & 0.985 & 0.0195 & 0.00540 & 0.968 & 0.0267 & 0.00840 \\ \hline
\rowcolor{gray60}\textbf{0.024625} & 1.000 & 0.0126 & 0.000405 & 0.986 & 0.0193 & 0.00534 & 0.968 & 0.0265 & 0.00833 \\ \hline
\textbf{0.024875} & 1.000 & 0.0126 & 0.000404 & 0.986 & 0.0191 & 0.00530 & 0.968 & 0.0262 & 0.00826 \\ \hline
\rowcolor{gray50}\textbf{0.025000} & 1.000 & 0.0126 & 0.000404 & 0.986 & 0.0190 & 0.00528 & 0.968 & 0.0261 & 0.00822 \\ \hline
\textbf{0.025125} & 1.000 & 0.0126 & 0.000405 & 0.986 & 0.0189 & 0.00527 & 0.968 & 0.0260 & 0.00819 \\ \hline
\rowcolor{gray60}\textbf{0.025375} & 1.000 & 0.0127 & 0.000407 & 0.986 & 0.0188 & 0.00525 & 0.968 & 0.0258 & 0.00812 \\ \hline
\textbf{0.025625} & 1.000 & 0.0128 & 0.000410 & 0.986 & 0.0187 & 0.00523 & 0.968 & 0.0255 & 0.00805 \\ \hline
\rowcolor{gray60}\textbf{0.025875} & 1.000 & 0.0129 & 0.000414 & 0.986 & 0.0187 & 0.00522 & 0.968 & 0.0253 & 0.00799 \\ \hline
\textbf{0.026125} & 1.000 & 0.0131 & 0.000419 & 0.986 & 0.0186 & 0.00522 & 0.968 & 0.0251 & 0.00793 \\ \hline
\rowcolor{gray50}\textbf{0.02625}  & 1.000 & 0.0132 & 0.000422 & 0.986 & 0.0186 & 0.00522 & 0.968 & 0.0250 & 0.00790 \\ \hline
\end{tabular}
}
\end{table}

While the present test set spans 60 days, future work will examine longer horizons and additional strategies to stabilize autoregressive rollouts. Further improvements will also target shallow regions and intervals with rapidly varying boundary conditions.

\subsection{Discussion}\label{sec:discussion}

This work introduces a neural operator–based emulator that delivers fast predictions and robust performance on out-of-distribution data, when evaluated with two distinct shallow-water flow regimes. The proposed MITONet framework uses the same inputs as traditional numerical models, such as the initial condition (IC), boundary conditions (BC), and relevant domain parameters. We demonstrate performance across distinct domain geometries, mesh resolutions, and discretizations. Once trained, however, MITONet does not transfer to unseen geometries or have the ability to spatially interpolate/extrapolate the solution to arbitrary points in the domain. Across benchmarks, MITONet outperforms popular DeepONet variants in model accuracy and stability by almost an order of magnitude, especially in longer-term prediction horizons. MITONet shows remarkable robustness over a $175$-day rollout forecast ($8,400$ time steps) in the Shinnecock example. The framework also achieves substantial speed-ups, up to $1,000\text{x}$ relative to the finite element-based numerical solver, AdH in the Red River example. The ability to emulate tide-driven coastal circulation patterns and advection-dominated riverine shallow-water dynamics with vastly different geometries, spatio-temporal scales, driving forces, provides strong evidence that the architecture is flexible and can be extended to other PDE-governed systems.

A key strength of MITONet is its ability to generalize in both time and parameter space while remaining computationally efficient. The model adapts to missing or random ICs, unseen values of $r$, and long forecast horizons, enabling rapid analyses and parametric sweeps that can be valuable for design and safety. That said, physics-based models like ADCIRC and AdH may retain advantages for extreme events that lie far outside the training distribution. As with most machine-learning models, performance can degrade when encountered dynamics differ substantially from those represented in the training data. Improving extrapolation requires a deeper understanding of the relationship between training coverage and forecast skill, as well as incorporating uncertainty quantification to improve confidence under distribution shift. These directions, along with strategies for geometry transfer and mesh-aware modeling, constitute important avenues for future work.

Moreover, a key advantage of many neural operator architectures such as DON, M-DON, and MIONet is the property of discretization invariance in the target function space. However, similar to Latent DeepONet, the proposed MITONet framework lacks the ability to spatially interpolate/extrapolate the solution to any arbitrary spatial points in the domain, due to the introduction of the autoencoder-generated latent space. While learning an operator map in the latent space has proven benefits both in terms of improved convergence during training as well as reduced computational demand in comparison to architectures that directly learn from data in high-dimensional physical spaces \cite{kontolati2023learning}, the inability to interpolate the solution spatially can be a limitation in some applications. Therefore, developing suitable extensions of this architecture that are capable of combining the advantages of latent space learning with the flexibility of discretization invariance, while also enabling the use of physics-based residual losses during training would be an exciting direction of future work.

\section{Conclusions}\label{sec:conclusion}

In this work, we introduced MITONet, a data-driven, physics-aware neural operator framework, that is capable of approximating PDE operator maps from a space spanned by multiple input functions and parameters to a specific target function space such as that of the PDE solution function. We demonstrated MITONet's capabilities in autoregressively approximating the long-term evolution of the solution operator of a complex nonlinear PDE, by considering two realistic computational examples of tidal and riverine flows governed by the parametric, time-dependent, 2D shallow water equations, while incorporating the effect of multiple input functions, such as variable initial conditions, time-dependent boundary conditions, and variable scalar domain parameters. 

The significant computational efficiency over traditional numerical solvers in predicting hydrodynamics in real-world domains, such as the Shinnecock Inlet in NY and a part of the Red River in LA, showcased MITONet's capability in providing accurate, real-time predictions of complex flow phenomena with multiscale features. The improved prediction accuracy and stability over long-time forecast horizons in comparison to other well-known DeepONet-based architectures, and the ability to extrapolate to both unseen parameter values and boundary forcings as well as generate autoregressive predictions from random or zero initial conditions, highlighted its potential as an effective neural emulator for different shallow-water applications, and possibly even for systems governed by other PDEs.  

Nevertheless, there are several design limitations of the current framework that provide exciting avenues of future work. For instance, exploring other relatively modern architectures such as graph convolutional networks and implicit neural representations would potentially lead to more memory and computationally efficient autoencoder models for simulation data on unstructured meshes and irregular geometries. Moreover, developing an integrated training workflow that trains such autoencoder models with the corresponding MITONet models would address the lack of discretization invariance in the spatial domain of the PDE solution function. Also, developing techniques to seamlessly model extreme dynamics like coastal storm surge or flood events, and exploring methods to enhance MITONet’s robustness for conditions beyond the training distribution, including strategies for improved extrapolation and uncertainty quantification will significantly enhance the applicability and reliability of the proposed MITONet framework.

\section*{Open Research Section}
The ADCIRC data for the Shinnecock Inlet example \cite{casillas2024designsafe} and the AdH data for the Red River example \cite{RedRiverData2024} are publicly available on DesignSafe \cite{rathje2017designsafe}. The source code for the MITONet framework is available at \url{https://github.com/erdc/SW_MITONet}. The repository will be made public upon publication of this article.

\acknowledgments
The authors would like to acknowledge the valuable support from the U.S. Army Engineer Research and Development Center (ERDC), in part through the Long Term Training program and also under the Laboratory University Collaboration Initiative (LUCI) with Brown University. The authors would like to thank Prof.~George Karniadakis for his guidance throughout this work, and Dr.~Eirik Valseth for insightful discussions and constructive suggestions that have  contributed to the improvement of this work. This work was supported in part by high-performance computer time and resources from the DoD High Performance Computing Modernization Program (HPCMP) and the Texas Advanced Computing Center (TACC) at The University of Texas at Austin. Permission was granted by the Chief of Engineers to publish this information.

\nocite{luettich2004formulation}
\nocite{savant2011adh}
\nocite{dutta_farthing_2021}
\nocite{akiba2019optuna}
\nocite{hutter2014fanova}
\nocite{kontolati2023learning}
\nocite{dutta2021data}

\bibliography{references}

@article{raissi2019physics,
  title={Physics-informed neural networks: A deep learning framework for solving forward and inverse problems involving nonlinear partial differential equations},
  author={Raissi, Maziar and Perdikaris, Paris and Karniadakis, George E},
  journal={Journal of Computational physics},
  volume={378},
  pages={686--707},
  year={2019},
  publisher={Elsevier}
}

@article{chen1995universal,
  title={Universal approximation to nonlinear operators by neural networks with arbitrary activation functions and its application to dynamical systems},
  author={Chen, Tianping and Chen, Hong},
  journal={IEEE transactions on neural networks},
  volume={6},
  number={4},
  pages={911--917},
  year={1995},
  publisher={IEEE}
}

@article{lu2021learning,
  title={Learning nonlinear operators via DeepONet based on the universal approximation theorem of operators},
  author={Lu, Lu and Jin, Pengzhan and Pang, Guofei and Zhang, Zhongqiang and Karniadakis, George Em},
  journal={Nature machine intelligence},
  volume={3},
  number={3},
  pages={218--229},
  year={2021},
  publisher={Nature Publishing Group UK London}
}

@inproceedings{akiba2019optuna,
  title={Optuna: A next-generation hyperparameter optimization framework},
  author={Akiba, Takuya and Sano, Shotaro and Yanase, Toshihiko and Ohta, Takeru and Koyama, Masanori},
  booktitle={{Proceedings of the 25th ACM SIGKDD International Conference on Knowledge Discovery \& Data Mining}},
  pages={2623--2631},
  year={2019}
}

@techreport{luettich1992adcirc,
    author = {Luettich, Richard Albert and Westerink, Joannes J and Scheffner, Norman W},
    title = {{ADCIRC}: An advanced three-dimensional circulation model for shelves, coasts, and estuaries. {Report 1, Theory and methodology of }{ADCIRC-2DD1} and {ADCIRC-3DL}},
    institution = {US Army Engineer Waterways Experimentation Station},
    address = {Vicksburg, MS},
    year = {1992}, 
    type = {Technical Report {DRP-92-6}}
}

@book{morang1999shinnecock,
  title={Shinnecock Inlet, New York, Site Investigation, Report 1: Morphology and Historical Behaviour},
  author={Morang, Andrew},
  year={1999},
  publisher={US Army Corps of Engineers Waterways Experiment Station}
}

@article{militello2001shinnecock,
  title={Shinnecock Inlet, New York, site investigation, Report 4, Evaluation of flood and ebb shoal sediment source alternatives for the west of Shinnecock Interim Project},
  author={Militello, Adele and Kraus, Nicholas C},
  journal={Coastal Inlets Research Program Technical Report ERDC-CHL-TR-98-32. US Army Engineer Research and Development Center, Vicksburg, MS},
  year={2001}
}

@book{luettich2004formulation,
  title={Formulation and numerical implementation of the {2D/3D ADCIRC} finite element model version 44.XX},
  author={Luettich, Richard Albert and Westerink, Joannes J},
  volume={20},
  year={2004},
  publisher={R. Luettich Chapel Hill, NC, USA}
}

@article{jin2022mionet,
  title={{MIONet: Learning multiple-input operators via tensor product}},
  author={Jin, Pengzhan and Meng, Shuai and Lu, Lu},
  journal={SIAM Journal on Scientific Computing},
  volume={44},
  number={6},
  pages={A3490--A3514},
  year={2022},
  doi={10.1137/22M1477751},
  publisher={SIAM}
}

@misc{ovadia2023ditto,
      title={Real-time Inference and Extrapolation via a Diffusion-inspired Temporal Transformer Operator {(DiTTO)}}, 
      author={Oded Ovadia and Vivek Oommen and Adar Kahana and Ahmad Peyvan and Eli Turkel and George Em Karniadakis},
      year={2023},
      eprint={2307.09072},
      archivePrefix={arXiv},
      primaryClass={cs.LG},
      url={https://arxiv.org/abs/2307.09072}, 
}

@article{wang2022improved,
  title={Improved architectures and training algorithms for deep operator networks},
  author={Wang, Sifan and Wang, Hanwen and Perdikaris, Paris},
  journal={Journal of Scientific Computing},
  volume={92},
  number={2},
  pages={35},
  year={2022},
  publisher={Springer}
}

@article{kontolati2023learning,
  title={Learning in latent spaces improves the predictive accuracy of deep neural operators},
  author={Kontolati, Katiana and Goswami, Somdatta and Karniadakis, George Em and Shields, Michael D},
  journal={arXiv preprint arXiv:2304.07599},
  year={2023}
}

@article{patel2024variationally,
  title={Variationally mimetic operator networks},
  author={Patel, Dhruv and Ray, Deep and Abdelmalik, Michael RA and Hughes, Thomas JR and Oberai, Assad A},
  journal={Computer Methods in Applied Mechanics and Engineering},
  volume={419},
  pages={116536},
  year={2024},
  publisher={Elsevier}
}

@INPROCEEDINGS{10337380,
  author={Rajagopal, Ellery and Babu, Anantha N. S. and Ryu, Tony and Haley, Patrick J. and Mirabito, Chris and Lermusiaux, Pierre F. J.},
  booktitle={OCEANS 2023 - MTS/IEEE U.S. Gulf Coast}, 
  title={Evaluation of Deep Neural Operator Models Toward Ocean Forecasting}, 
  year={2023},
  volume={},
  number={},
  pages={1-9},
  doi={10.23919/OCEANS52994.2023.10337380}}

@article{choi2024applications,
  title={Applications of the Fourier neural operator in a regional ocean modeling and prediction},
  author={Choi, Byoung-Ju and Jin, Hong Sung and Lkhagvasuren, Bataa},
  journal={Frontiers in Marine Science},
  volume={11},
  pages={1383997},
  year={2024},
  publisher={Frontiers Media SA}
}

@article{brandstetter2022message,
  title={Message passing neural PDE solvers},
  author={Brandstetter, Johannes and Worrall, Daniel and Welling, Max},
  journal={arXiv preprint arXiv:2202.03376},
  year={2022}
}

@inbook{oppenheimer2019sea, 
place={Cambridge, UK and New York, NY, USA}, 
title={Sea Level Rise and Implications for Low-Lying Islands, Coasts and Communities}, 
booktitle={{IPCC Special Report on the Ocean and Cryosphere in a Changing Climate}}, 
publisher={Cambridge University Press}, 
author={Oppenheimer, M. and B.C. Glavovic and J. Hinkel and R. van de Wal and A.K. Magnan and A. Abd-Elgawad and R. Cai and M. Cifuentes-Jara and R.M. DeConto and T. Ghosh and J. Hay and F. Isla and B. Marzeion and B. Meyssignac and Z. Sebesvari}, 
year={2019}, 
pages={321–445},
doi = {10.1017/9781009157964.006}
}

@article{karim2023review,
  title={A review of hydrodynamic and machine learning approaches for flood inundation modeling},
  author={Karim, Fazlul and Armin, Mohammed Ali and Ahmedt-Aristizabal, David and Tychsen-Smith, Lachlan and Petersson, Lars},
  journal={Water},
  volume={15},
  number={3},
  pages={566},
  year={2023},
  publisher={MDPI}
}

@article{abouhalima2024machine,
  title={Machine Learning in Coastal Engineering: Applications, Challenges, and Perspectives},
  author={Abouhalima, Mahmoud and das Neves, Luciana and Taveira-Pinto, Francisco and Rosa-Santos, Paulo},
  journal={Journal of Marine Science and Engineering},
  volume={12},
  number={4},
  pages={638},
  year={2024},
  publisher={MDPI}
}

@article{li2020fourier,
  title={Fourier neural operator for parametric partial differential equations},
  author={Li, Zongyi and Kovachki, Nikola and Azizzadenesheli, Kamyar and Liu, Burigede and Bhattacharya, Kaushik and Stuart, Andrew and Anandkumar, Anima},
  journal={arXiv preprint arXiv:2010.08895},
  year={2020}
}

@article{patel2018nonlinear,
  title={Nonlinear integro-differential operator regression with neural networks},
  author={Patel, Ravi G and Desjardins, Olivier},
  journal={arXiv preprint arXiv:1810.08552},
  year={2018}
}

@article{li2020neural,
  title={Neural operator: Graph kernel network for partial differential equations},
  author={Li, Zongyi and Kovachki, Nikola and Azizzadenesheli, Kamyar and Liu, Burigede and Bhattacharya, Kaushik and Stuart, Andrew and Anandkumar, Anima},
  journal={arXiv preprint arXiv:2003.03485},
  year={2020}
}

@article{cao2021choose,
  title={Choose a transformer: Fourier or galerkin},
  author={Cao, Shuhao},
  journal={Advances in neural information processing systems},
  volume={34},
  pages={24924--24940},
  year={2021}
}

@article{azizzadenesheli2024neural,
  title={Neural operators for accelerating scientific simulations and design},
  author={Azizzadenesheli, Kamyar and Kovachki, Nikola and Li, Zongyi and Liu-Schiaffini, Miguel and Kossaifi, Jean and Anandkumar, Anima},
  journal={Nature Reviews Physics},
  pages={1--9},
  year={2024},
  publisher={Nature Publishing Group UK London}
}

@article{alkin2024universal,
  title={Universal physics transformers},
  author={Alkin, Benedikt and F{\"u}rst, Andreas and Schmid, Simon and Gruber, Lukas and Holzleitner, Markus and Brandstetter, Johannes},
  journal={arXiv preprint arXiv:2402.12365},
  year={2024}
}

@article{lim2021time,
  title={Time-series forecasting with deep learning: a survey},
  author={Lim, Bryan and Zohren, Stefan},
  journal={Philosophical Transactions of the Royal Society A},
  volume={379},
  number={2194},
  pages={20200209},
  year={2021},
  publisher={The Royal Society Publishing}
}

@article{10.1145/3533382,
author = {Benidis, Konstantinos and Rangapuram, Syama Sundar and Flunkert, Valentin and Wang, Yuyang and Maddix, Danielle and Turkmen, Caner and Gasthaus, Jan and Bohlke-Schneider, Michael and Salinas, David and Stella, Lorenzo and Aubet, Fran\c{c}ois-Xavier and Callot, Laurent and Januschowski, Tim},
title = {Deep Learning for Time Series Forecasting: Tutorial and Literature Survey},
year = {2022},
issue_date = {June 2023},
publisher = {Association for Computing Machinery},
address = {New York, NY, USA},
volume = {55},
number = {6},
issn = {0360-0300},
url = {https://doi.org/10.1145/3533382},
doi = {10.1145/3533382},
journal = {ACM Comput. Surv.},
month = dec,
articleno = {121},
numpages = {36}
}

@article{lam2022graphcast,
  title={GraphCast: Learning skillful medium-range global weather forecasting},
  author={Lam, Remi and Sanchez-Gonzalez, Alvaro and Willson, Matthew and Wirnsberger, Peter and Fortunato, Meire and Alet, Ferran and Ravuri, Suman and Ewalds, Timo and Eaton-Rosen, Zach and Hu, Weihua and others},
  journal={arXiv preprint arXiv:2212.12794},
  year={2022}
}

@article{bodnar2024aurora,
  title={Aurora: A foundation model of the atmosphere},
  author={Bodnar, Cristian and Bruinsma, Wessel P and Lucic, Ana and Stanley, Megan and Brandstetter, Johannes and Garvan, Patrick and Riechert, Maik and Weyn, Jonathan and Dong, Haiyu and Vaughan, Anna and others},
  journal={arXiv preprint arXiv:2405.13063},
  year={2024}
}

@article{chattopadhyay2024oceannet,
  title={OceanNet: A principled neural operator-based digital twin for regional oceans},
  author={Chattopadhyay, Ashesh and Gray, Michael and Wu, Tianning and Lowe, Anna B and He, Ruoying},
  journal={Scientific Reports},
  volume={14},
  number={1},
  pages={21181},
  year={2024},
  publisher={Nature Publishing Group UK London}
}

@article{giaremis2024storm,
  title={{Storm surge modeling in the AI era: Using LSTM-based machine learning for enhancing forecasting accuracy}},
  author={Giaremis, Stefanos and Nader, Noujoud and Dawson, Clint and Kaiser, Carola and Nikidis, Efstratios and Kaiser, Hartmut},
  journal={Coastal Engineering},
  volume={191},
  pages={104532},
  year={2024},
  publisher={Elsevier}
}

@article{pachev2023framework,
  title={A framework for flexible peak storm surge prediction},
  author={Pachev, Benjamin and Arora, Prateek and del-Castillo-Negrete, Carlos and Valseth, Eirik and Dawson, Clint},
  journal={Coastal Engineering},
  volume={186},
  pages={104406},
  year={2023},
  publisher={Elsevier}
}

@inproceedings{kingma2015adam,
  author = {Kingma, Diederik P. and Ba, Jimmy},
  booktitle = {{3rd International Conference for Learning Representations (ICLR)}}, 
  address = {San Diego},
  editor = {Bengio, Yoshua and LeCun, Yann},
  title = {Adam: A Method for Stochastic Optimization.},
  year = 2015
}

@inproceedings{stanzione2020frontera,
author = {Stanzione, Dan and West, John and Evans, R. Todd and Minyard, Tommy and Ghattas, Omar and Panda, Dhabaleswar K.},
title = {Frontera: The Evolution of Leadership Computing at the National Science Foundation},
year = {2020},
publisher = {Association for Computing Machinery},
address = {New York, NY, USA},
doi = {10.1145/3311790.3396656},
booktitle = {{Practice and Experience in Advanced Research Computing 2020: Catch the Wave}},
pages = {106–111},
numpages = {6},
location = {Portland, OR, USA},
series = {PEARC '20}
}

@inproceedings{hutter2014fanova, 
author = {F. Hutter and H. Hoos and K. Leyton-Brown}, 
title = {An Efficient Approach for Assessing Hyperparameter Importance}, 
booktitle = {{Proceedings of International Conference on Machine Learning 2014 (ICML 2014)}}, 
year = {2014}, 
pages = {754–762}, 
month = jun, }

@misc{casillas2024designsafe,
  doi = {10.17603/DS2-W533-0N80},
  author = {Rivera-Casillas, Peter and Dutta, Sourav and Cai, Shukai and Loveland, Mark and Nath, Kamaljyoti and Shukla, Khemraj and Trahan, Corey and Lee, Jonghyun and Farthing, Matthew and Dawson, Clinton N},
  title = {{2D ADCIRC simulation of tidal flow in Shinnecock Inlet, NY parameterized by bottom friction coefficient}},
  publisher = {Designsafe-CI},
  year = {2024}
}

@article{lin2022numerical,
  title={NUMERICAL MODELING OF SHINNECOCK INLET, NEW YORK, FOR COASTAL EROSION CONTROL SUPPORT AND INLET SEDIMENT MANAGEMENT},
  author={Lin, Lihwa and Demirbilek, Zeki and Greenblatt, Elisheva R and Rice, Suzana S and Buonaiuto, Frank},
  journal={Coastal Engineering Proceedings},
  number={37},
  pages={51--51},
  year={2022}
}

@techreport{RTI2015_adcirc,
  author      = "{Riverside Technology, Inc.} and {AECOM}",
  title       = "{Mesh Development, Tidal Validation, and Hindcast Skill Assessment of an Adcirc Model for the Hurricane Storm Surge Operational Forecast System On the US Gulf-Atlantic Coast}",
  institution = "Report prepared for the National Oceanic and Atmospheric Administration/National Ocean Service, Coast Survey Development Laboratory, Office of Coast Survey",
  doi = "10.17615/4z19-y130",
  year = "2015"
}

@article{dutta2021data,
  title={Data-driven reduced order modeling of environmental hydrodynamics using deep autoencoders and neural ODEs},
  author={Dutta, Sourav and Rivera-Casillas, Peter and Cecil, Orie M and Farthing, Matthew W and Perracchione, Emma and Putti, Mario},
  journal={arXiv preprint arXiv:2107.02784},
  year={2021}
}

@article{karniadakis_review_2021, 
title={Physics-informed machine learning}, 
volume={3}, 
DOI={10.1038/s42254-021-00314-5}, 
number={6}, journal={Nature Reviews Physics}, 
publisher={Springer Science and Business Media LLC}, 
author={Karniadakis, George Em and Kevrekidis, Ioannis G. and Lu, Lu and Perdikaris, Paris and Wang, Sifan and Yang, Liu}, 
year={2021}, 
pages={422–440} }

@article{brunton_review_2019, 
title={Machine Learning for Fluid Mechanics}, 
volume={52}, 
DOI={10.1146/annurev-fluid-010719-060214}, 
number={1}, 
journal={Annual Review of Fluid Mechanics}, 
publisher={Annual Reviews}, 
author={Brunton, Steven L and Noack, Bernd R and Petros Koumoutsakos}, 
year={2019},
pages={477–508} }

@article{dutta_farthing_2021, 
title={A greedy non-intrusive reduced order model for shallow water equations}, 
volume={439},
DOI={10.1016/j.jcp.2021.110378}, 
journal={Journal of Computational Physics}, 
publisher={Elsevier BV}, 
author={Dutta, Sourav and Farthing, Matthew W and Perracchione, Emma and Savant, Gaurav and Putti, Mario}, 
year={2021}, 
pages={110378–110378} }

@article{dutta_rc_2021, 
title={{pyNIROM—A suite of python modules for non-intrusive reduced order modeling of time-dependent problems}}, 
volume={10},
DOI={10.1016/j.simpa.2021.100129}, 
journal={Software Impacts}, 
publisher={Elsevier BV}, 
author={Dutta, Sourav and Rivera-Casillas, Peter and Cecil, Orie M and Farthing, Matthew W}, 
year={2021}, 
pages={100129–100129} }

@article{alifu2022enhancement,
  title={Enhancement of river flooding due to global warming},
  author={Alifu, Haireti and Hirabayashi, Yukiko and Imada, Yukiko and Shiogama, Hideo},
  journal={Scientific Reports},
  volume={12},
  number={1},
  pages={20687},
  year={2022},
  publisher={Nature Publishing Group UK London}
}

@article{arnell2016impacts,
  title={The impacts of climate change on river flood risk at the global scale},
  author={Arnell, Nigel W and Gosling, Simon N},
  journal={Climatic Change},
  volume={134},
  number={3},
  pages={387--401},
  year={2016},
  publisher={Springer}
}

@article{holmberg2025accelerating,
  title={Accelerating HEC-RAS: A Recurrent Neural Operator for Rapid River Forecasting},
  author={Holmberg, Edward and Pokhrel, Pujan and Zoch, Maximilian and Ioup, Elias and Pathak, Ken and Sloan, Steven and Niles, Kendall and Ratcliff, Jay and Flanagin, Maik and Guetl, Christian and others},
  journal={arXiv preprint arXiv:2507.15614},
  year={2025}
}

@article{pang2024efficient,
  title={Efficient river hydrodynamics modelling in realistic river systems using a Fourier neural operator-based network},
  author={Pang, Min},
  journal={Journal of Hydrology},
  volume={637},
  pages={131345},
  year={2024},
  publisher={Elsevier}
}

@article{chen2025graph,
  title={Graph-enhanced neural operator for missing velocities reconstruction in river surface velocimetry},
  author={Chen, Xueqin and Winsemius, Hessel and Taormina, Riccardo},
  journal={Machine Learning: Earth},
  volume={1},
  number={1},
  pages={015006},
  year={2025},
  publisher={IOP Publishing}
}

@inproceedings{sun2023rapid,
  title={Rapid flood inundation forecast using Fourier neural operator},
  author={Sun, Alexander Y and Li, Zhi and Lee, Wonhyun and Huang, Qixing and Scanlon, Bridget R and Dawson, Clint},
  booktitle={Proceedings of the IEEE/CVF international conference on computer Vision},
  pages={3733--3739},
  year={2023}
}

@article{sun2024bridging,
  title={Bridging hydrological ensemble simulation and learning using deep neural operators},
  author={Sun, Alexander Y and Jiang, Peishi and Shuai, Pin and Chen, Xingyuan},
  journal={Water Resources Research},
  volume={60},
  number={10},
  pages={e2024WR037555},
  year={2024},
  publisher={Wiley Online Library}
}

@article{trahan2018adh,
title = {Formulation and application of the {Adaptive Hydraulics} three-dimensional shallow water and transport models},
journal = {Journal of Computational Physics},
volume = {374},
pages = {47-90},
year = {2018},
doi = {https://doi.org/10.1016/j.jcp.2018.04.055},
author = {C.J. Trahan and G. Savant and R.C. Berger and M. Farthing and T.O. McAlpin and L. Pettey and G.K. Choudhary and C.N. Dawson},
}

@article{martin2012adh,
author = {S. Keith Martin  and Gaurav Savant  and Darla C. McVan },
title = {Two-Dimensional Numerical Model of the Gulf Intracoastal Waterway near {New Orleans}},
journal = {Journal of Waterway, Port, Coastal, and Ocean Engineering},
volume = {138},
number = {3},
pages = {236-245},
year = {2012},
doi = {10.1061/(ASCE)WW.1943-5460.0000119},
URL = {https://ascelibrary.org/doi/abs/10.1061/%28ASCE%29WW.1943-5460.0000119},
}

@misc{loshchilov2019adamw,
      title={Decoupled Weight Decay Regularization}, 
      author={Ilya Loshchilov and Frank Hutter},
      year={2019},
      eprint={1711.05101},
      archivePrefix={arXiv},
      primaryClass={cs.LG},
      url={https://arxiv.org/abs/1711.05101}, 
}

@article{pachev2023adcirce3sm,
author = {Pachev, B. and Leung, L.R. and Zhou, T. and Dawson, C.},
title  = {{One-way coupling of E3SM with ADCIRC demonstrated on Hurricane Harvey}},
journal= {Natural Hazards},
volume = {119},
pages  = {2063-2087},
year   = {2023},
doi    = {10.1007/s11069-023-06192-7},
}

@TECHREPORT{pevey2020adhcolumbia,
  title     = "Lower {Columbia} River {Adaptive Hydraulics} ({AdH}) model :
               development, water surface elevation validation, and sea level
               rise analysis",
  author    = "Pevey, Kimberley and Savant, Gaurav and Moritz, Hans and Childs,
               Elvon",
  publisher = "Engineer Research and Development Center (U.S.)",
  month     =  Apr,
  year      =  2020,
  doi       =  "10.21079/11681/36295",
}

@article{savant2011adh,
author = {Gaurav Savant  and Charlie Berger  and Tate O. McAlpin  and Jennifer N. Tate },
title = {Efficient Implicit Finite-Element Hydrodynamic Model for Dam and Levee Breach},
journal = {Journal of Hydraulic Engineering},
volume = {137},
number = {9},
pages = {1005-1018},
year = {2011},
doi = {10.1061/(ASCE)HY.1943-7900.0000372},
}

@article{butler2015definition,
  title={Definition and solution of a stochastic inverse problem for the {Manning’s} n parameter field in hydrodynamic models},
  author={Butler, Troy and Graham, L and Estep, D and Dawson, C and Westerink, JJ},
  journal={Advances in Water Resources},
  volume={78},
  pages={60--79},
  year={2015},
  doi = {https://doi.org/10.1016/j.advwatres.2015.01.011},
  publisher={Elsevier}
}

@article{song2023surrogate,
  title={A surrogate model for shallow water equations solvers with deep learning},
  author={Song, Yalan and Shen, Chaopeng and Liu, Xiaofeng},
  journal={Journal of Hydraulic Engineering},
  volume={149},
  number={11},
  pages={04023045},
  year={2023},
  publisher={American Society of Civil Engineers}
}

@article{liu2024bathymetry,
  title={Bathymetry inversion using a deep-learning-based surrogate for shallow water equations solvers},
  author={Liu, Xiaofeng and Song, Yalan and Shen, Chaopeng},
  journal={Water Resources Research},
  volume={60},
  number={3},
  pages={e2023WR035890},
  year={2024},
  publisher={Wiley Online Library}
}

@misc{RedRiverData2024,
  doi = {10.17603/DS2-3W1M-HT51},
  url = {https://www.designsafe-ci.org/data/browser/public/designsafe.storage.published/PRJ-6207/#detail-5855ec2a-cd4d-423b-80a4-93769c023f29},
  author = {Rivera-Casillas, Peter and Dutta, Sourav and Cai, Shukai and Loveland, Mark and Nath, Kamaljyoti and Shukla, Khemraj and Trahan, Corey and Lee, Jonghyun and Farthing, Matthew and Dawson, Clinton N},
  language = {en},
  title = {2D AdH simulation of riverine hydrodynamics in a section of the Red River in Louisiana, USA, parameterized by the bottom friction coefficient. , in A neural operator emulator for coastal and riverine shallow water dynamics},
  publisher = {Designsafe-CI},
  year = {2025},
  copyright = {Open Data Commons Attribution}
}

@article{rathje2017designsafe,
  title={DesignSafe: New cyberinfrastructure for natural hazards engineering},
  author={Rathje, Ellen M and Dawson, Clint and Padgett, Jamie E and Pinelli, Jean-Paul and Stanzione, Dan and Adair, Ashley and Arduino, Pedro and Brandenberg, Scott J and Cockerill, Tim and Dey, Charlie and others},
  journal={Natural hazards review},
  volume={18},
  number={3},
  pages={06017001},
  year={2017},
  publisher={American Society of Civil Engineers}
}

% --- Prepares the environment for the supplementary document ---
\newpage
\appendix

% === This is the fix to get "S" numbering for your figures and tables ===
\renewcommand{\thefigure}{S\arabic{figure}}
\renewcommand{\thetable}{S\arabic{table}}
\setcounter{figure}{0}
\setcounter{table}{0}
\begin{flushleft}
    \Large\textbf{Supporting Information for ``A Neural Operator Emulator for Coastal and Riverine Shallow Water Dynamics''}
\end{flushleft}

\vspace{1cm}
\noindent\textbf{Contents of this file}
%%%Remove or add items as needed%%%
\begin{enumerate}

\item Text S1 to S4
\item Tables S1 to S5
\item Figures S1 to S11
\end{enumerate}

\clearpage

\noindent\textbf{Introduction}

This supporting information document provides additional details on the problem setup and supplementary results to reinforce the findings presented in the results and discussion section of the main paper. A detailed visual comparison of MITONet against the benchmark models is presented in Figures S1-S4. These figures provide three types of analysis: side-by-side snapshots of the full solution fields, zoomed-in comparisons of key regions, and maps showing the spatial distribution of the Mean Absolute Error (MAE) over time. Figure S5-S7 show comparisons of MITONet predictions and ADCIRC simulation output at the three sensors over a $10$-day forecast period for three unseen test values of the bottom friction coefficient, $r$. Figures S8 to S10 provide visualizations to compare the snapshots of MITONet predictions and the ADCIRC simulation outputs on day $60$, following 55 days of autoregressive predictions, over the full domain as well as zoomed-in over the inlet area, for three unseen test values of $r$. Figure S11 presents violin plots of the RMSE for MITONet's Shinnecock Inlet predictions. These plots illustrate the performance for all variables over a 55-day prediction window (days 5-60), using models that were trained on all bottom friction coefficient ($r$) values.

\clearpage

\noindent\textbf{Text S1. Numerical Models}

\noindent\textbf{Advanced Circulation modeling suite (ADCIRC)}\\
The first numerical model used in this study is ADCIRC, which solves the vertically-integrated Generalized Wave Continuity Equation (GWCE) for water surface elevation \cite{luettich2004formulation} given by,

\begin{equation} \frac{\partial{H}}{\partial{t}} + \frac{\partial}{\partial{x}}(UH) + \frac{\partial}{\partial{y}}(VH)=0, \label{eq:2DSW1} \end{equation}

\noindent where

\noindent $U, V \equiv \frac{1}{H} \int_{-b}^{\zeta} u,v , dz$: depth-averaged velocities in the $x$ and $y$ directions,\\
$u, v$: vertically-varying velocities in the $x$ and $y$ directions,\\
$H=\zeta + b$: total water column thickness,\\
$\zeta$: free surface departure from the geoid,\\
$b$: bathymetric depth.\\

In addition to the continuity equation, ADCIRC solves the depth-averaged momentum equations, which describe the transport of momentum in both horizontal directions. These equations are written in non-conservative form as,

\begin{equation} \frac{\partial{U}}{\partial{t}} + U \frac{\partial{U}}{\partial{x}} + V \frac{\partial{U}}{\partial{y}} - fV=-g \frac{\partial{\zeta}}{\partial{x}} + \frac{\tau_{sx}}{\rho H} - \frac{\tau_{bx}}{\rho H}, \label{eq:2DSW2
} \end{equation}

\begin{equation} \frac{\partial{V}}{\partial{t}} + U \frac{\partial{V}}{\partial{x}} + V \frac{\partial{V}}{\partial{y}} + fU=-g \frac{\partial{\zeta}}{\partial{y}} + \frac{\tau_{sy}}{\rho H} - \frac{\tau_{by}}{\rho H}, \label{eq:2DSW3
} \end{equation}

\noindent where

\noindent$f$: Coriolis parameter,\\
$g$: gravitational constant,\\
$\tau_{sx}, \tau_{sy}$: surface stresses in the $x$ and $y$ directions,\\
$\tau_{bx}, \tau_{by}$: bottom stresses in the $x$ and $y$ directions,\\
$\rho$: density of water.\\

For the bottom stress term, ADCIRC utilizes a generalized slip formulation given by: 

\begin{equation}
\frac{\tau_{bx}}{\rho} = K_{slip} U; \quad   \frac{\tau_{by}}{\rho}=K_{slip} V
\end{equation}

\noindent where

\noindent$K_{slip} = C_d \sqrt{U^2+V^2} $: a quadratic slip bottom boundary condition parameterized by the quadratic drag coefficient $C_d$ (referred to as $r$ in the article).\\

\noindent\textbf{Adaptive Hydraulics suite (AdH)}\\
AdH is the second numerical model used in the study, which solves the standard 2D depth-averaged shallow water equations, written in conservative form as \cite{savant2011adh,dutta_farthing_2021}

\begin{equation}
    \mathcal{R} \equiv \frac{\partial \mathbf{Q}}{\partial t} + \frac{\partial \mathbf{P}_x}{\partial x} +\frac{\partial \mathbf{P}_y}{\partial y} + \mathbf{R} = 0,
\end{equation}

\noindent where 

\noindent $\mathbf{Q} = [H, UH, VH]^T$: the conservative state variable.\\ 
$\mathbf{P}_x$, $\mathbf{P}_y$: flux vectors given by,

\begin{align}
    \mathbf{P}_x &= 
            \begin{pmatrix}
                U\,H \\[6pt]
                U^2\,H + \tfrac{1}{2}g H^{2} - H\!\left(\dfrac{\sigma_{xx}}{\rho}\right) \\[6pt]
                U \,V\,H - H\!\left(\dfrac{\sigma_{yx}}{\rho}\right)
            \end{pmatrix}, \label{eq:px}\\[6pt]
    \mathbf{P}_y &= 
            \begin{pmatrix}
                V\,H \\[6pt]
                U \,V\,H - H\!\left(\dfrac{\sigma_{xy}}{\rho}\right) \\[6pt]
                V^2\,H + \tfrac{1}{2}g H^{2} - H\!\left(\dfrac{\sigma_{yy}}{\rho}\right)
            \end{pmatrix}, \quad \text{and}\label{eq:py}\\[8pt]
    \mathbf{R} &=
            \begin{pmatrix}
                0 \\[6pt]
                g H\,\dfrac{\partial b}{\partial x}
                    + g H\!\left[\dfrac{n_{mn}^{2}\,U\sqrt{U^{2}+V^{2}}}{H^{4/3}}\right] - f\,H\,V \\[8pt]
                g H\,\dfrac{\partial b}{\partial y}
                    + g H\!\left[\dfrac{n_{mn}^{2}\,V\sqrt{U^{2}+V^{2}}}{H^{4/3}}\right] + f\,H\,U
            \end{pmatrix}. \label{eq:r}
\end{align}\\

In the above, $n_{mn}$ is the coefficient of the standard Manning's parametrization for bottom roughness (referred to as $r$ in the article), and $\sigma_{xx,xy,yy,yx}$ are Reynolds stresses due to turbulence, which are modeled using the Bousinessq approach for the gradient in the mean currents. AdH uses a residual-based SUPG stabilization scheme and a second-order backward Euler time-discretization for numerically solving the above equations.\\

\noindent\textbf{Text S2. Hyperparameter Optimization Setup}

As with any deep neural network, DeepONets require meticulous hyperparameter optimization to achieve optimal performance. The literature often cites standard hyperparameter configurations that, while common, do not ensure robust results across various problems. The manual configuration of hyperparameters poses a significant challenge, especially with a large hyperparameter space. To address this challenge,  we employ a hyperparameter optimization tool, Optuna \cite{akiba2019optuna}, to automatically optimize the hyperparameters of the autoencoder and MITONet models (Table \ref{tab:hyperparam_search_space}). This enables efficient exploration of the high-dimensional hyperparameter space, leading to improved model performance.\\

\begin{table}[h]
\caption{Hyperparameter optimization space for autoencoder and MITONet architectures.}
\label{tab:hyperparam_search_space}
\resizebox{\textwidth}{!}{
\centering
\begin{tabular}{p{.3cm} l l c}
\hline
\multicolumn{1}{c}{4cm} & \textbf{Network} & \textbf{Hyperparameter} & \textbf{Optimization Space} \\ \hline

\multirow{6}{4cm}{\rotatebox[origin=c]{90}{\textsc{Autoencoder}}}
 & \multirow{4}{4cm}{Autoencoder Network} 
 & Encoder Activation Function & tanh, elu, relu, swish \\
 && Decoder Activation Function & tanh, elu, relu, swish \\
 && Number of Layers & 2--5 \\
 && Latent Space Dimension & 30--60 \\ \hhline{~---}

 & \multirow{2}{4cm}{Autoencoder Training} 
 & Batch Size & 64, 128, 256, 512, 1024 \\
 && Initial Learning Rate & 1e-6, 1e-5, 5e-5, 1e-4, 5e-4, 1e-3 \\ \hline

\multirow{20}{4cm}{\rotatebox[origin=c]{90}{\textsc{Mitonet}}}
 & \multirow{4}{4cm}{Branch Network} 
 & Number of Layers & 2--5 \\
 && Activation Function & tanh, elu, relu, swish \\
 && Weight Initializer & he normal, he uniform, glorot normal, glorot uniform \\
 && Regularizer & L1, L2, none \\ \hhline{~---}

 & \multirow{4}{4cm}{Trunk Network} 
 & Number of Layers & 2--5 \\
 && Activation Function & tanh, elu, relu, swish \\
 && Weight Initializer & he normal, he uniform, glorot normal, glorot uniform \\
 && Regularizer & L1, L2, none \\ \hhline{~---}

 & \multirow{4}{4cm}{Branch Encoder Network} 
 & Number of Layers & 2--4 \\
 && Activation Function & tanh, elu, relu, swish \\
 && Weight Initializer & he normal, he uniform, glorot normal, glorot uniform \\
 && Regularizer & L1, L2, none \\ \hhline{~---}

 & \multirow{4}{4cm}{Trunk Encoder Network} 
 & Number of Layers & 2--4 \\
 && Activation Function & tanh, elu, relu, swish \\
 && Weight Initializer & he normal, he uniform, glorot normal, glorot uniform \\
 && Regularizer & L1, L2, none \\ \hhline{~---}

 & \multirow{2}{4cm}{Shared Parameters} 
 & $\mathcal{L}_{factor}$ & 2--7 \\
 && $\mathcal{L}_{encoder-factor}$ & 1--5 \\ \hhline{~---}

 & \multirow{2}{4cm}{MITONet Training} 
 & Batch Size & 512, 1024, 2048 \\
 && Initial Learning Rate & 1e-6, 1e-5, 5e-5, 1e-4, 5e-4, 1e-3 \\ \hline
\end{tabular}
}
\end{table}

Note that the number of neurons per layer in the network architectures was not directly optimized. For the autoencoder, the input layer size corresponds to the degrees of freedom of the computational mesh ($3070$ for Shinnecock Inlet and $12291$ for Red River). For the Shinnecock Inlet example the subsequent hidden layer has half the number of neurons as the preceding layer. For the Red River example, however, this reduction factor was increased to approximately $32$ to ensure a comparable number of trainable parameters between the two distinct examples. The decoder architecture mirrors this design in reverse; therefore, we optimize (a) the number of layers in the encoder (and decoder) model, and (b) the dimension of the latent space, which is equal to the size of the encoder output layer as well as the size of the decoder input layer. For the operator network (MITONet), we optimize a hyperparameter (\(L_{factor}\)) that determines the number of neurons in the hidden layers based on the latent space dimension $(\text{neurons} = L_{factor} \times \text{latent space dimension})$; this ensures compatibility between the latent space dimension and the dimension of MITONet's output.\\

For the Shinnecock Inlet example, the optimization study was configured to run $100$ trials for the autoencoder and $200$ trials for the MITONet, with each trial running for $2,000$ epochs. This process was performed individually for each hydrodynamic variable ($H$, $U$, and $V$), and to ensure a fair comparison, the benchmark models (DeepONet, L-DeepONet, M-DeepONet, and MIONet) were also optimized using the same number of trials and epochs. For the Red River example, the settings were adjusted to account for its larger mesh size. The study consisted of $50$ trials for the autoencoder and $100$ trials for the MITONet, with each trial running for $3,000$ epochs. The total computational cost for these optimization studies was approximately $74–96$ wall-clock hours per variable on a single NVIDIA A40 GPU.\\

\noindent\textbf{Text S3. Hyperparameter Optimization Results and Discussion}

The best configurations from the  hyperparameter optimization for the Shinnecock Inlet example are shown in Table \ref{tab:best_hyperparams}. Although no consistent patterns are observed in the resulting hyperparameters for $H$, $U$ and $V$, certain similarities exist.  Notably, similarities exist between $H$ and $U$, as well as between $U$ and $V$, whereas few hyperparameters are shared between $H$ and $V$, except for a few that are common across all variables, such as some regularizers, the number of layers in the autoencoder and the autoencoder training hyperparameters.\\

\begin{table}[h]
\centering
\caption{Best Configurations for autoencoder and MITONet architectures corresponding to the Shinnecock Inlet example.}
\label{tab:best_hyperparams}
\resizebox{\textwidth}{!}{
\begin{tabular}{p{.3cm} l l c c c}
\hline
\multicolumn{1}{c}{4cm} & \textbf{Network} & \textbf{Hyperparameter} & $\boldsymbol{H}$ & $\boldsymbol{U}$ & $\boldsymbol{V}$ \\ 
\hline

\multirow{6}{4cm}{\rotatebox[origin=c]{90}{\textsc{Autoencoder}}}
 & \multirow{4}{4cm}{Autoencoder Network} 
 & Encoder Activation Function & tanh & swish & swish \\
 && Decoder Activation Function & tanh & swish & tanh \\
 && Number of Layers & 3 & 2 & 2 \\
 && Latent Space Dimension & 60 & 60 & 60 \\ \hhline{~-----}

 & \multirow{2}{4cm}{Autoencoder Training} 
 & Batch Size & 64 & 64 & 64 \\
 && Initial Learning Rate & 5.00E-05 & 5.00E-05 & 5.00E-05 \\ \hline

\multirow{20}{4cm}{\rotatebox[origin=c]{90}{\textsc{Mitonet}}}
 & \multirow{4}{4cm}{Branch Network} 
 & Number of Layers & 3 & 5 & 5 \\
 && Activation Function & elu & elu & relu \\
 && Weight Initializer & he uniform & he uniform & glorot normal \\
 && Regularizer & none & none & none \\ \hhline{~-----}

 & \multirow{4}{4cm}{Trunk Network} 
 & Number of Layers & 4 & 4 & 2 \\
 && Activation Function & elu & swish & tanh \\
 && Weight Initializer & he normal & he normal & glorot normal \\
 && Regularizer & none & l1 & l2 \\ \hhline{~-----}

 & \multirow{4}{4cm}{Branch Encoder Network} 
 & Number of Layers & 4 & 3 & 3 \\
 && Activation Function & relu & tanh & elu \\
 && Weight Initializer & he uniform & glorot uniform & glorot normal \\
 && Regularizer & l1 & none & none \\ \hhline{~-----}

 & \multirow{4}{4cm}{Trunk Encoder Network} 
 & Number of Layers & 3 & 2 & 3 \\
 && Activation Function & relu & tanh & swish \\
 && Weight Initializer & glorot uniform & glorot uniform & he uniform \\
 && Regularizer & none & none & none \\ \hhline{~-----}

 & \multirow{2}{4cm}{Shared Parameters} 
 & $\mathcal{L}_{factor}$ & 4 & 5 & 6 \\
 && $\mathcal{L}_{encoder factor}$ & 3 & 5 & 5 \\ \hhline{~-----}

 & \multirow{2}{4cm}{MITONet Training} 
 & Batch Size & 512 & 1024 & 512 \\
 && Initial Learning Rate & 5.00E-04 & 5.00E-04 & 5.00E-05 \\ \hline
\end{tabular}
}
\end{table}

The best hyperparameter configurations were continuously updated throughout the optimization, suggesting that optimizing over more trials could produce better models. Additionally, more epochs per trial, expanding the optimization space, or making it more flexible, could improve the top configurations. In this context, studying the relative importance of hyperparameters could offer valuable insights to guide future hyperparameter optimization efforts. \\

For a more detailed analysis, Table \ref{tab:hyperparams_imp_shin} presents the hyperparameter importance values based on the completed trials, as computed by Optuna using the fANOVA hyperparameter importance evaluation algorithm \cite{hutter2014fanova}. This algorithm fits a random forest regression (RFR) model that predicts the objective values of completed Optuna trials given their parameter configurations such that the most influential hyperparameters yield higher importance values. However, the reliability of this importance score is dependent upon the accuracy of the underlying RFR model, which, in turn, relies on a thorough exploration of the hyperparameter search space.\\

\begin{table}[h]
\centering
\caption{Hyperparameter importance for autoencoder and MITONet architectures corresponding to the Shinnecock Inlet example, with most important values highlighted in gray.}
\label{tab:hyperparams_imp_shin}
\resizebox{\textwidth}{!}{
\begin{tabular}{p{.3cm} l l c c c}
\hline
\multicolumn{1}{c}{4cm} & \textbf{Network} & \textbf{Hyperparameter} & $\boldsymbol{H}$ & $\boldsymbol{U}$ & $\boldsymbol{V}$ \\ 
\hline

\multirow{6}{4cm}{\rotatebox[origin=c]{90}{\textsc{Autoencoder}}}
 & \multirow{4}{4cm}{Autoencoder Network} 
  & Encoder Activation Function & \cellcolor{lightgray} 1.33E-01 & 9.43E-02 & \cellcolor{lightgray} 1.34E-01 \\
 && Decoder Activation Function & \cellcolor{lightgray} 1.72E-01 & 6.30E-02 & 5.94E-03 \\
 && Number of Layers            & 5.33E-02 & 4.00E-04 & 6.81E-02 \\
 && Latent Space Dimension      & 2.80E-03 & 3.70E-04 & 1.21E-02 \\ \hhline{~-----}

 & \multirow{2}{4cm}{Autoencoder Training} 
  & Batch Size                  & 3.03E-06 & \cellcolor{lightgray} 5.00E-01 & 2.40E-03 \\
 && Initial Learning Rate       & 7.47E-02 & \cellcolor{lightgray} 2.74E-01 & 2.23E-02 \\ \hline

\multirow{20}{4cm}{\rotatebox[origin=c]{90}{\textsc{Mitonet}}}
 & \multirow{4}{4cm}{Branch Network} 
  & Number of Layers            & 5.33E-02 & 6.64E-02 & 1.71E-02 \\
 && Activation Function         & 4.02E-02 & 1.75E-02 & 3.90E-04 \\
 && Weight Initializer          & 2.18E-02 & 2.36E-03 & \cellcolor{lightgray} 1.08E-01 \\
 && Regularizer                 & \cellcolor{lightgray} 2.10E-01 & \cellcolor{lightgray} 3.84E-01 & 3.70E-06 \\ \hhline{~-----}

 & \multirow{4}{4cm}{Trunk Network} 
  & Number of Layers            & 8.01E-02 & 1.81E-03 & 1.77E-02 \\
 && Activation Function         & 6.19E-02 & 3.32E-02 & 1.08E-03 \\
 && Weight Initializer          & 4.73E-02 & 5.91E-03 & 3.64E-02 \\
 && Regularizer                 & 6.84E-02 & 5.24E-02 & 1.62E-02 \\ \hhline{~-----}

 & \multirow{4}{4cm}{Branch Encoder Network} 
  & Number of Layers            & 1.00E-04 & 1.86E-03 & 4.22E-02 \\
 && Activation Function         & 8.56E-02 & 4.69E-03 & \cellcolor{lightgray} 2.61E-01 \\
 && Weight Initializer          & 1.00E-01 & 4.36E-03 & 2.29E-02 \\
 && Regularizer                 & 1.00E-04 & 9.35E-02 & 9.24E-02 \\ \hhline{~-----}

 & \multirow{4}{4cm}{Trunk Encoder Network} 
  & Number of Layers            & 6.67E-03 & 2.45E-05 & 6.40E-02 \\
 && Activation Function         & 3.18E-02 & 3.37E-03 & \cellcolor{lightgray} 1.22E-01 \\
 && Weight Initializer          & 4.73E-02 & 3.30E-03 & 2.29E-02 \\
 && Regularizer                 & 6.84E-02 & 9.35E-02 & 1.62E-02 \\ \hhline{~-----}

 & \multirow{2}{4cm}{Shared Parameters} 
  & $\mathcal{L}_{factor}$          & 7.84E-02 & 4.69E-02 & 3.93E-02 \\
 && $\mathcal{L}_{encoder\ factor}$ & 1.61E-02 & 1.03E-02 & 3.73E-03 \\ \hhline{~-----}

 & \multirow{2}{4cm}{MITONet Training} 
  & Batch Size                  & \cellcolor{lightgray} 2.03E-01 & \cellcolor{lightgray} 6.38E-01 & \cellcolor{lightgray} 1.14E-01 \\
 && Initial Learning Rate       & \cellcolor{lightgray} 4.36E-01 & \cellcolor{lightgray} 2.04E-01 & \cellcolor{lightgray} 6.65E-01 \\ \hline
\end{tabular}
}
\end{table}

The results indicate that the two hyperparameters controlling the training of the MITONet models (i.e., batch size and initial learning rate) are the most influential for each solution variable. Additionally, the regularization used for the branch networks has a relatively higher impact on the MITONet models for $H$ and $U$, whereas the activation functions of the branch and trunk encoder networks as well as the weight initialization algorithm of the branch networks have a relatively higher effect on the performance of the MITONet model for $V$. While it is well-known that the predictive performance of any neural network model is heavily dependent on choices of batch size, initial learning rate, and the activation functions, the effect of regularization and weight initialization schemes is typically not explored during training. However, this study provides an important insight that the diversity in the types and amount of input features that are being incorporated using the multiple branch architecture of the MITONet framework is potentially a reason why using appropriate regularization and weight initialization algorithms might have an elevated impact on convergence during training, as well as on controlling overfitting to improve generalization capabilities during inference. On the other hand, the hyperparameter importance analysis for the autoencoder models reveals a different trend. The training-related hyperparameters of the autoencoder model for $U$ appear to have a higher impact compared to the architecture-related hyperparameters, whereas the opposite trend can be observed for the autoencoder models of $H$ and $V$.\\

It is worth noting that in the proposed framework, for a given variable, the training is performed sequentially, where the autoencoder model is trained first, followed by the MITONet model, which is trained using the latent space representation generated by the pre-trained autoencoder. Hence, the hyperparameter importance studies are conducted separately for the autoencoder and the MITONet models of each solution variable, and as such, the importance values of the autoencoder hyperparameters are independent of those of the MITONet hyperparameters.\\

A key challenge of hyperparameter optimization under sequential training is that the latent space dimension is optimized for reconstruction quality alone, without accounting for its effectiveness in the downstream task. Since the latent space dimension had low importance in the autoencoder optimization, but significantly affected MITONet’s accuracy, the selection made during hyperparameter optimization did not necessarily yield the best downstream performance. For instance, the latent space dimension was found to be 60, 40, and 60 for $H$, $U$, and $V$, respectively. However, evaluation of the trained MITONet model for $U$ revealed that a latent space dimension of 60 significantly improved accuracy, leading us to override the optimized value of 40 with 60. While a larger latent space generally improves reconstruction accuracy and performance in downstream tasks \cite{kontolati2023learning, dutta2021data}, an excessively large latent space can introduce computational bottlenecks and increase the risk of overfitting.\\ 

To fully capture the interactions between the autoencoder and MITONet hyperparameters, a concurrent training framework could be considered. This approach would inherently optimize the latent space for both reconstruction and downstream performance simultaneously. However, such a framework would involve a significantly larger joint parameter space that would lead to training complexities and potentially prohibitive computational and memory costs, and is, therefore, not considered in this work. An interesting avenue of future work would involve exploring such ideas to estimate the optimal latent space dimension through a broader hyperparameter search, particularly for scenarios where computational efficiency is a key constraint.\\

While the preceding analysis focused on the optimization process and hyperparameter importance for our proposed architecture, a fair comparison to existing architectures, as presented in Section 4.1.1, required applying the same rigorous tuning to the benchmark models. Accordingly, the results of these optimization studies are presented in Table \ref{tab:model_comparison_final}, detailing the final configurations for DeepONet (DON), L-DeepONet (L-DON), M-DeepONet (M-DON), and MIONet.\\

\begin{table}[h]
\centering
\caption{Optimal hyperparameter configurations for the DeepONet (DON), L-DeepONet (L-DON), M-DeepONet (M-DON), and MIONet architectures, applied to the Shinnecock Inlet example. Comma-separated values correspond to the ($H$, $U$ and $V$) variables.}
\label{tab:model_comparison_final}

\newcolumntype{C}[1]{>{\centering\arraybackslash}p{#1}}
\newcolumntype{L}[1]{>{\raggedright\arraybackslash}p{#1}}

% The \resizebox command will force the table to fit the page width.
\resizebox{\textwidth}{!}{

% Column format is preserved and extended for the four model columns
\begin{tabular}{l L{2.5cm} l C{3.4cm} C{3.4cm} C{3.4cm} C{3.4cm}}
\hline
\multicolumn{1}{c}{4cm} & \textbf{Network} & \textbf{Hyperparameter} & \textbf{DON} & \textbf{L-DON} & \textbf{M-DON} & \textbf{MIONet} \\
\hline

% === Autoencoder Section ===
\multirow{6}{4cm}{\rotatebox[origin=c]{90}{\textsc{Autoencoder}}}
 & \multirow{4}{3cm}{Autoencoder Network}
 & Encoder Activation Function & --- & swish, tanh, elu & --- & --- \\
 && Decoder Activation Function & --- & tanh, elu, elu & --- & --- \\
 && Number of Layers & --- & 3, 2, 2 & --- & --- \\
 && Latent Space Dimension & --- & 40, 60, 50 & --- & --- \\ \hhline{~------}
 & \multirow{2}{3cm}{Autoencoder Training}
 & Batch Size & --- & 64, 64, 64 & --- & --- \\
 && Initial Learning Rate & --- & 1E-5, 5E-5, 5E-5 & --- & --- \\ \hline

% === Operator Network Section ===
\multirow{32}{4cm}{\rotatebox[origin=c]{90}{\textsc{Operator Network}}}
 & \multirow{7}{3cm}{Branch Network}
 & Number of Layers & 2, 5, 4 & 4, 3, 3 & 2, 4, 2 & 2, 2, 2 \\
 && Neurons per Layer & 128, 256, 128 & 384, 512, 512 & 192, 256, 448 & 256, 128, 256 \\
 && Activation Function & relu, swish, elu & swish, swish, swish & elu, elu, tanh & swish, relu, relu \\
 && Weight Initializer & he normal, glorot normal, glorot uniform & glorot uniform, glorot uniform, he uniform & glorot uniform, glorot normal, he normal & he normal, he uniform, he uniform \\
 && Regularizer & none, none, none & none, none, none & none, none, none & none, none, none \\ \hhline{~------}
 & \multirow{7}{3cm}{Trunk Network}
 & Number of Layers & 5, 3, 5 & 2, 4, 4 & 4, 5, 2 & 3, 5, 5 \\
 && Neurons per Layer & 512, 512, 512 & 256, 384, 320 & --- & 512, 384, 448 \\
 && Activation Function & relu, relu, relu & tanh, elu, tanh & relu, relu, relu & relu, relu, swish \\
 && Weight Initializer & glorot normal, glorot normal, glorot normal & glorot normal, glorot normal, glorot normal & he normal, he uniform, he normal & glorot uniform, he uniform, he uniform \\
 && Regularizer & none, none, none & none, none, none & none, none, none & none, none, none \\ \hhline{~------}
 & \multirow{7}{3cm}{Branch Encoder}
 & Number of Layers & --- & --- & 4, 3, 4 & --- \\
 && Neurons per Layer & --- & --- & 384, 448, 256 & --- \\
 && Activation Function & --- & --- & relu, tanh, tanh & --- \\
 && Weight Initializer & --- & --- & glorot normal, glorot normal, glorot uniform & --- \\
 && Regularizer & --- & --- & l1, l1, l1 & --- \\ \hhline{~------}
 & \multirow{7}{3cm}{Trunk Encoder}
 & Number of Layers & --- & --- & 3, 3, 4 & --- \\
 && Neurons per Layer & --- & --- & 384, 384, 320 & --- \\
 && Activation Function & --- & --- & swish, swish, relu & --- \\
 && Weight Initializer & --- & --- & he uniform, he uniform, glorot normal & --- \\
 && Regularizer & --- & --- & none, none, none & --- \\ \hhline{~------}
 & \multirow{2}{3cm}{Shared Parameters}
 & {Output Shape} & 384, 512, 512 & --- & 64, 320, 448 & 192, 320, 320 \\
 && $\mathcal{L}_{\text{factor}}$ & --- & 6, 6, 5 & --- & --- \\ \hhline{~------}
 & \multirow{2}{3cm}{Training}
 & Batch Size & 64, 64, 64 & 512, 2048, 1024 & 512, 512, 512 & 64, 64, 64 \\
 && Initial Learning Rate & 5E-4, 5E-4, 5E-4 & 1E-3, 1E-3, 1E-3 & 1E-3, 1E-3, 1E-3 & 5E-4, 1E-3, 1E-3 \\ \hline
\end{tabular}
}
\end{table}

Finally, the optimal hyperparameter configurations for the Red River case study are presented in Table \ref{tab:best_hyperparams_rr}. Consistent with the findings from the Shinnecock Inlet example, these results show no single, dominant pattern across the different hydrodynamic variables. Instead, each variable ($H$, $U$, and $V$) often converges to a unique set of hyperparameters, reinforcing the need for individual tuning for each component of the physical system.\\

\begin{table}[h]
\centering
\caption{Best Configurations for autoencoder and MITONet architectures for the Red River example.}
\label{tab:best_hyperparams_rr}
\resizebox{\textwidth}{!}{
\begin{tabular}{p{.3cm} l l c c c}
\hline
\multicolumn{1}{c}{4cm} & \textbf{Network} & \textbf{Hyperparameter} & $\boldsymbol{H}$ & $\boldsymbol{U}$ & $\boldsymbol{V}$ \\ 
\hline

\multirow{6}{4cm}{\rotatebox[origin=c]{90}{\textsc{Autoencoder}}}
 & \multirow{4}{4cm}{Autoencoder Network} 
  & Encoder Activation Function & swish    & elu      & elu \\
 && Decoder Activation Function & swish    & swish    & elu \\
 && Number of Layers            & 2        & 2        & 2 \\
 && Latent Space Dimension      & 96       & 64       & 96 \\ \hhline{~-----}

 & \multirow{2}{4cm}{Autoencoder Training} 
  & Batch Size                  & 256      & 64       & 64 \\
 && Initial Learning Rate       & 1.00E-04 & 1.00E-05 & 5.00E-05 \\ \hline

\multirow{20}{4cm}{\rotatebox[origin=c]{90}{\textsc{Mitonet}}}
 & \multirow{4}{4cm}{Branch Network} 
  & Number of Layers            & 3        & 2        & 3 \\
 && Activation Function         & elu      & tanh     & elu \\
 && Weight Initializer & he normal & glorot normal & he uniform \\
 && Regularizer & none & none & none \\ \hhline{~-----}

 & \multirow{4}{4cm}{Trunk Network} 
 & Number of Layers & 4 & 4 & 5 \\
 && Activation Function & tanh & tanh & tanh \\
 && Weight Initializer & glorot normal & he normal & he uniform \\
 && Regularizer & none & l1 & none \\ \hhline{~-----}

 & \multirow{4}{4cm}{Branch Encoder Network} 
 & Number of Layers & 4 & 2 & 3 \\
 && Activation Function & relu & elu & relu \\
 && Weight Initializer & he normal & he uniform & he normal \\
 && Regularizer & l1 & l2 & l2 \\ \hhline{~-----}

 & \multirow{4}{4cm}{Trunk Encoder Network} 
 & Number of Layers & 4 & 2 & 2 \\
 && Activation Function & relu & tanh & tanh \\
 && Weight Initializer & glorot uniform & glorot uniform & glorot normal \\
 && Regularizer & none & l2 & l1 \\ \hhline{~-----}

 & \multirow{2}{4cm}{Shared Parameters} 
 & $\mathcal{L}_{factor}$ & 3 & 3 & 3 \\
 && $\mathcal{L}_{encoder factor}$ & 2 & 1 & 2 \\ \hhline{~-----}

 & \multirow{2}{4cm}{MITONet Training} 
 & Batch Size & 2048 & 512 & 512 \\
 && Initial Learning Rate & 5.00E-04 & 1.00E-05 & 1.00E-04 \\ \hline
\end{tabular}
}
\end{table}

\clearpage

\noindent\textbf{Text S4. Supporting Figures}

\begin{figure}[h]
    \centering
    \includegraphics[width=.75\textwidth]{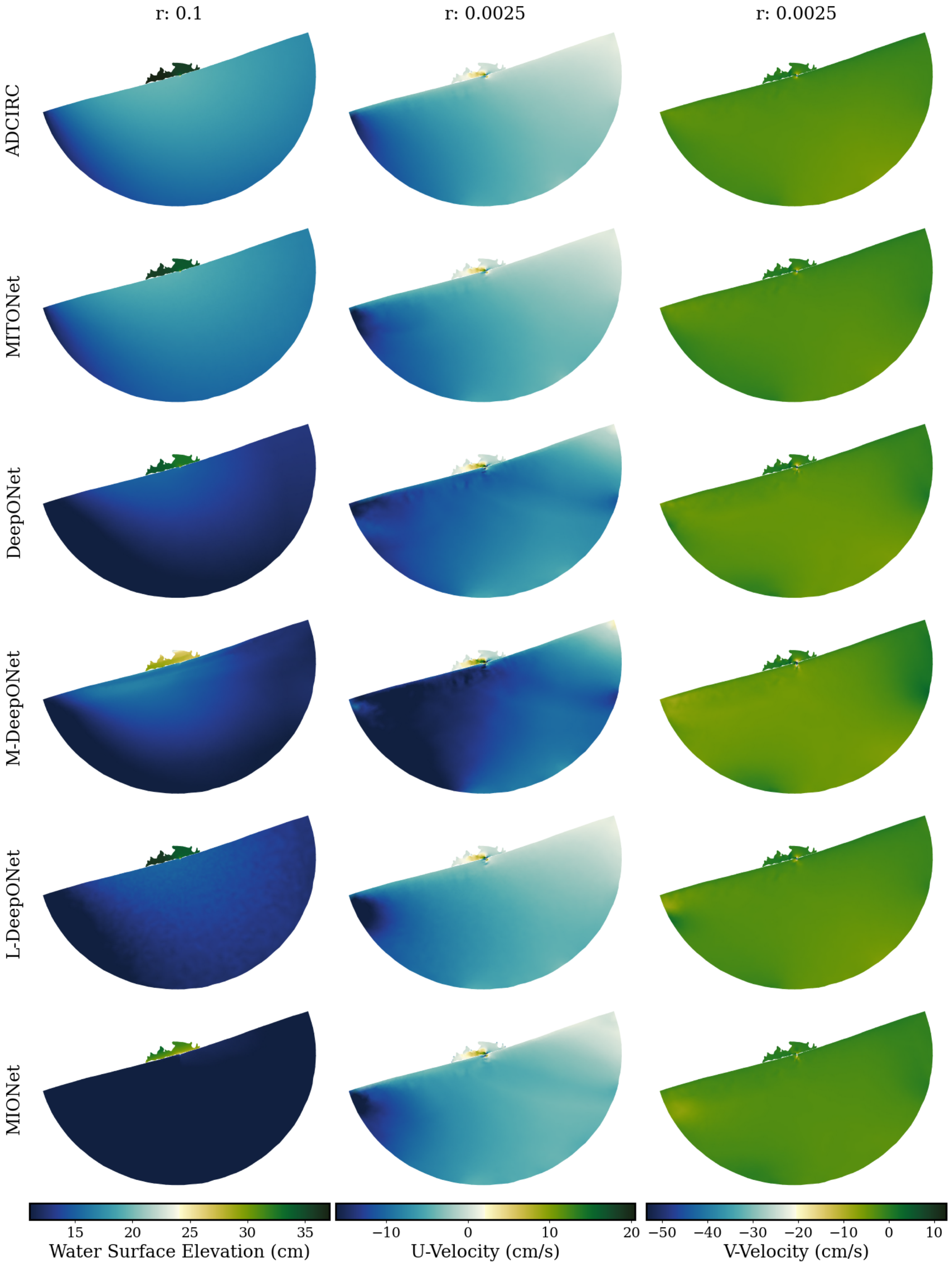}
    \caption{Full-domain comparison of ADCIRC and all five models for the Shinnecock Inlet example on day 45, following 40 days of autoregressive predictions. Columns display the variables $H$, $U$, and $V$, while rows correspond to solutions from: (1) ADCIRC, (2) MITONet, (3) DeepONet, (4) L-DeepONet, (5) M-DeepONet, and (6) MIONet. The bottom friction coefficient used are $r = 0.1$ for $H$ and $r = 0.0025$ for $U$ and $V$.}
    \label{fig:benchmark_full_snaps}
\end{figure}

\begin{figure}[h]
    \centering
    \includegraphics[width=.75\textwidth]{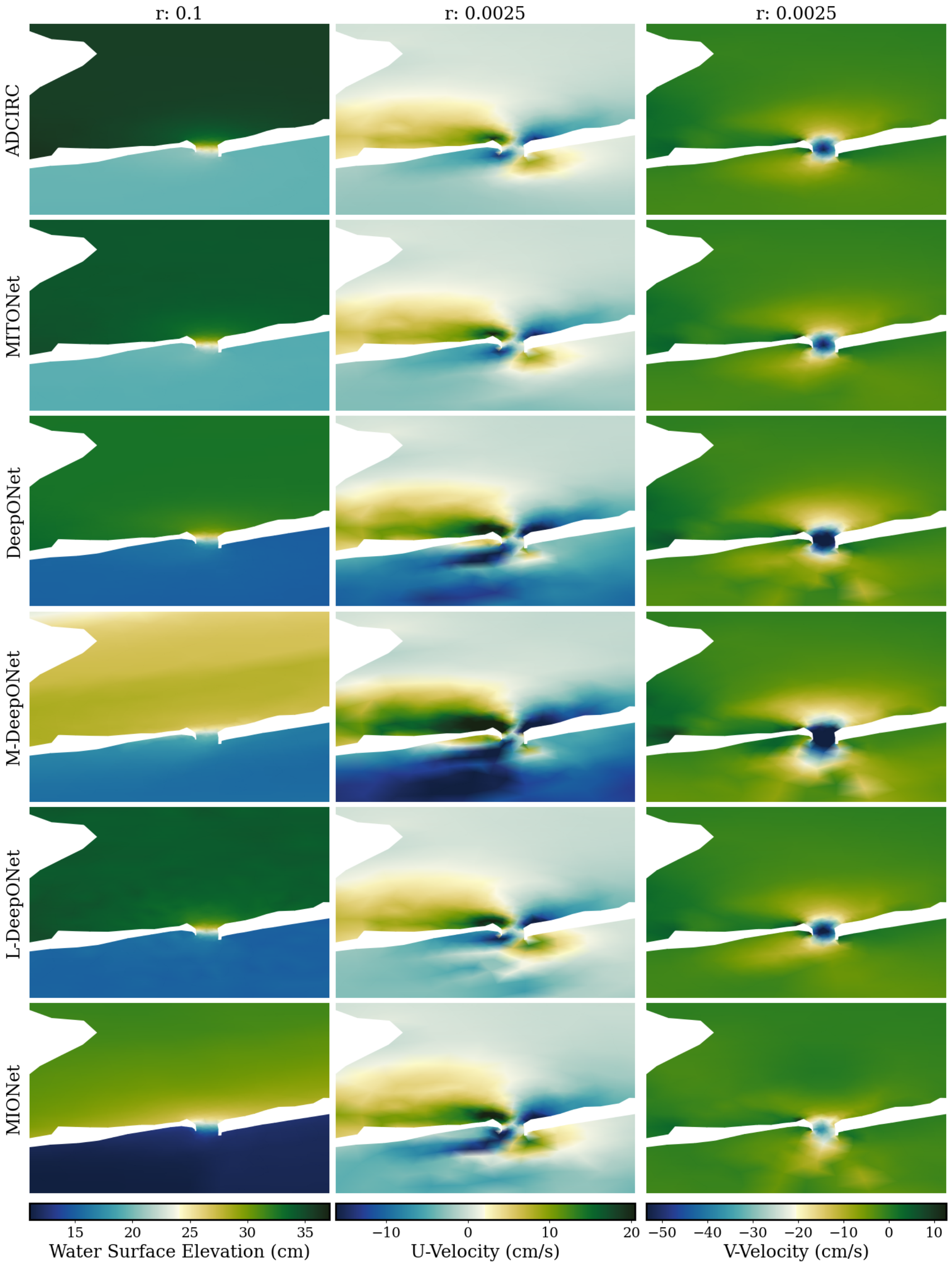}
    \caption{Zoomed-in comparison of ADCIRC and all five models for the Shinnecock Inlet example on day 45, following 40 days of autoregressive predictions. Columns display the variables $H$, $U$, and $V$, while rows correspond to solutions from: (1) ADCIRC, (2) MITONet, (3) DeepONet, (4) L-DeepONet, (5) M-DeepONet, and (6) MIONet. The bottom friction coefficient used are $r = 0.1$ for $H$ and $r = 0.0025$ for $U$ and $V$.}
    \label{fig:benchmark_zoom_snaps}
\end{figure}

\begin{figure}[h]
    \centering
    \includegraphics[width=.85\textwidth]{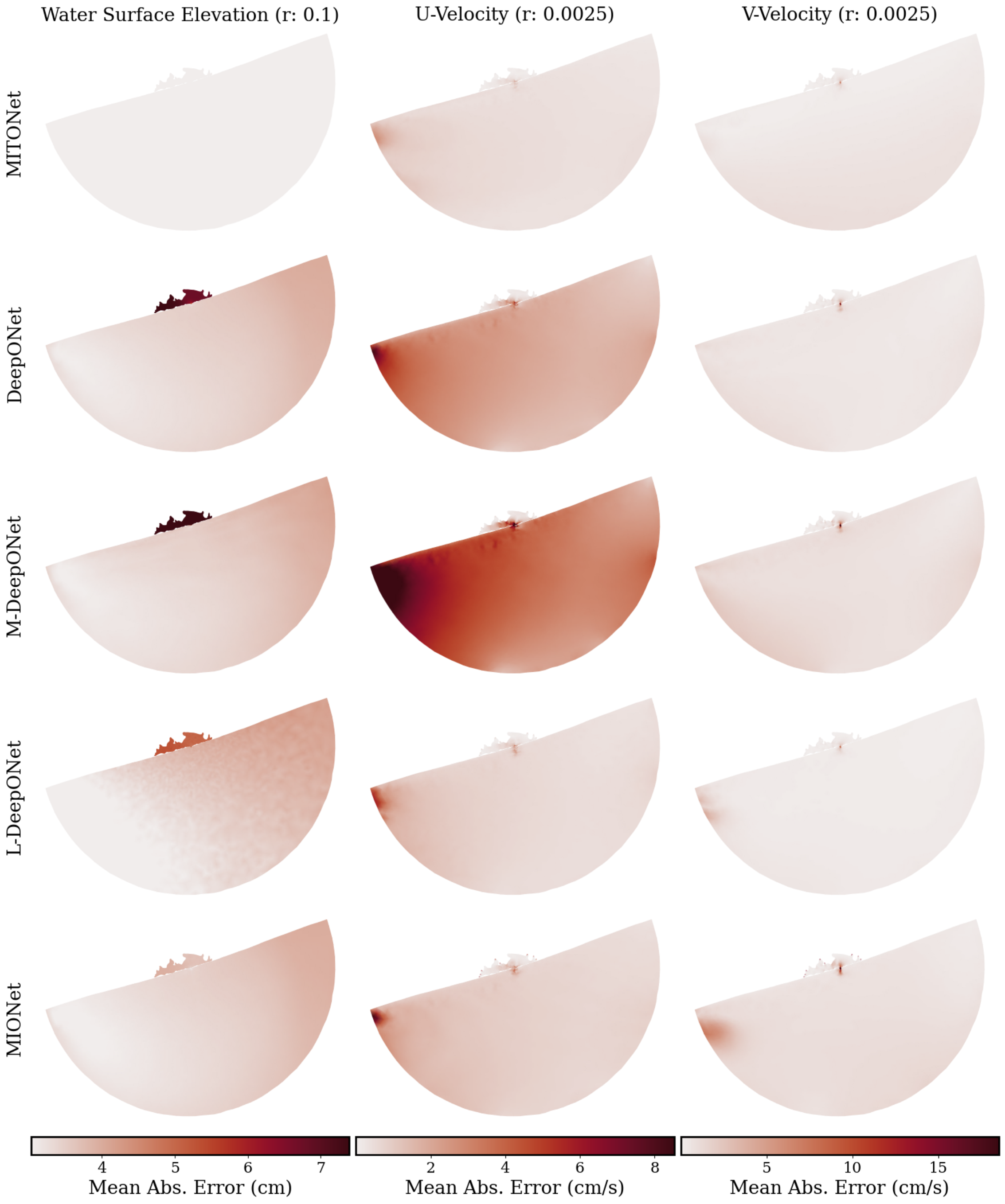}
    \caption{Full-domain spatial distribution of Mean Absolute Error (MAE) for the five competing models, calculated with respect to the ADCIRC simulation over 40 days of autoregressive predictions. Columns display the error for variables $H$, $U$, and $V$, while rows correspond to the results for: (1) MITONet, (2) DeepONet, (3) L-DeepONet, (4) M-DeepONet, and (5) MIONet. The bottom friction coefficient used are $r = 0.1$ for $H$ and $r = 0.0025$ for $U$ and $V$.}
    \label{fig:MAE_benchmark_full_snaps}
\end{figure}

\begin{figure}[h]
    \centering
    \includegraphics[width=.85\textwidth]{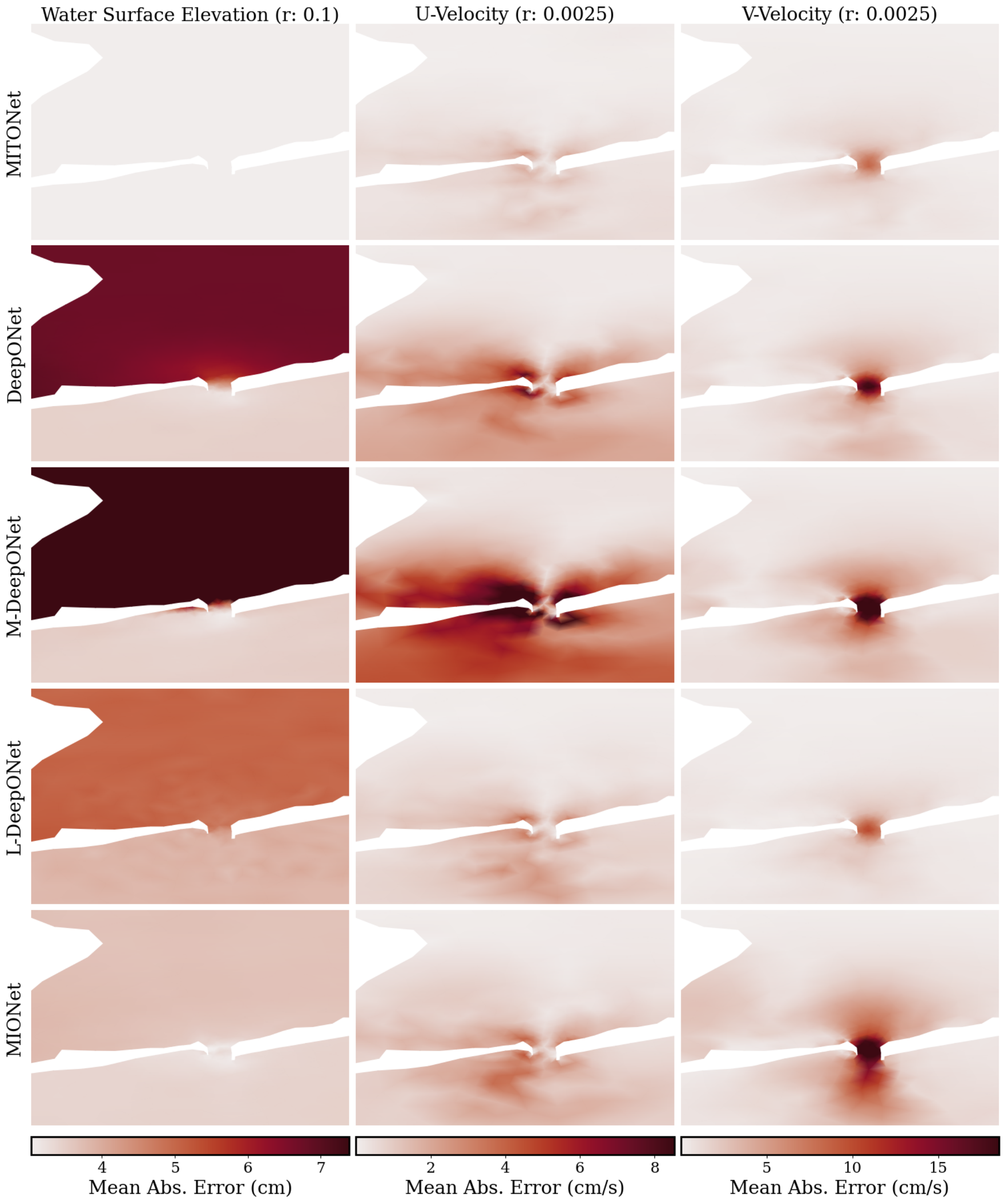}
    \caption{Zoomed-in spatial distribution of Mean Absolute Error (MAE) for the five competing models, calculated with respect to the ADCIRC simulation over 40 days of autoregressive predictions. Columns display the error for variables $H$, $U$, and $V$, while rows correspond to the results for: (1) MITONet, (2) DeepONet, (3) L-DeepONet, (4) M-DeepONet, and (5) MIONet. The bottom friction coefficient used are $r = 0.1$ for $H$ and $r = 0.0025$ for $U$ and $V$.}
    \label{fig:MAE_benchmark_zoom_snaps}
\end{figure}

\begin{figure}[h]
    \centering
    \includegraphics[width=\textwidth]{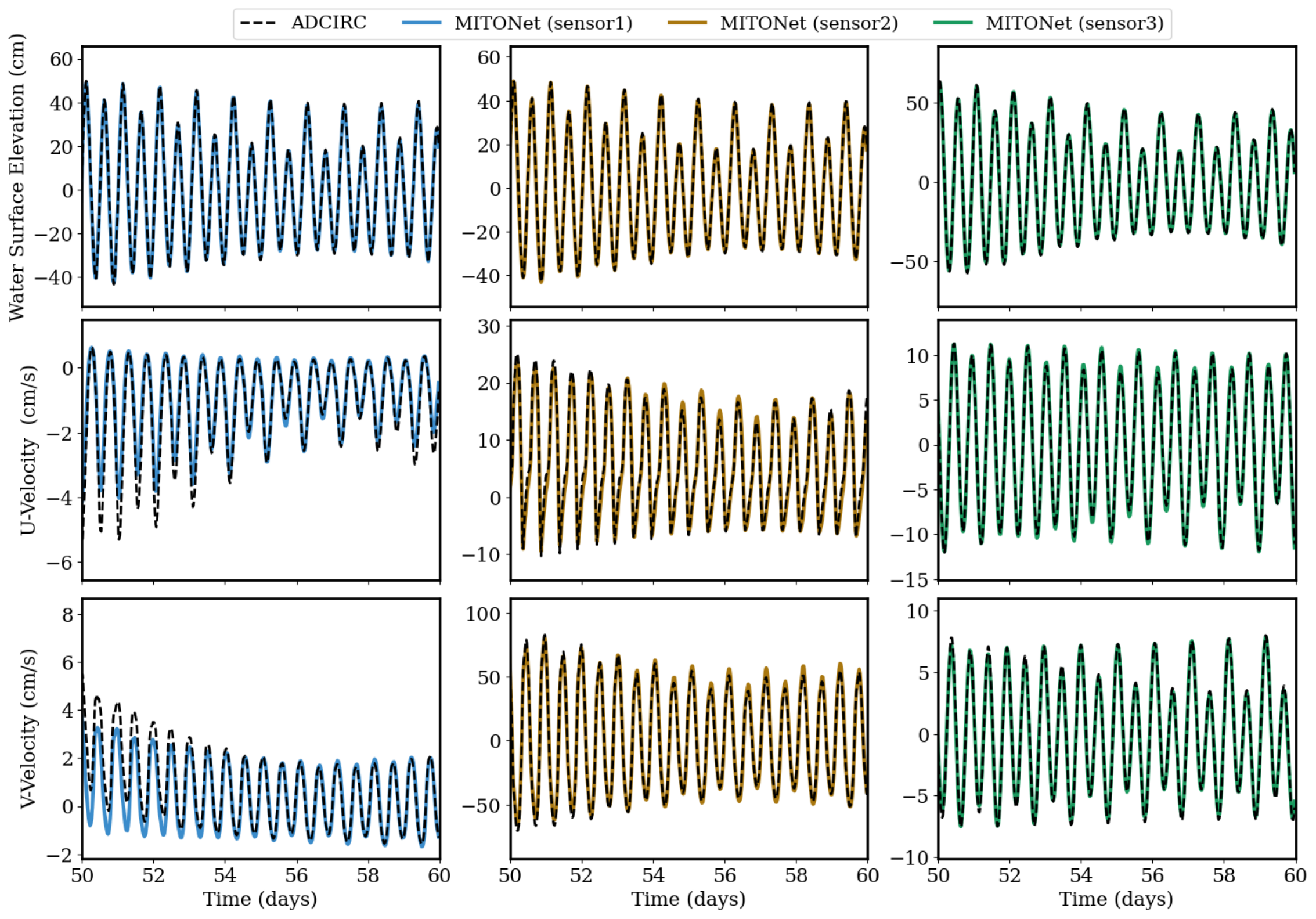}
    \caption{Comparison of the ADCIRC solution (black dashed lines) with the MITONet predictions (blue, yellow, and green lines) for the Shinnecock Inlet example at three different sensor locations (see Figure 2 in the article) between days $50$ to $60$ using test parameter $r=0.0025$ for $H$, $U$, and $V$ (Rows 1, 2, and 3, respectively)}
    \label{fig:shinnecock_sensors_r0025}
\end{figure}

\begin{figure}[h]
    \centering
    \includegraphics[width=\textwidth]{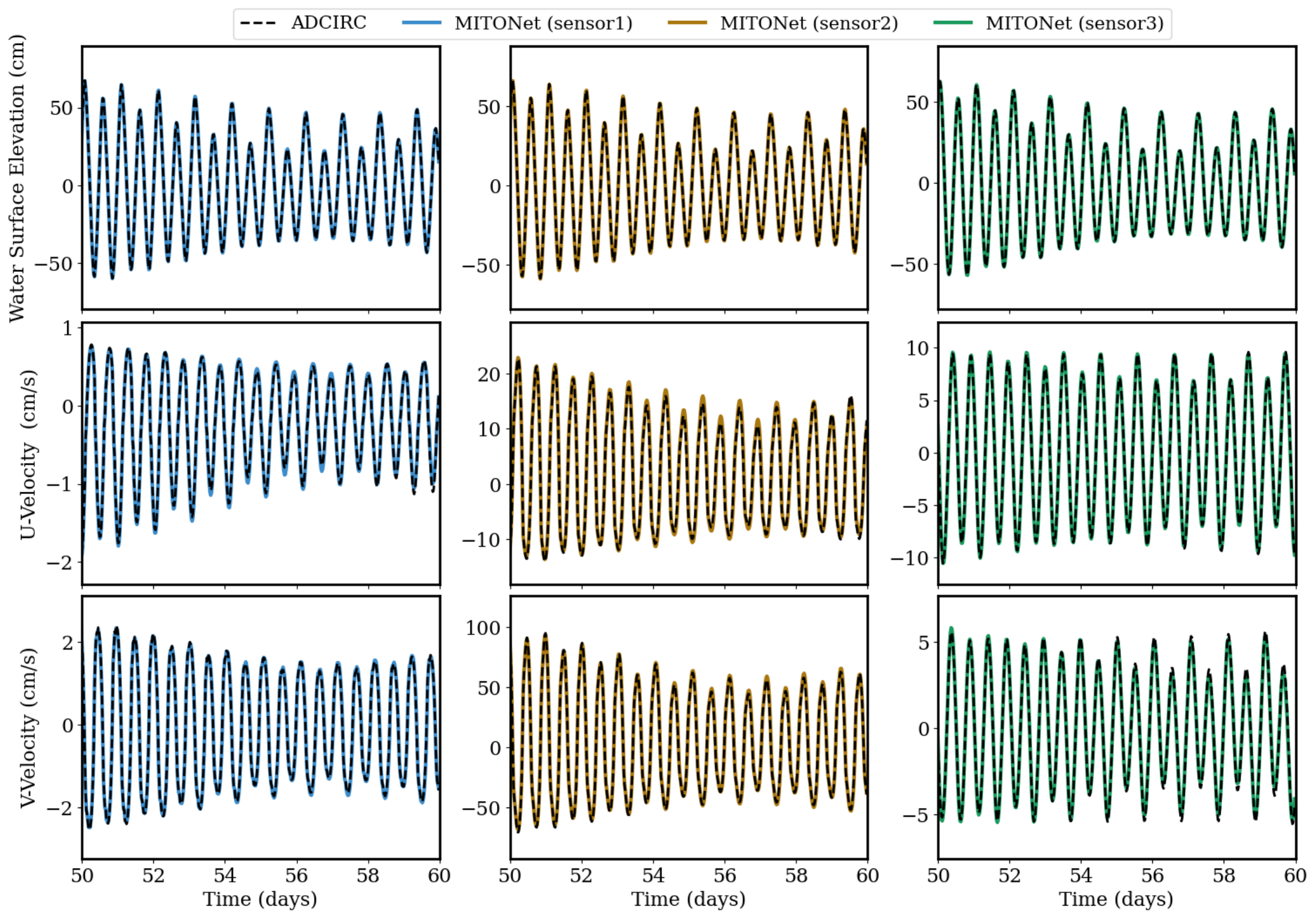}
    \caption{Comparison of the ADCIRC solution (black dashed lines) with the MITONet predictions (blue, yellow, and green lines) for the Shinnecock Inlet example at three different sensor locations (see Figure 2 in the article) between days $50$ to $60$ using test parameter $r=0.015$ for $H$, $U$, and $V$ (Rows 1, 2, and 3, respectively)}
    \label{fig:shinnecock_sensors_r015}
\end{figure}

\begin{figure}[h]
    \centering
    \includegraphics[width=\textwidth]{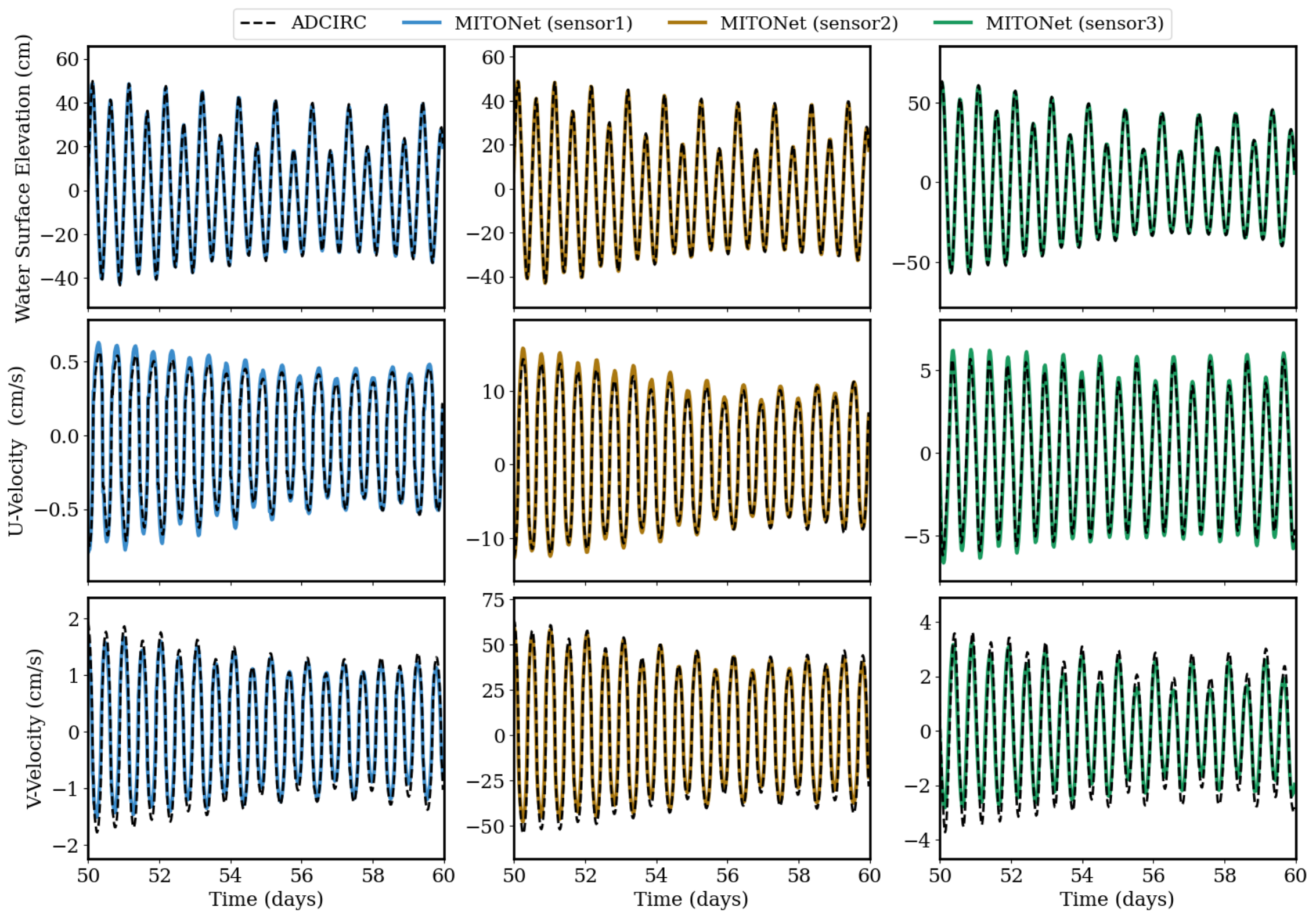}
    \caption{Comparison of the ADCIRC solution (black dashed lines) with the MITONet predictions (blue, yellow, and green lines) for the Shinnecock Inlet example at three different sensor locations (see Figure 2 in the article) between days $50$ to $60$ using test parameter $r=0.1$ for $H$, $U$, and $V$ (Rows 1, 2, and 3, respectively)}
    \label{fig:shinnecock_sensors_r1}
\end{figure}

\begin{figure}[h]
    \centering
    \includegraphics[width=\textwidth]{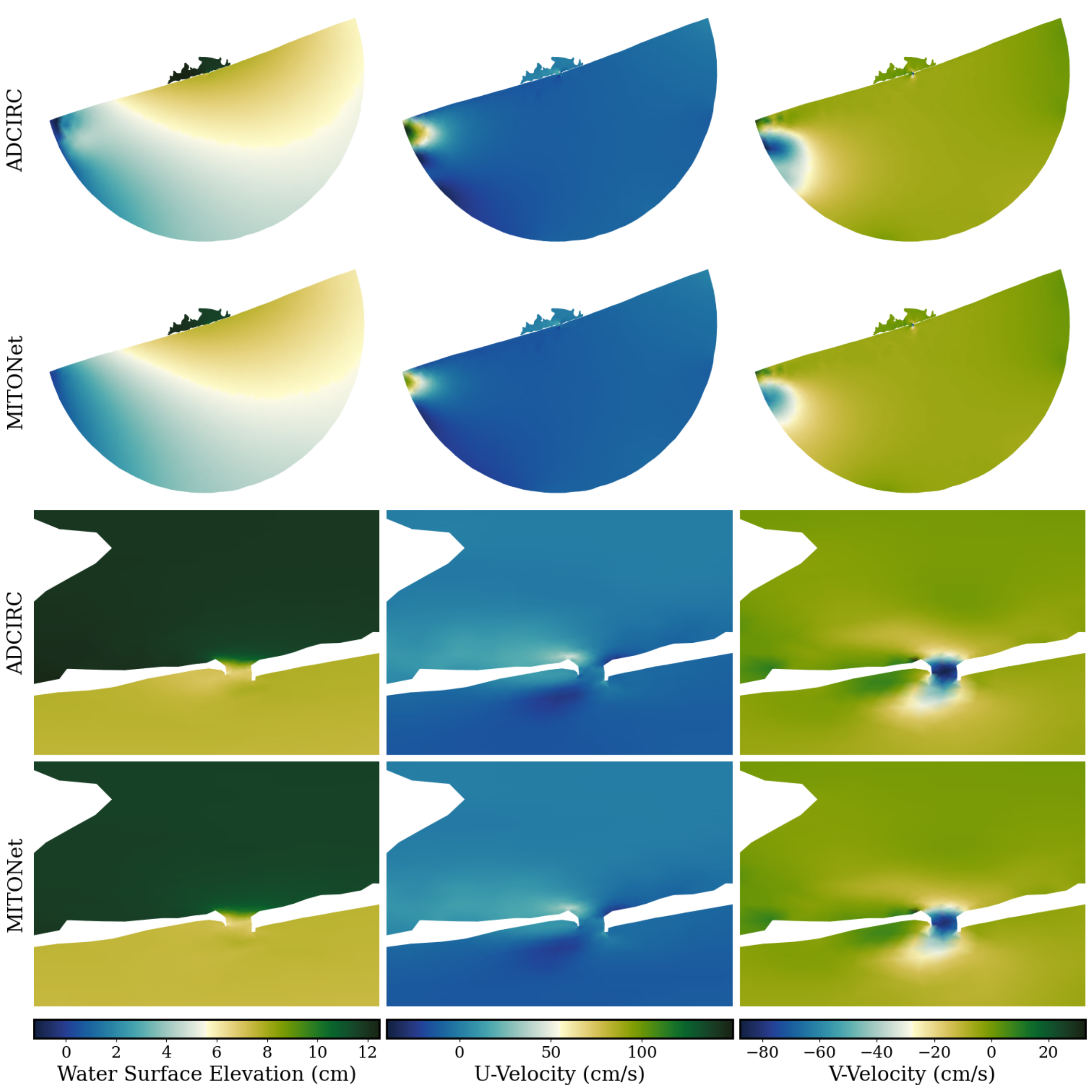}
    \caption{Full-domain and zoomed-in snapshots of ADCIRC solutions (Rows 1 and 3) and MITONet predictions (Rows 2 and 4) of variables $H$, $U$, and $V$ (Columns 1, 2, and 3) for the Shinnecock Inlet example. These results correspond to $r = 0.0025$ on day 60, following 55 days of autoregressive predictions.}
    \label{fig:rmse_snaps_0025_2}
\end{figure}

\begin{figure}[h]
    \centering
    \includegraphics[width=\textwidth]{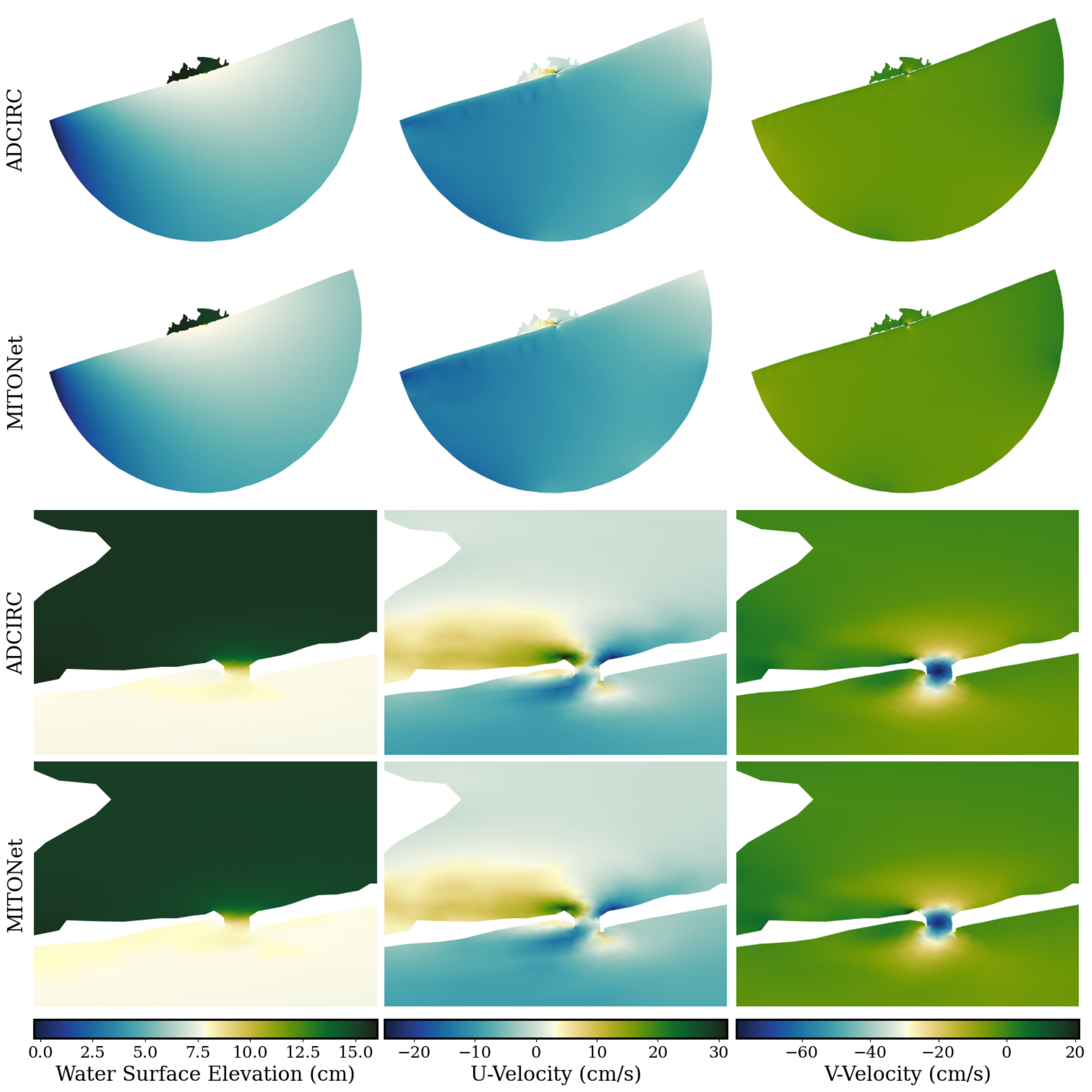}
    \caption{Full-domain and zoomed-in snapshots of ADCIRC solutions (Rows 1 and 3) and MITONet predictions (Rows 2 and 4) of variables $H$, $U$, and $V$ (Columns 1, 2, and 3) for the Shinnecock Inlet example. These results correspond to $r = 0.015$ on day 60, following 55 days of autoregressive predictions.}
    \label{fig:rmse_snaps_015_2}
\end{figure}

\begin{figure}[h]
    \centering
    \includegraphics[width=\textwidth]{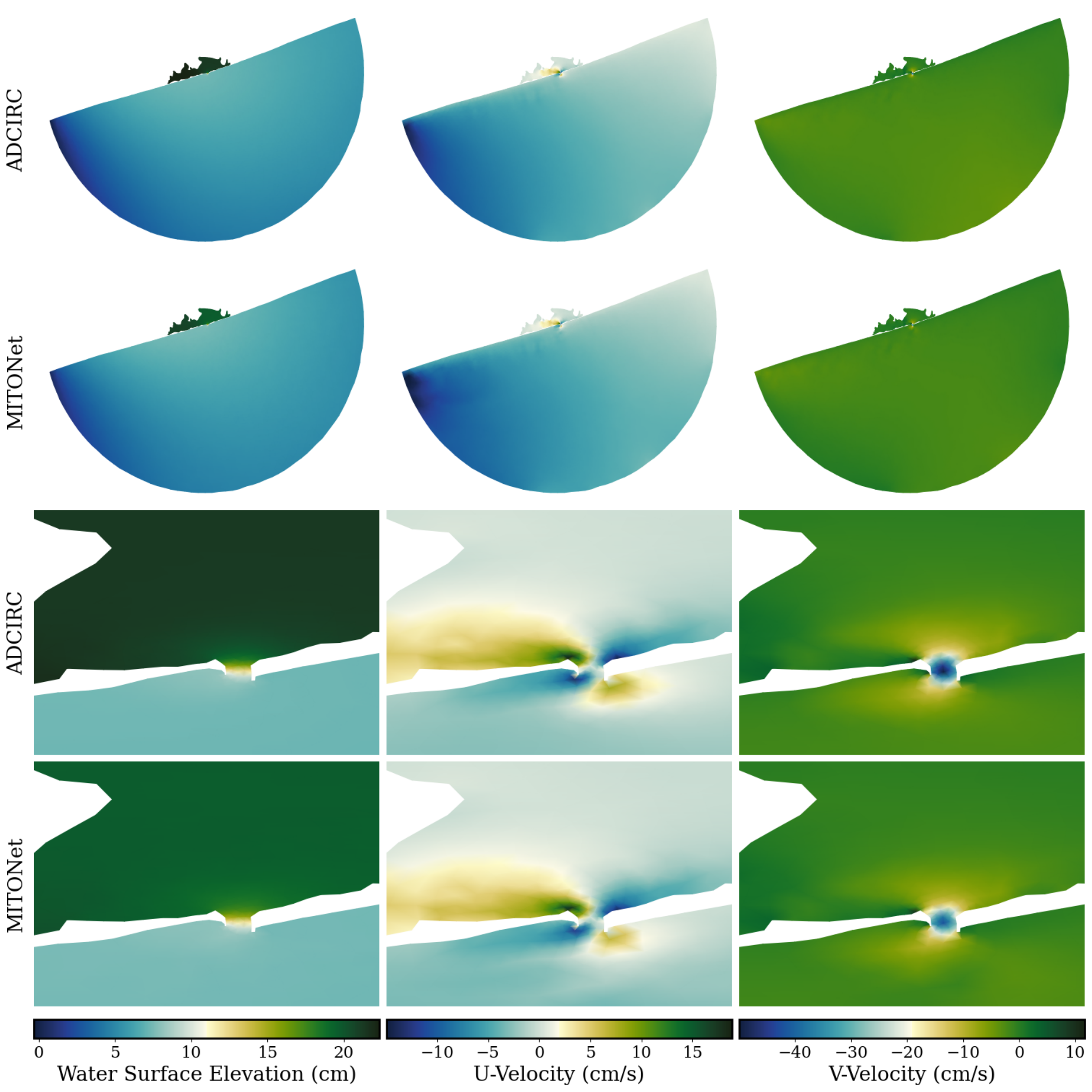}
    \caption{Full-domain and zoomed-in snapshots of ADCIRC solutions (Rows 1 and 3) and MITONet predictions (Rows 2 and 4) of variables $H$, $U$, and $V$ (Columns 1, 2, and 3) for the Shinnecock Inlet example. These results correspond to $r = 0.1$ on day 60, following 55 days of autoregressive predictions.}
    \label{fig:rmse_snaps_1_2}
\end{figure}

\begin{figure}[h]
    \centering
    \includegraphics[width=\textwidth]{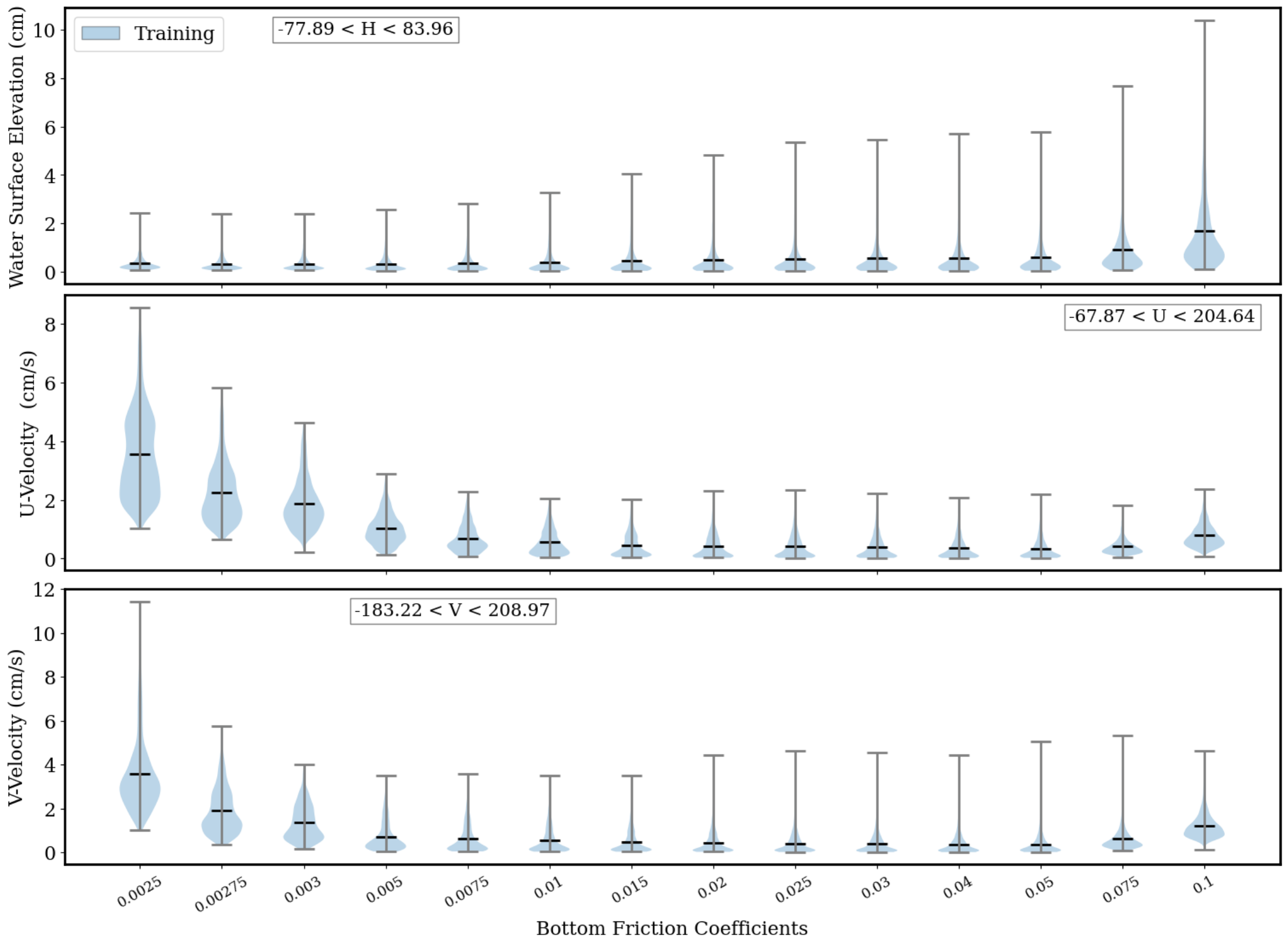}
    \caption{Violin plots of the $RMSE$ for MITONet models trained on all $r$ values of the Shinnecock Inlet example. The RMSE values are computed using predictions of $H$ (Row 1), $U$ (Row 2), and $V$ (Row 3)  between days $5-60$ and for all bottom friction coefficient ($r$) values. Each subplot includes a textbox indicating the range of the corresponding ground truth solution over the spatio-temporal domain.}
    \label{fig:violin_all}
\end{figure}

\end{document}